\documentclass[12pt]{article}
\usepackage{amssymb}
\usepackage{graphics,graphicx}
\usepackage{amssymb,epsfig,amsmath,euscript,array}
\usepackage{subfigure, placeins, float}
\usepackage{url}
\usepackage{pstricks}
\usepackage{color, dsfont}

\usepackage[nosort, nocompress]{cite}

\makeatletter \@addtoreset{equation}{section} \makeatother

\newcounter{multieqs}

\newcommand{\be}{\begin{equation}}
\newcommand{\ee}{\end{equation}}
\newcommand{\bea}{\begin{eqnarray}}
\newcommand{\eea}{\end{eqnarray}}

\begin{document}

\begin{flushright}
\end{flushright}

\vspace{8pt}

\begin{center}

{\Large \bf Y-system for form factors at strong coupling in $AdS_5$ and with multi-operator insertions in $AdS_3$}

\vspace{40pt}

{\mbox {\bf  Zhiquan Gao$^{^{a, b}}$ and   Gang Yang$^{^c}$}}

\vspace{10pt}

\begin{quote}
{\small \em
\begin{itemize}
\item[\ \ \ \ \ \ $^a$]
\begin{flushleft}
State Key Laboratory of Theoretical Physics,\\
Institute of Theoretical Physics,
Chinese Academy of Sciences, \\
Beijing 100190, China%
\footnote{ {\tt  zhiquan@itp.ac.cn} }
\end{flushleft}
\item[\ \ \ \ \ \ $^b$]
Kavli Institute for Theoretical Physics China, \\
Beijing 100190, China
\item[\ \ \ \ \ \ $^c$]
II. Institut f\"ur Theoretische Physik, Universit\"at
Hamburg\\ Luruper Chaussee 149, D-22761 Hamburg, Germany%
\footnote{
{\tt gang.yang@desy.de}
}

\end{itemize}
}
\end{quote}

\end{center}

\begin{center}
\vspace{60pt} {\bf Abstract}

\end{center}

\noindent
We study form factors in ${\cal N}$=4 SYM at strong coupling in general kinematics and with multi-operator insertions by using gauge/string duality and integrability techniques. This generalizes the $AdS_3$ results of Maldacena and Zhiboedov in two non-trivial aspects. The first generalization to $AdS_5$ space was motivated by its potential connection to strong coupling Higgs-to-three-gluons amplitudes in QCD which was observed recently at weak coupling. The second generalization to multi-operator insertions was motivated as a step towards applying on-shell techniques to compute correlation functions at strong coupling. In this picture, each operator is associated to a monodromy condition on the cusp solutions.
We construct Y-systems for both cases. The $Y$-functions are related to the spacetime (cross) ratios. Their WKB approximations based on a rational function $P(z)$ are also studied.
We focus on the short operators, while the prescription is hopefully also applicable for more general operators.

\noindent

\setcounter{page}{0} \thispagestyle{empty}
\newpage


\setcounter{tocdepth}{4}
\hrule height 0.75pt
\tableofcontents
\vspace{0.8cm}
\hrule height 0.75pt \vspace{0.5cm}

\setcounter{tocdepth}{2}

\setcounter{footnote}{0}

\setlength{\parskip}{2pt}
\section{Introduction}

One of the most challenging problems of modern theoretical physics is to understand the dynamics of strong coupling QCD analytically. While this is still very difficult, lots of studies have been focused on simpler models such as theories with supersymmetry. The general philosophy is that a good knowledge of these theories may finally help us to understand the real QCD.
A particularly interesting theory that has drawn much attention is the ${\cal N}$=4 super Yang-Mills theory. There has been some evidence that ${\cal N}$=4 SYM results are important building blocks of QCD quantities, see for example \cite{BDK04, KLOV04}.
By the gauge/string duality, it becomes also possible to study the ${\cal N}$=4 SYM in the strong coupling regime where it is dual to a perturbative or semi-classical string theory in an $AdS$ background \cite{Maldacena, GKP98, Witten98}.

An impressive achievement is that we can now compute anomalous dimensions in ${\cal N}=4$ SYM to any reasonable order in practice, see for example \cite{BES06, GKV09,BFT09,AF09,GKLV11}, where the integrability of the theory plays a fundamental role \cite{MSW02, MZ02, BPR03} (for a review on many aspects of integrability see \cite{IntegrabilityReview}).
It is expected that similar achievement may also be made for other more complicated observables, such as scattering amplitudes and correlation functions.
Indeed surprising dualities and integrable structures have been found for amplitudes and null Wilson loops \cite{AM, DKS07, BHT07, DHKS07,DHP09, MS10, CH11}, and also correlation functions in a special light-like limit \cite{AEKMS10, EKS10}.

One remarkable development is the computation of scattering amplitudes in ${\cal N}=4$ SYM at strong coupling \cite{AM}. It was shown that the problem can be dual to a string minimal surface problem in $AdS$. The solving of this non-trivial geometrical problem was developed based on the integrability of the classical worldsheet theory \cite{AM09, AGM09, AMSV10}\,%
\footnote{See also \cite{Yang10} for a treatment of tricky $4K$-gluon cases, \cite{BKS11, BSS12} for the study of Regge limit, and \cite{HISS10, HIS12} for the connection to CFT in the regular-polygon limit.}.
Hopefully, these classical results will be useful to solve the full quantum problem such as in the study of operator dimensions \cite{KMMZ04, AFS04}.

In this paper, we will focus on a more general class of observables, the so-called form factors.
They are observables involving both on-shell particles and off-shell operators, therefore are in some sense hybrids of amplitudes and correlation functions
\be \langle \textrm{Out-states} \, | \, \prod_i{\cal O}_i(x_i) \, | \, \textrm{In-states} \rangle \, . \nonumber \ee
We will consider form factors in pure momentum space
\be F(q_1, \cdots, q_l; p_1, \cdots, p_n) \ = \ \prod_{k=1}^l  \, \int d^4 x_l \, e^{iq_k \cdot x_k} \langle {\cal O}(x_1) \cdots {\cal O}(x_l) \, | \, p_1 \cdots p_n \rangle \, , \ee
where $p_i^2=0$ and $q_k$ is arbitrary.

Most studies so far have been focused on form factors with one operator inserted.  Form factors in string theory in $AdS$ were first studied in
\cite{PS01}. Based on the recent developments of strong coupling amplitudes, a T-dual picture of form factors was proposed in \cite{AM07}, and the problem was solved in the $AdS_3$ case by using integrability techniques in \cite{MZ10}. At weak coupling, form factors in ${\cal N}=4$ SYM were first studied in \cite{vanNeerven}, and have
received attention  recently, see for example \cite{BSTY10,
BKV10, BGMTY11, BKV11, GHH11}.
One  surprising observation in \cite{BTY} is that the remainder function of a two-loop three-point  form factor in ${\cal N}=4$ SYM matches exactly with the maximally transcendental part of the two-loop Higgs-to-3-gluon amplitudes in QCD \cite{GJGK11}\,%
\footnote{The relation between form factors and Higgs-gluons
amplitudes may be understood by noting that the operator in form
factor \cite{BTY} is equivalent to the Higgs-gluon effective vertex obtained
by integrating out a quark loop.}.

As this correspondence looks very intriguing, one may think that this is an accidental coincidence. However, this two-loop coincidence is already rather non-trivial, which may be appreciated  by a simple look at the very different perturbative structures of Feynman diagrams in ${\cal N}=4$ SYM and QCD. It may be therefore reasonable to expect that there could be some hidden relations which will explain this coincidence and might play further roles for other situations, at least for the three-point case due to its particularly simple kinematics\,%
\footnote{It would therefore be interesting to study three-loop
case. Hopefully the progress can be made in ${\cal N}=4$ side, as in
\cite{BTY} (also with the techniques developed in \cite{BKTY12}), while the computation in QCD
seems much more challenging.}.
If the two-loop coincidence is going to be true for higher loops,
one may expect that strong coupling form factors in ${\cal N}=4$  would carry a
non-trivial piece of information of strong coupling QCD. Considering
that there are very few tools to study strong coupling QCD amplitudes,
this possibility provides us enough motivation to study  strong
coupling form factors seriously.

The computation of form factors at strong coupling in \cite{MZ10}
was restricted to two dimensional kinematics. In such case  non-trivial
quantities start at four-point.  In order to study the three-point
form factor, one needs to consider more general kinematics. In this
paper we will consider the form factors in full $R^{1,3}$
kinematics, corresponding to string in $AdS_5$.
As is usually happened, the generalization from $AdS_3$ to $AdS_5$
is a nontrivial step. Although the underlying picture is similar to
the $AdS_3$ case, the monodromy structure in $AdS_5$ is more
complicated. In particular, the truncation conditions involve small
solution contractions which are not $T$-functions. 
This complexity also makes it much more difficult to construct the Y-system, which is a main new challenge of the $AdS_5$ problem. We will describe the  general construction, and the Y-system for the three-point form factor will be explicitly given.

Another interesting generalization in this paper is to compute
form factors with multi-operator insertions.
The main motivation is to study correlation functions at strong coupling with the help of on-shell techniques. Similar idea has been used at weak coupling in \cite{ER12}.
Although the observables we consider contain on-shell structures, they involve multiple operators, and in principle should contain all kinds of information of correlation functions. In particular, one should be able to extract the OPE coefficients from form factors containing two or more operators.

The basic idea we propose may be illustrated by the following flow chart\,%
\footnote{One may note that this is different from the logic used in
computing anomalous dimensions via Y-system. Here it is important to
obtain the Y-system, where the Y-functions are interpreted as the
spacetime cross ratios, and for which the boundary condition can be
conveniently introduced. }
\begin{center}
$\begin{matrix}
\begin{tabular}{| c |}
\hline $\begin{matrix}  \, \\ \,  \end{matrix}$  Cusps  \cr \hline
\end{tabular}
& \Rightarrow &
\begin{tabular}{|c|}
\hline $\begin{matrix}  \, \\ \,  \end{matrix}$  Small solutions  \cr \hline
\end{tabular}
& \hskip-0.5cm \Rightarrow &
 \begin{tabular}{|c|}
\hline $\begin{matrix}  \, \\ \,  \end{matrix}$  Hirota system  \cr \hline
\end{tabular}
\Rightarrow
 \begin{tabular}{|c|}
\hline $\begin{matrix}  \, \\ \,  \end{matrix}$  Y-system  \cr \hline
\end{tabular}
\\ & & \Uparrow & & \\
\begin{tabular}{| c |}
\hline $\begin{matrix}  \, \\ \,  \end{matrix}$  Operators  \cr \hline
\end{tabular}
& \Rightarrow &
\begin{tabular}{|c|}
\hline $\begin{matrix}  \, \\ \,  \end{matrix}$  Monodromy matrices  \cr \hline
\end{tabular} \,
& &
 \end{matrix}$ 
 \end{center}
The main picture is that  for each operator one can define a
corresponding monodromy matrix, which will give a linear relation for
the small solutions. These small solutions are related to
the cusps and are the same building blocks for calculating amplitudes,
therefore the known method of computing amplitudes can be applied to
these more general class of observables.  It is in this sense that
we can compute off-shell observables by using on-shell techniques.

We derive explicitly the Y-system for form factors with
multi-operator insertions in $AdS_3$, while in principle
it should be possible to generalize to the $AdS_5$ case.
The construction proposed in this paper is expected to be in principle applicable to arbitrary operators, although the study will be focused on light operators\,%
\footnote{These include short BPS operators such as the stress tensor supermultiplets which are studied in form factors at weak coupling, and also non-protected light operators with dimensions $\propto \lambda^{1/4}$, for example the Konish operator.}, for which the monodromy can be given explicitly.

This paper is organized as follows. In Section \ref{sec-review}, we
review the main physical pictures and the general strategy of strong
coupling computation via AdS/CFT and integrability.  We then review
form factors in the $AdS_3$ case in section \ref{sec-ffads3}.  Form
factors in general $AdS_5$ kinematics are developed in section
\ref{sec-ffads5}, and the three-point case is discussed  explicitly in section \ref{sec-3ptFF}. The generalization to multi-operator
insertions is given in section \ref{sec-ffmulti}. In section
\ref{sec-ppoly}, the $P(z)$ function and WKB approximation are
studied. Section  \ref{sec-summary} contains a summary and some discussions. There are three appendices.  Appendix \ref{app-funeqs} is a collection of the definition of $T$- and $Y$-functions and their corresponding equations.  A  review of (momentum)
twistor variables is given in appendix \ref{app-twistor}. Appendix \ref{app-omegabar} is a brief discussion of the monodromy in a different basis.

\section{Classical string and integrable system \label{sec-review} }

Due to the nature of the problem which involves quite a few different
stories and intermediate steps, in this section we give a
brief review of the whole picture. The
discussion here is not supposed to be self-contained, but  we hope
to cover the key physical pictures and central ideas. Interested reader is referred to the original papers (in
particular \cite{AM09, AGM09}) for more details.

\subsection{Form factor as a classical string solution}

As a first step to set up the problem, we explain how to map the computation of amplitudes and form factors at strong coupling in ${\cal N}=4$ SYM  to a classical string problem in an $AdS$ background  \cite{AM, AM07}.

We first consider the picture for gluon states. Recall the $AdS$ space in Poincar\'{e} coordinate
\be d s^2 \ = \ {dy^\mu dy_\mu + d z^2 \over z^2} \, .\ee
Gluon states in ${\cal N}=4$ SYM are dual to open strings on the IR D3 branes (as an IR regulator) at the horizon (i.e. $z\rightarrow\infty$) \cite{AM}. One important property of the open strings on  IR branes is that they carry very large proper momenta. Because high energy scattering is dominated by a saddle point approximation \cite{GM88}, the computation of open string amplitude becomes a classical string problem.

\begin{figure}[t]
\begin{center}

\includegraphics[height=5.2cm]{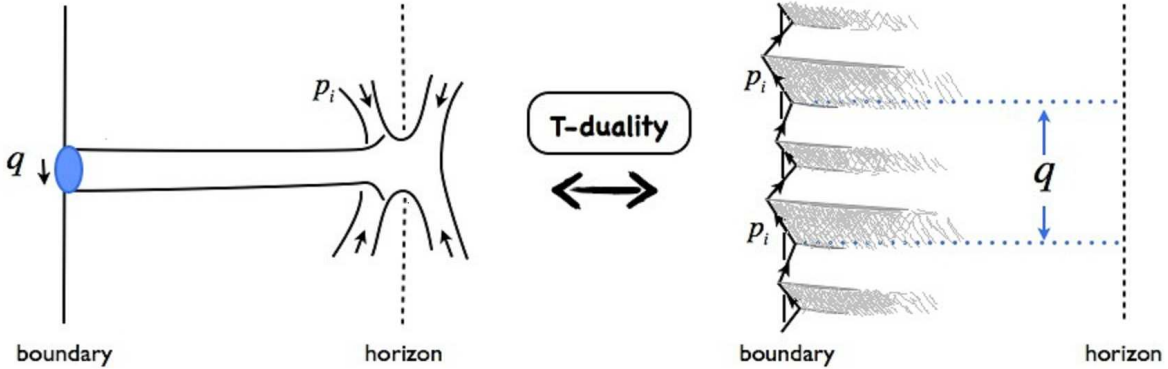}
\caption{\it The picture of T-duality for form factor. $q$ is the momentum of the operator which corresponds to a closed string state in the bulk. $q= \sum_i p_i$ due to momentum conservation. After T-duality, the picture becomes a minimal surface ending on a periodic null Wilson line at the boundary and extending to horizon. The period is given by $q$.}
\label{Tdual}

\end{center}
\end{figure}

Form factors also contain operators, which are dual to closed string states in the bulk with boundary condition at $z \rightarrow0$ \cite{GKP98, Witten98}. Therefore form factors correspond to scattering open and closed strings which are  from the horizon and the boundary respectively, as shown on the left-hand side of Figure \ref{Tdual}\,%
\footnote{It is assumed that the scattering is still dominated by the classical saddle point.}.

To simplify the problem, one important trick is to apply a {\it formal} T-duality along $y^\mu$ directions \cite{KT98, AM}\,%
\footnote{This is in the sense of using Buscher's formalism defined at action level \cite{Buscher}, in which it is also straightforward to generalize to fermionic directions \cite{BM08, BRTW08}.}. The T-dual space is still an $AdS$ space
\be d s_{\textrm{T-dual}}^2 \ = \ {dx^\mu dx_\mu + d r^2 \over r^2} \, , \ee
where $r = {1/ z}$. The boundary and horizon reverse their roles in the T-dual space. The momenta of strings become the ``windings" of strings. For amplitudes the problem becomes a type of Wilson loop problem \cite{Maldacena98, RY98},  with a null polygonal boundary. For form factors, the boundary becomes a periodic null Wilson line \cite{AM07}, where the period is determined by the momentum of the closed string $q$. The minimal surface extends to the horizon, as illustrated on the right-hand side of Figure \ref{Tdual}\,%
\footnote{It is also obvious that a mixing of Wilson loop and operators such as studied in \cite{ABT11} is very different from the form factors we consider here.}.

Therefore, the form factor problem becomes to find the area of the minimal surface over one period, with boundary conditions at both the boundary and the horizon of the T-dual $AdS$ space.
The general structure of the strong coupling results is
\be {\rm Observable} \ = \ e^{-{\sqrt{\lambda}\over 2\pi} \, {\rm Area}} \times ({\rm string} \ \alpha' \ {\rm corrections}) \,, \qquad \alpha' \sim {1\over\sqrt{\lambda}} \, . \ee
The string corrections in principle may be computed by considering string fluctuations where the classical solution is taking as a background, along the line of \cite{FT02, KRTT07}\,%
\footnote{It seems no such computation has been done for any solutions corresponding to amplitudes, even for the simplest four-point case, where both the classical solution \cite{AM} and the result (given by ABDK/BDS ansatz \cite{ABDK,BDS}) are  known. The pure spinor formalism \cite{Berkovits} might be useful for such computations.}.

\subsection{String in $AdS$ as a classical integrable system}

Because of  the non-trivial boundary conditions,  it is very hard to solve the string equations. The idea, proposed in \cite{AM09} (see also for example \cite{DS93, JJKV07, DJW09}), is that rather than solving the string equations directly, one can apply the Pohlmeyer's reduction \cite{Pohlmeyer} to reformulate the string equations to a Hitchin like system and then use the techniques of integrability. Here we briefly review the main strategy.

Since Pohlmeyer reduction is a well-understood procedure, we only point out that after the reduction, the string equations of motion and Virasoro constraints take a form of flat equation
\be \label{consis-A} \partial {\cal A}_{\bar z} - \bar\partial {\cal
A}_z + [{\cal A}_z , {\cal A}_{\bar z}] \ = \ 0 \, . \ee
If one decomposes ${\cal A}$ into two parts ${\cal A} = A + \Phi$,  the equations form a Hitchin like system
\be \label{Hitchin} D_z \Phi_{\bar z} \, = \, 0 \, , \qquad D_{\bar z} \Phi_z \, = \, 0 \, , \qquad [D_z, D_{\bar z}] + [\Phi_z, \Phi_{\bar z}]  \, = \, 0 \, , \ee
where $D_z := \partial_z + [A_z , \ ]$.
For $AdS_5$ case it is a $SU(4)$ system\,%
\footnote{Here we have changed to the spinor representation of $SO(2,4)$ \cite{AGM09}. This change of representation is equivalent to the using of momentum twistor variables at weak coupling \cite{Hodges}.}, while for $AdS_3$ it can be reduced to $SL(2)$. The flat connection is not arbitrary but  satisfies a $Z_4$ automorphism
\be A \, = \, -C \, A^T \, C^{-1} \, , \qquad  \Phi_z \, = \, -iC \, \Phi_z^T \, C^{-1} \, , \qquad
\Phi_{\bar{z}} \, = \, iC \, \Phi_{\bar{z}}^T \, C^{-1} \, , \ee
where $C$ is a constant matrix whose explicit form is not important here. This $Z_4$ constraint plays an important role in the construction as we will see later.  One can solve the linear equation
\be (d + {\cal A}) \, \psi \, = \, 0 \, ,
\ee
where the solution $\psi$ is related to the target space coordinates and therefore to the string solutions.

A natural logic would be to first find the solution for ${\cal A}$ which solves the Hitchin equations,  then solve the linear problem to find the solution $\psi$ which gives the classical string solution.  However, the strategy used here is different. 
Roughly speaking, we will use the properties of the linear solution and the flat connection
to construct the area directly without knowing the explicit solution.

The key idea is to use integrability. The integrability can be understood by the fact that one can lift the flat connection to a family of connections
\be \label{Azeta}  {\cal A} \ \ \rightarrow \ \ {\cal A}(\zeta) \ = \  \left( A_z+{1 \over \zeta}\Phi_z \right) d z + \left( A_{\bar{z}}+ \zeta\Phi_{\bar{z}} \right) d \bar z \, , \ee
while the Hitchin equations are still satisfied. The new parameter $\zeta$ is called {\it spectral parameter}. We also use another variable $\theta$ where $\zeta = e^{i \theta}$. If one solves the linear problem with ${\cal A}(\zeta)$, one obtains a one-parameter family of solutions $\psi(\zeta)$, and the original physical solution can be obtained by taking $\zeta=1$.

With this extra parameter it seems one is dealing with a more general problem. However, new powerful techniques are available based on this new parameter\,%
 \footnote{We would like to point out that the idea of introducing new parameters has played many other important roles in theoretical physics, such as the $\Omega$-deformation in localization techniques and the orbifold generalization in ABJM theory, see for example the talk by John Schwarz \cite{Schwarz}. It would be very interesting to study their possible connection to integrability.}.
The main result is that a set of functional equations, so-called
Y-system, can be constructed.
The non-trivial part of the area can be extracted from the solution of Y-system, which turns out to be the free energy in a thermodynamic Bethe ansatz (TBA)
form \cite{YY69, Zamolodchikov90}. For amplitudes in $AdS_5$ it is
like \cite{AMSV10}
\be A_{\rm free} \ = \ \sum_s { m_s \over 2\pi}
\int d\theta \cosh\theta \log \left[(1+
Y_{1,s})(1+ Y_{3,s})(1+ Y_{2,s})^{\sqrt{2}}\right] \, . \ee
In this form,  area is a function of mass parameters which are implicitly related to the physical cross ratios. An important observation later in \cite{AGMSV10} is that the area can also be written as the critical value of Yang-Yang functional, and in the new form area can be expressed directly as a function of cross ratios.

\subsection{Boundary conditions and function $P(z)$ \label{sec2-pfun}}

In this section we explain an important aspect of the story: how to formulate the boundary conditions. This will involve an important holomorphic function $P(z)$. We also discuss the special feature of form factors in which an operator is inserted.

One particular equation of the Hitchin system is the generalized sinh-Gorden equation \cite{AGM09}
\be  \partial\bar\partial \alpha - e^\alpha - e^{-\alpha}  |P(z)| \, = \, 0 \, ,  \ee
where $P$ and $\alpha$ are given as
\be \label{defP} P \, = \, \partial^2 X \cdot \partial^2 X \, , \qquad \alpha \, = \, \log(\partial X \cdot \bar\partial X) \, . \ee
$P(z)$ is a holomorphic function. By making a field redefinition and introducing a new coordinate $w$ by a worldsheet conformal transformation as
\be \alpha(z,\bar z)  \ =\ \hat\alpha(z,\bar z) + {1\over4} \log
P(z)\bar P(\bar z) \, , \qquad  d w \ = \ (P(z))^{1/4} d z \ , \ee
one can simplify the generalized sinh-Gordon equation as
\be \label{sG-alphahat} \partial_w \, \bar\partial_{\bar w} \, \hat\alpha -
(e^{\hat\alpha} + e^{-\hat\alpha}) \, = \, 0 \, , \ee
which is a simple sinh-Gordon equation.
One should note that the change of worldsheet coordinate is only well-defined locally.

\begin{figure}[t]
\begin{center}

\includegraphics[height=3.5cm]{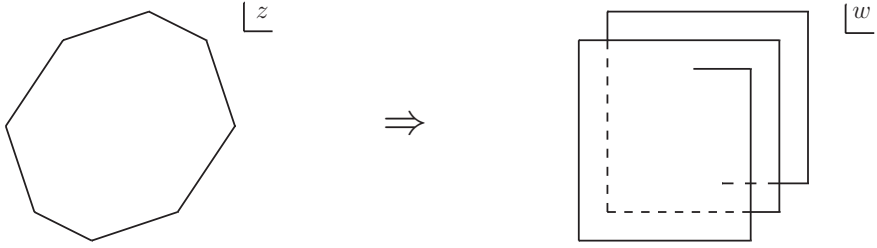}
\caption{\it $z$-plane and $w$-plane.}
\label{zandwplane}

\end{center}
\end{figure}

An important fact is that the four-cusp solution (first found in \cite{AM}) is simply the solution $P=1$ and  $\alpha = \hat\alpha=0$ of the generalized sinh-Gordon equation \cite{DJW09}. This is an important reason of doing the above transformation\,%
\footnote{It is also simpler to introduce cut-off and compute regularized area in $w$-plane \cite{AM09, AGM09}. }. Asymptotically, the solution near each cusp should be the same as the four-cusp solution. This implies that near boundary where $z \rightarrow \infty$, we should have $\hat\alpha \rightarrow0$. It also implies that each cover of $w$-plane contains four cusps. Therefore, $P(z)$ should be a polynomial, and the degree of the polynomial would depend on the number of cusps. The corresponding picture is shown in Figure \ref{zandwplane}.
The coefficients in the polynomial would encode the shape of the polygon.

\begin{figure}[t]
\begin{center}

\includegraphics[height=3.5cm]{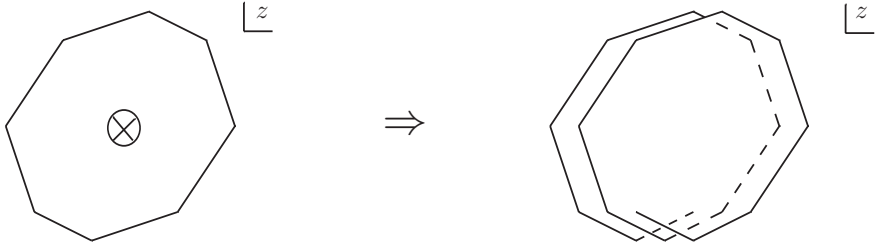}
\caption{\it The picture for form factor. An insertion of a operator introduces a singularity in the $z$-plane, and corresponds to a multi-cover of $z$-plane.}
\label{ffzplane}

\end{center}
\end{figure}

A new feature of form factors is that there are also operators. As observed in \cite{MZ10}, an insertion of an operator will introduce a pole term in $P(z)$. This requires a study of the boundary condition near the horizon, which will be discussed and generalized to $AdS_5$  in section \ref{sec-ppoly}. Due to the insertion of operator, the $z$-plane is no longer smooth. One can however smooth the $z$-plane with the sacrifice of introducing a multi-branch-cover of $z$-plane, as illustrated in Figure \ref{ffzplane}\,%
\footnote{This is in some sense similar to what happened in $w$-plane in the $4K$-cusp case \cite{AM09, Yang10}.}. This picture is consistent with the periodic null Wilson line picture in the target space. We will use this picture later to introduce the monodromy for small solutions.

\subsection{Small solutions}

Now we introduce a very important building block, the so-called {\it small solution}. Consider again the linear problem
\be \big(d + {\cal A}(\zeta,z) \big)\, \psi(\zeta,z) \, = \, 0 \, .
\ee
Because of the special null-cusp boundary conditions, the solution has different asymptotic behaviors near different cusps. This displays the so-called Stocks phenomenon. The asymptotic behavior of the solution when $z\rightarrow\infty$ is only valid within a given Stokes sector. Let us consider the simple $AdS_3$ case (similar picture applies for $AdS_5$). While approaching an edge $i$, the solution can be approximated as
\be \psi \ \sim \ c_i^{\rm big} \, S_i + c_i^{\rm small} \, s_i \, . \ee
Small solutions $s_i$ are the solutions which decay fastest while approaching the boundary. They are unique up to a normalization. At first sight, it may be confusing why it is the small solution rather than the big solution that is important.
However, it is not the big solution, but the {\it coefficient} of the big solution that
contains the boundary information. This coefficient can be extracted by
contracting the full solution with the small solution as $c_i^{\rm big}
\sim \langle \psi , s_i \rangle$. In this way, all non-trivial
boundary information can be obtained in terms of the contraction of
small solutions.

The coefficient $c_i^{\rm big}$ is related to target-space variable, the momentum twistor $\lambda_i$ in the general $AdS_5$ case. $\lambda_i^a$ carries spacetime indices $a$. On the other hand, the small solution $s_i^\alpha$ is a solution of worldsheet theory and carry internal-space indices $\alpha$.  This change of variables from target space coordinates to worldsheet solutions plays a very important role in the strong coupling story. 

We now consider the relation between small solutions
and the spectral parameter. One important fact is that: the change
of the phase of spectral parameter corresponds to the rotation of
small solutions (i.e. cusps)\,%
\footnote{This implies an intriguing correspondence between the worldsheet $z$-plane and spectral $\zeta$-plane.}. The $Z_4$ automorphism mentioned
before plays a very important role. For example, the $Z_2$ automorphism in $AdS_3$ gives the relation
$s_{i+1} (\zeta) \propto i \sigma_3 \, s_i(e^{i\pi} \zeta)$. The
contraction of small solutions can be defined as T-function and 
Y-functions,  see Appendix \ref{app-funeqs}. Using $Z_4$ property
and some other identities, it is possible to construct a finite set
of difference equations between these functions.

While a Y-system is basically a set of algebraic equations given by a set of determinant identities, one needs to provide further information such as the asymptotic behavior of the corresponding functions, so that the obtained solution is corresponding to the observables being studied. This important information can be obtained by WKB approximation, in the limits of the spectral parameter: $|\zeta| \rightarrow 0$ or $\infty$. In such limits, the contraction of small solutions is dominated by an integral along WKB lines that connect different edges, for example in $AdS_3$\,%
\footnote{The reason that there is a path connecting different edges can be understood that when computing the contraction one
needs to bring the small solutions to a same point in the $z$-plane.
In the limit $\zeta \rightarrow 0$, the solution $s_i(z)$ is
determined by the integrand $\int_i^z {\sqrt{p} \, dz \over \zeta}$,
where the subindex $i$ should be understood as the point where the
cusp lives.}:
\be \langle s_i, s_j \rangle|_{\zeta\rightarrow0}  \ \sim \ \exp\Big({\int_i^j  {\sqrt{p} \, dz \over \zeta}}\Big) \ \sim \begin{tabular}{c}{\includegraphics[height=1.3cm]{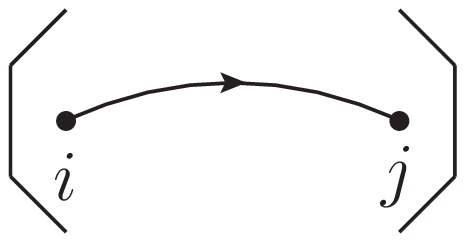}}\end{tabular}.  \ee
The integrand ${\sqrt{p} \, dz / \zeta}$ is obtained as the dominant term of flat connection (\ref{Azeta}) in the limit $\zeta\rightarrow0$. The WKB lines can be obtained as the parametric curves $z(t)$ which solve the equation ${\rm Im}( { \dot z \, \sqrt{p}/\zeta}) = 0$. Therefore they are determined by the function $P(z)$ which is related to the boundary conditions. These will be discussed further in section \ref{sec-ppoly}.

One can see that the problem is set up as a Riemann-Hilbert problem (see for example \cite{GMN09}): finding the exact functions from their discontinuities (provided by Y-system equations) and asymptotic behaviors (WKB approximations).
In this paper we will construct the Y-systems for form factors in $AdS_5$ and  with multi-operator insertions in $AdS_3$. We also study the WKB approximations.

\subsection{Conventions}

The basic definitions of $T$- and $Y$-functions and how to obtain Hirota and Y-system equations are summarized in Appendix \ref{app-funeqs}. For the reader who is not familiar with the definitions it may be necessary to have a look at the appendix before reading the following sections. Below we mention a few important relations and conventions.

We will often assume the normalization conditions \cite{AMSV10}
\bea  AdS_3\ {\rm case}: && \quad  \langle s_i , s_{i+1} \rangle \ = \ 1  \, , \\ AdS_5\ {\rm case}:  && \quad  \langle s_i , s_{i+1}, s_{i+2}, s_{i+3} \rangle \ =\ 1\, , \eea
unless indicated otherwise. The $Z_4$ automorphism imposes the following relations \cite{AMSV10}
\bea
\label{Z2-relation}  AdS_3 \ {\rm case}: && \ s_{i+1} (\zeta) \ = \ i \sigma_3 \, s_i(e^{i\pi} \zeta) \, , \\
 \label{Zautomorphism}   AdS_5 \ {\rm case}: && \  \bar s_{i+1}(\zeta) \ = \ C^{-1} s_i(e^{i \pi/2 } \zeta) \, , \qquad
s_{i+1}(\zeta) \ = \ C^T \bar s_i(e^{i\pi/2} \zeta) \, ,\eea
where
\be \bar s_i \ := \ s_{i-1} \wedge s_i \wedge s_{i+1} \, . \ee
$\sigma_3$ and $C$ are some constant matrices whose explicit forms are not important in this paper.

There are two different conventions used for $AdS_3$ and $AdS_5$ cases:
\bea \label{ads3-convention}
  AdS_3 \ {\rm case}: && \   f^\pm := f(e^{\pm i{\pi\over2} }\zeta) , \qquad f^{[k]} := f(e^{i{k\pi\over2}} \zeta) \, , \\
  AdS_5 \ {\rm case}: && \   f^\pm := f(e^{\pm i{\pi\over4} }\zeta) \, , \qquad f^{[k]} := f(e^{i{k\pi\over4}} \zeta) \, . \eea
Since the number of cusps is always even in  the $AdS_3$ case, for convenience we define
\be {\hat n} \ := \ n/2 \, , \ee
where $n$ is the number of cusps.

\section{Review of form factors in $AdS_3$ \label{sec-ffads3} }

In this section we review the  Y-system for form factors in $AdS_3$ \cite{MZ10}. The construction in this case is relatively simple, but the basic picture in later generalizations is similar.

\subsection{A look at amplitudes}

We first look at the case of scattering amplitudes.
For amplitudes, the corresponding minimal surface is smooth. The small solutions are single-valued on the $z$ plane
 \be \label{singles} s_j(e^{i2\pi} z, \zeta) \ = \ s_j(z, \zeta) \ . \ee
By definition $s_{j+\hat n}$ is  the small solution in
the same sector as $s_j$ but after going around the complex $z$
plane once. Because they are the solutions in the same sector, they should be proportional to each other
 \be \label{ads3amp_sn} s_{j+{\hat n}}(e^{i2\pi} z , \zeta) \ \propto \ s_j(z, \zeta) \, .\ee
This may be also understood from the periodic condition. Note that an arbitrary
proportionality constant is allowed. 

To do the contraction of small solutions, one needs to bring two small solutions to the {\it same} worldsheet point. Using (\ref{singles}) and (\ref{ads3amp_sn}), one gets that
 \be s_{i+{\hat n}}(z , \zeta) \ \propto \ s_i(z, \zeta) \, , \ee
which implies  $\langle s_i, s_{i+{\hat n}} \rangle = 0$, or equivalently $T_{\hat n-1}=0$. This provides a natural truncation for Hirota equations. The corresponding Y-system is given in terms of $\hat n-3$ $Y$-function: $Y_{m}$, $m=1,\ldots, \hat n-3$ \cite{AMSV10}.

\subsection{Operator as a monodromy}

\begin{figure}[t]
\begin{center}

\includegraphics[height=4cm]{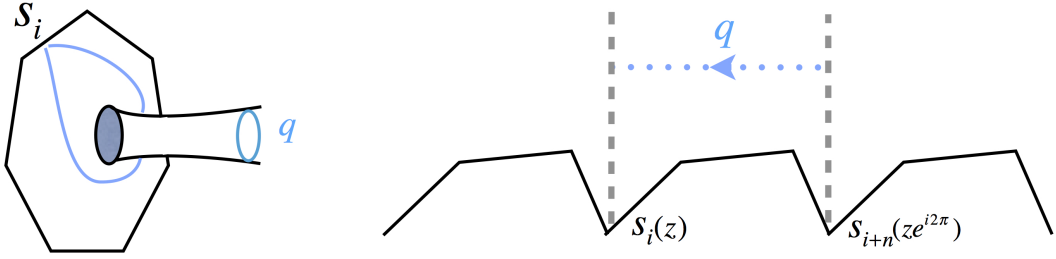}
\caption{\it Small solution and monodromy.}
\label{monodromy}

\end{center}
\end{figure}

Now we consider form factors. Since there is an operator inserted,
the worldsheet is not smooth but contains a singular point. The small
solutions are therefore no longer single-valued on $z$ plane. In other words, they change their values after going around the complex $z$
plane, or more exactly, going around the singular point where the operator is inserted, as shown in Figure \ref{monodromy}.

This effect can be characterized by introducing a {\it monodromy
matrix}.  One can firstly choose two linearly independent small solutions
as a basis. To be explicit, one can choose $\{s_0, s_1\}$. The monodromy is defined as a $2$ by
$2$ matrix $\Omega(\zeta)$ satisfying
\be \label{ads3_worldsheet_monodromydef}
\begin{pmatrix} s_1 \\
s_0
\end{pmatrix} (z e^{2\pi i} , \zeta) \, = \, \Omega(\zeta) \begin{pmatrix} s_1 \\
s_0
\end{pmatrix} (z , \zeta) \ .
\ee
Using the $Z_2$ automorphism relation (\ref{Z2-relation}), this also
fixes the monodromy relations for other small solutions. By taking the wedge
of the small solutions, one can obtain
\be \det [\Omega(\zeta)] \, = \, 1 \, . \ee
The exact property of $\Omega$ is determined by the corresponding operator, which can be taken as an input of the system.

By definition, as discussed for amplitudes above, $s_{j+\hat n}$ is  the small solution in
the same sector as $s_j$ but after going around the  $z$-plane once (see Figure \ref{monodromy}). Since they are in the same sector, their relation of proportionality does not change: $s_{j+\hat n}(e^{i2\pi} z) \propto s_j(z)$. We  introduce a proportionality constant $B(\zeta)$ so that
\be s_{\hat n}(z, \zeta) = B(\zeta) \, s_0(z e^{-2\pi i}, \zeta) \, .
\ee
By $Z_2$ relation (\ref{Z2-relation}), this also determines the proportionality constants for other small solutions, in particular
\be s_{\hat n+1}(z, \zeta) \, = \, B^{[2]}(\zeta) s_1(z e^{-2\pi i} ,
\zeta) \, . \ee
Taking the wedge of small solutions and using the normalization condition one  can get a constraint on $B$:
\be B^{-1} \, = \, B^{[2]}  \, . \ee

At the {\it same} point of worldsheet, one obtains 
\be \label{ads3-basic} \begin{pmatrix} s_{\hat n+1} \\
s_{\hat n} \end{pmatrix} (z, \zeta) \ = \ {\cal B}(\zeta) \cdot \Omega^{-1}(\zeta) \begin{pmatrix} s_1 \\
s_0 \end{pmatrix} (z, \zeta) \, , \ee
where the proportionality constants are written into a diagonal
matrix
\be {\cal B}(\zeta) \, = \, \begin{pmatrix} B^{[2]}(\zeta) & 0 \\ 0 & B(\zeta) \end{pmatrix} , \qquad \det[ {\cal B}(\zeta) ] \, = \, 1 \, . \ee
Unlike amplitudes, $s_i$ and $s_{i+\hat n}$ are not proportional to each other and $\langle s_i, s_{i+{\hat n}} \rangle \neq 0$, so the truncation becomes much more non-trivial for form factors.

\subsection{Truncation and Y-system for form factors}

We consider now how to use the above monodromy relation to truncate the Hirota equations, and then how to write them into a Y-system, following \cite{MZ10}.

Firstly, a useful relation is from the trace of $\Omega$. From (\ref{ads3-basic}) one can obtain
\be \label{TrOmega-ads3_1} {\rm Tr}[\Omega(\zeta)] \  = \ B(\zeta) \langle s_0, s_{\hat n+1} \rangle (\zeta) - B^{-1}(\zeta) \langle s_1, s_{\hat n} \rangle (\zeta) \, , \ee
where we have used the fact that  for a $2\times2$ unitary matrix (${\rm det}[\Omega]=1$), ${\rm Tr}[\Omega] = {\rm Tr}[\Omega^{-1}]$. This can be further written as
\be \label{TrOmega-ads3} {\rm Tr}[\tilde\Omega(\zeta)] \ := \ {\rm Tr}[\Omega(\tilde\zeta)] = B(\tilde\zeta) \,T_{\hat n}(\zeta) - B^{-1}(\tilde\zeta)\, T_{\hat n -2}(\zeta) \, ,\ee
where $ \tilde\zeta = e^{-i ({\hat n}+1) \pi/2} $.

One can see that the trace relation provides a truncation for the Hirota equations, since one can solve for $T_{\hat n}$ in terms of $T_{\hat n-2}$ and ${\rm Tr}[\Omega]$. As mentioned before,  ${\rm Tr}[\Omega]$ can be taken as an input of the system.

While Hirota equations are not gauge invariant, it is necessary to write the system in a conformally invariant way, i.e. in a form of Y-system. This can be done by defining a new $Y$-function as
\be \label{defofYbar} {\overline Y}(\zeta) \ := \ B^{-1}(\tilde\zeta)\, T_{\hat n -2}(\zeta)  \, . \ee
Then using (\ref{TrOmega-ads3}), $T_{\hat n}(\zeta)$ can be solved as
\be T_{\hat n}(\zeta) \ = \ B^{-1}(\tilde\zeta) \left[ {\rm Tr}[\tilde\Omega](\zeta) + \overline Y(\zeta) \right]  , \ee
and furthermore for $Y_{\hat n-1}$
\be Y_{\hat n-1} \ = \ T_{\hat n-2} \, T_{\hat n} \ = \ {\rm Tr}[\tilde\Omega] \overline Y + \overline Y^2 \ . \ee
The equation for $\overline Y$ is simply $\overline Y^+ \overline Y^- = 1 + Y_{{\hat n}-2}$.

\begin{figure}[t]
\begin{center}

\includegraphics[height=4.2cm]{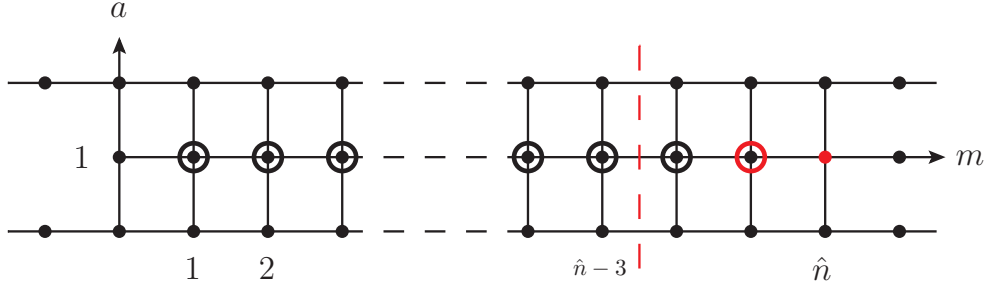}
\caption{\it Lattice picture for the Hirota and Y-system of $AdS_3$ form
factors. Black dots are for functions $T_{a,m}$. Dots with black
circles correspond to Y-functions $Y_{a,m}$, while the dot with red
circle is for $\overline Y$. Red dot is for $T_{\hat n}$ which can be solved via the trace relation. For  amplitudes, the Y-functions are
truncated to the left hand side of the red dashed line.}
\label{ads3_lattice}

\end{center}
\end{figure}

In this way, one obtains a set of equations in terms of $\hat n-1$ Y-functions:
\bea  Y_s^+Y_s^- &=& (1+Y_{s+1})(1+Y_{s-1}) \, ,  \\  Y_{{\hat n}-2}^+ Y_{{\hat
n}-2}^- &=& (1+
Y_{{\hat n}-3}) ( 1 + {\rm Tr}[\tilde\Omega] \overline Y + \bar Y^2 ) , \\
\overline Y^+ \overline Y^- &=& 1 + Y_{{\hat n}-2}  \, , \eea
where $s=1, \ldots, \hat n-3$. 
A lattice structure of T- and Y-functions is shown in Figure \ref{ads3_lattice}.

Comparing to amplitudes, the system contains two new functions $Y_{\hat n-2}$ and $\overline Y$. This matches with the counting of the degrees of freedom. The operator is  associated to the $2\times2$ monodromy matrix. Because ${\rm det}[\Omega]=1$, there are three independent components, which correspond to three functions: ${\rm Tr}[\Omega]$, $Y_{\hat n-2}$ and $\overline Y$, while ${\rm Tr}[\Omega]$ is taken as an input in the Y-system\,%
\footnote{This is not necessary to be true. In some special cases the trace of the monodromy may also depend on the spectral parameter, then one can introduce another $Y$-function while the trace does not appears in the Y-system, see section 5.4 of \cite{MZ10}. In the periodic case that we consider in this paper, the trace of monodromy will be fixed to be a pure number. }. These $\hat n-1$ $Y$-functions also match the number of degrees of the T-dual Wilson line picture, which is $2(\hat n-1)$.

Because of the simple structure of the above Y-system, it can be easily written in terms of integrable equations. The free energy part of the area can be extracted from the solution and takes a TBA form \cite{MZ10}\,%
\footnote{Here for simplicity $m_s, \overline m$ are choose to be real.},
\be A_{\rm free} \ = \ \sum_s { m_s \over 2\pi}
\int d\theta \cosh\theta \log (1+ Y_{s})  + 2 { \overline m \over 2\pi}
\int d\theta \cosh\theta \log (1+ \overline Y) \, .
\ee

For the $AdS_3$ case here, the construction of Y-system looks straightforward and simple. As we will see later,  the generalizations to $AdS_5$ and to multi-operator insertions are much more involving.  However,  the underlying picture is similar. 

We would  like to emphasize again how the picture of operator enters the story.  
The basic building blocks are small solutions, which are related to the null-cusp structure.  The operator is introduced by a monodromy matrix which imposes some linear relations on the small solutions. The method of computing null Wilson loops can then be applied in a similar way to compute form factors. 
The detailed information of the monodromy is  taken as an input. In principle one may consider arbitrary operators, depending on the choice of corresponding monodromy matrices.


\subsection{Spacetime picture}

Finally, in this subsection we review the spacetime picture \cite{MZ10}. We first clarify the difference between the worldsheet monodromy and spacetime monodromy.  Then we solve the monodromy and also write the $\overline Y$-functions in terms of target-space variables which specify the spacetime boundary configuration.

There are two kinds of monodromy. One is defined in terms of small solutions, as discussed above.  It characterizes the non-single-valuedness of small solutions while going around the path that surrounds the operator.  We call it {\it worldsheet monodromy} and denote it as $\Omega$.
The other one is defined in terms of spacetime variables. It will be called {\it spacetime monodromy} and denoted by $\hat\Omega$.

The spacetime monodromy can be taken as a spacetime conformal transformation. It can be given by mapping $\{ X_{n}, X_{n+1}, X_{n+2} \}$ to
$\{ X_0, X_1, X_2 \}$.  Using spinor decomposition $X_{i} = \lambda^L_i \lambda^R_i$, it is enough to consider left-hand spinor $\lambda^L_i$\,%
\footnote{Note that in the $AdS_5$ case, $\lambda$ is used for momentum twistor.}. Explicitly, the monodromy can be defined as
\be \label{spacetime_omega} \lambda^L_{i+n} \ \propto \ \hat\Omega \,  \lambda^L_i \ = \ \begin{pmatrix} \hat\Omega_{11} \lambda^L_{i,1} + \hat\Omega_{12} \lambda^L_{i,2} \\ \hat\Omega_{21} \lambda^L_{i,1} + \hat\Omega_{22} \lambda^L_{i,2} \end{pmatrix}  , \ee
where $ i = 0, 1, 2$.
Because  $X_i$'s are embedding coordinates, the mapping is only {\it projectively}, and an arbitrary proportionality constant is allowed for each $i$.  One also has $\det(\hat\Omega)=1$, because the conformal group is $SL(2,R)$.

The trace truncation equation (\ref{TrOmega-ads3_1}) in terms of spacetime variables can be written as
\be {\rm Tr}[\hat\Omega] \langle \lambda^L_0, \lambda^L_1 \rangle \ = \ \langle \lambda^L_0, \hat\Omega \lambda^L_1 \rangle + \langle \hat\Omega \lambda^L_0, \lambda^L_1 \rangle \, . \ee

Unlike the worldsheet picture where monodromy relates different small solutions, in spacetime the monodromy matrix operates on a single spinor variable.
The worldsheet monodromy $\Omega_{ij}$ carries the indices of small solution basis, while the spacetime monodromy $\hat\Omega_{ab}$ carries spacetime indices. For each component of $\hat\Omega_{ij}$ and $\Omega_{ab}$, they are in general different from each other. But for special  combinations like  ${\rm Tr}[\Omega]$, the worldsheet and spacetime monodromies take the same value \cite{MZ10}.

One can solve for $\hat\Omega$ more explicitly  in terms of Poincar\'{e} coordinate $x$. It is convenient to use  light-cone coordinates $x^\pm := x_0 \pm x_1$, which can be related to spinors as \cite{AM09}
\be  x^+_i \, = \, {(\lambda^L_i)_2 \over (\lambda^L_i)_1}  \, . \ee
The monodromy relations (\ref{spacetime_omega}) written in terms of $x^+$ variables are
\be \label{ads3_monodromy_x} {1 \over x_{i+\hat n}^+} \ = \ { \hat\Omega_{11} + \hat\Omega_{12} x_i^+ \over \hat\Omega_{21} + \hat\Omega_{22} x_i^+ } \, ,  \ee
where $i = 0, 1, 2$.
Together with the condition $\det(\hat\Omega)=1$, they uniquely fix the monodromy (up to a whole sign).

We focus on the short operators which is T-dual to a null Wilson line boundary condition.  One has 
\be x_{i+\hat n}^+ \ = \ x_i^+ + q \, , \ee
for all $i$.  Using (\ref{ads3_monodromy_x}), one can obtain the monodromy 
\be \hat\Omega \ = \ \begin{pmatrix} 1 & 0 \\ q & 1  \end{pmatrix}  , \ee
and ${\rm Tr}[\hat \Omega]=2$.  As shown in \cite{MZ10}, the trace of the worldsheet monodromy takes the same value
\be {\rm Tr}[\Omega] \ = \ {\rm Tr}[\hat \Omega]\ =\ 2 \, . \ee

The $\overline Y$-function can be also written in terms of spacetime variables. To do this one needs first to write the  $\overline Y$ function in a form which is independent of  normalization.  One can define $\tilde s_1(z) = s_1(z e^{-i2\pi})$ and therefore $\langle  s_{\hat n} , \tilde s_1 \rangle = B \langle  s_{\hat n} , s_{\hat n+1} \rangle$. Recall the definition (\ref{defofYbar}), one obtains
\be \overline Y (\zeta) \ = \ {\langle s_1, s_{\hat n} \rangle \over \langle  s_{\hat n} , \tilde s_1 \rangle} (\zeta e^{-i\pi(n+1)/2}) \, . \ee
This form makes it obvious that the WKB approximation of $\overline Y$ forms a closed path which contains the singular point corresponding to the insertion of operator \cite{MZ10}.

Then ${\overline Y}$ can be written in terms of spacetime coordinates
\be \overline Y (\zeta= i^{\hat n+1}) \ = \ {\langle \lambda^L_1, \lambda^L_{\hat n} \rangle \over \langle  \lambda^L_{\hat n} , \hat\Omega \lambda^L_1 \rangle} = - {\lambda^L_{1,1} \lambda^L_{\hat n,2} - \lambda^L_{1,2} \lambda^L_{\hat n,1} \over \lambda^L_{1,1} \lambda^L_{\hat n,2} - (q \lambda^L_{1,1} + \lambda^L_{1,2}) \lambda^L_{\hat n,1} } = { x_{1,\hat n}^+ \over x_{\hat n,\hat n+1}^+} \ . \ \ee
One can see that $\overline Y$  is scale-invariant, but different from other $Y$-functions which are usual conformal cross ratios. This is related to the fact that form factor is not dual-conformally invariant, unlike scattering amplitudes. 
The nice point is that at strong coupling in the worldsheet picture, integrability techniques are still available. One can deal with $\overline Y$ exactly in the same way as for usual $Y$-functions.

\section{Form factors in $AdS_5$ \label{sec-ffads5} }

In this section, we study form factors in $AdS_5$. The construction will be parallel to the $AdS_3$ case, but as we will see, new problems will appear.

\subsection{Monodromy}

As in $AdS_3$, one can first choose four linearly
independent small solutions $\{s_{-2}, s_{-1}, s_0, s_1\}$ as a basis for the general
solutions of the linear problem.  The worldsheet monodromy is characterized by a 4 by 4 matrix
$\Omega(\zeta)$, which is defined by the relation%
\footnote{Note that compare to the $AdS_3$ case, here we choose $\Omega^{-1}$ in the definition for convenience.}
\be
\begin{pmatrix} s_1 \\
s_0 \\ s_{-1} \\ s_{-2}
\end{pmatrix} (z e^{2\pi i} , \zeta) \ = \ \Omega^{-1}(\zeta) \begin{pmatrix} s_1 \\
s_0 \\ s_{-1} \\ s_{-2}
\end{pmatrix} (z , \zeta) \, .
\ee
Taking the wedge of small solutions one has
$\det [\Omega(\zeta)] = 1$.

By definition, one has the same proportionality relation that $s_{j+n}(e^{i2\pi} z , \zeta) \propto s_j(z, \zeta)$. We introduce the proportionality constant $B(\zeta)$ such that
\be \label{AdS5_defofB}  s_{n+1}(z, \zeta) \ = \ B(\zeta) \, s_{1}(e^{-2\pi i} z, \zeta) \, . \ee
Using the $Z_4$ automorphism relations (\ref{Zautomorphism}),
proportionality constants are fixed for all other small solutions, in particular
\bea s_{n}(z, \zeta) &=& B^{-1}(e^{i\pi/2}\zeta)\, s_{0}(e^{ -2\pi i} z, \zeta) \, , \\
s_{n-1}(z, \zeta) &=& B(e^{-i\pi}\zeta)\, s_{-1}(e^{ -2\pi i} z, \zeta) \, , \\
s_{n-2}(z, \zeta) &=& B^{-1}(e^{-i\pi/2}\zeta)\, s_{-2}(e^{ -2\pi i} z, \zeta) \, . \eea
There are also extra constraints from  (\ref{Zautomorphism}):
\be B \,  B^{[4]} \, = \, B^{[2]} B^{[-2]} \, , \qquad B \, = \, B^{[8]} \, . \ee

One obtains that at the {\it same} point of the worldsheet $s_j$ and $s_{j+ n}$ are related to each other as
\be \label{AdS5_monodromy_relation}  \begin{pmatrix} s_{n+1} \\
s_n \\ s_{n-1} \\ s_{n-2} \end{pmatrix} (z, \zeta) \ = \ {\cal B}^{-1}(\zeta) \cdot \Omega(\zeta) \begin{pmatrix} s_1 \\
s_0 \\ s_{-1} \\ s_{-2} \end{pmatrix} (z, \zeta) \, , \ee
where the proportionality factors are written into a  diagonal matrix
\be {\cal B}^{-1} \ := \ {\rm diag} \big\{ B , (B^{[2]})^{-1}, B^{[-4]}, (B^{[-2]})^{-1} \big\} \, , \qquad \det[ {\cal B} ] = 1 \, . \ee
%

\subsection{Truncation of Hirota equations}

Similar to the $AdS_3$ case, one can apply trace conditions to truncate the Hirota equations. The traces of monodromy are conformally invariant quantities \cite{MZ10}. Their spacetime picture will be discussed later. We first consider the simple trace ${\rm Tr}[\Omega] = \sum_{i=1}^4 \Omega_{ii}$. Using (\ref{AdS5_monodromy_relation}), one can obtain
\bea {\rm Tr}[\Omega] &=& {\cal B}_{11}\,
\langle s_{-2}, s_{ -1} , s_0, s_{n+1} \rangle  +
 {\cal B}_{44}\,  \langle s_{n-2}, s_{ -1} , s_0, s_{1}\rangle  \nonumber\\ && + \,  {\cal B}_{33} \, \langle
s_{-2}, s_{ n-1} , s_0, s_{1} \rangle  +  {\cal B}_{22} \,\langle s_{-2}, s_{ -1} , s_n, s_{1}
\rangle \, . \eea
Using the definition of $T$-functions, this can be further written as
\be
{\rm Tr}[\Omega] \ = \ {\cal B}_{11}\  T^{[n]}_{1,n} - {\cal B}_{44}\ T^{[n-1]}_{3,n-3}  -  {\cal B}_{22} \,\langle s_{-2}, s_{ -1}, s_{1} , s_n \rangle  + \,  {\cal B}_{33} \, \langle
s_{-2} , s_0, s_{1} , s_{ n-1}\rangle  \, , \ee
which provides a truncation for the chain of Hirota equations by expressing $T_{1,n}$ in terms of other $T$ functions. One may worry about the other two terms which are not $T$ functions. However, they can be expressed in terms of $T$-functions as will be discussed in the next subsection.

To obtain a truncation relation for $T_{3,n}$, it is natural to consider the $\bar s_i$ variables and the corresponding $\overline \Omega, \overline{\cal B}$, which gives\,%
\footnote{Note that $T_{3,m}$ is dual to $T_{1,m}$ in the sense of $s_i \leftrightarrow \bar s_i$ \cite{AMSV10}.}
\be
{\rm Tr}[\overline\Omega] \ = \ {\overline {\cal B}}_{11}\ T^{[n]}_{3,n} - {\overline {\cal B}}_{44}\ T^{[n-1]}_{1,n-3}  -  {\overline {\cal B}}_{22} \,\langle \bar s_{-2}, \bar s_{ -1}, \bar s_{1} , \bar s_n \rangle  + \,  {\overline {\cal B}}_{33} \, \langle
\bar s_{-2} , \bar s_0, \bar s_{1} , \bar s_{ n-1}\rangle  \, . \ee
Note that $\bar \Omega, \bar{\cal B}$ are not new but related to $\Omega, {\cal B}$, see Appendix \ref{app-omegabar}.

Finally, we need a truncation relation for $T_{2,n}$. One can consider the double trace (see also \cite{Zhiboedov})
\be {\rm Tr}^{(2)}[\Omega] \ := \ \sum_{1\leq i<j \leq 4} \Omega_{ii}  \,  \Omega_{jj} =  {1\over2} \left( {\rm Tr}[\Omega]^2 - {\rm Tr}[\Omega^2] \right) \, . \ee
Using (\ref{AdS5_monodromy_relation}) and the definition of $T$ functions, one obtains
\bea \label{ads5-trace2}  {\rm Tr}^{(2)}[\Omega] &=& {\cal B}_{11}\, {\cal B}_{22}\ T_{2,n}^{[n-1]} + {\cal B}_{33}\, {\cal B}_{44}\ T_{2,n-4}^{[n-1]} \nonumber\\
&& - {\cal B}_{11}\, {\cal B}_{33}\, \langle s_{-2}, s_0, s_{n-1}, s_{n+1} \rangle  - {\cal B}_{22}\, {\cal B}_{44}\, \langle s_{-1}, s_1, s_{n-2}, s_{n} \rangle \nonumber\\
&& + {\cal B}_{11}\, {\cal B}_{44}\, \langle s_{-1}, s_0, s_{n-2}, s_{n+1} \rangle  + {\cal B}_{22}\, {\cal B}_{33}\, \langle s_{-2}, s_1, s_{n-1}, s_{n} \rangle \, . \eea
As will be discussed in next subsection, all small solution contractions can be written in terms of $T$ functions.
One may consider further a relation from the triple-trace
\be
 {\rm Tr}^{(3)}[\Omega] := \sum_{i<j<k} \Omega_{ii}  \,  \Omega_{jj}  \, \Omega_{kk} = {1\over6} \left( {\rm Tr}[\Omega]^3 - 3\, {\rm Tr}[\Omega^2] \, {\rm Tr}[\Omega] + 2 \, {\rm Tr}[\Omega^3] \right) \, , \ee
but it can be shown that it gives  an equivalent
truncation relation for  $T_{1,n}$ function.

\subsection{Recursion relations \label{subsec-recursion}}

In the above truncation relations, several new small solution contractions appear.
In this subsection we show that they can be expressed in terms of $T$-functions by using some recursion relations. A few new functions will be defined  for convenience.

We define the contractions appearing in simple-trace relations as $R$- and $S$-functions
\bea
& & R_{1,m} \ := \ \langle s_{-2}, s_{-1} , s_{1}, s_{m+2}\rangle^{[-m]} \, ,  \qquad
R_{3,m} \ := \ \langle \bar s_{-2}, \bar s_{-1} , \bar s_{1}, \bar s_{m+2}\rangle^{[-m]} \, , \qquad \\
& & S_{1,m} \ := \ \langle s_{-2}, s_{0} , s_{1}, s_{m+2}\rangle^{[-m]} \, , \qquad \ \ 
S_{3,m} \ := \ \langle \bar s_{-2}, \bar s_{0} , \bar s_{1}, \bar s_{m+2}\rangle^{[-m]} \, .
\eea
Using the Wronskian relations reviewed in Appendix \ref{app-funeqs},
one can obtain the following recursion relations
\bea &&  R_{1,m} \ = \ R^-_{1,m-1} { T^+_{1,m+1} \over T_{1,m} } +{ T_{2,m+1} \over T_{1,m} } \, , \qquad R_{3,m} \ = \ R^-_{3,m-1} { T^+_{3,m+1} \over T_{3,m} } +{ T_{2,m+1} \over T_{3,m} } \, , \qquad \\ &&
S_{1,m} \ = \ S_{1,m-1}^- { T_{3,m}^{[2]} \over T_{3,m-1}^{+} } - { T_{2,m-1}^{[2]} \over T_{3,m-1}^+ } \, , \qquad \ S_{3,m} \ = \ S_{3,m-1}^- { T_{1,m}^{[2]} \over T_{1,m-1}^{+} } - { T_{2,m-1}^{[2]} \over T_{1,m-1}^+ } \, .  \eea
Together with the initial conditions
\be R_{1,0} = R_{3,0} = T_{2,1} \, , \qquad S_{1,0} = T_{1,1}, S_{3,0} = T_{3,1} \, , \ee
all $R$- and $S$-functions can be expressed in terms of $T$-functions.

To consider the contractions appearing in the double-trace relation, we first define
\bea
U_{1,m} \ :=\ \langle s_{-2}, s_{-1} , s_{m}, s_{m+2}\rangle^{[-m]} \, ,  & \quad &
U_{3,m} \ :=\ \langle s_{-1}, s_{0} , s_{m+1}, s_{m+3}\rangle^{[-m-2]} \, ,  \\
V_{1,m} \ :=\ \langle s_{-2}, s_{0} , s_{m+1}, s_{m+2}\rangle^{[-m]} \, ,  & \quad &
V_{3,m} \ :=\ \langle s_{-1}, s_{1} , s_{m+2}, s_{m+3}\rangle^{[-m-2]} \, ,
\eea
which satisfy
\bea \label{eqforU}  U_{1,m} =  { T^-_{1,m-1} T_{2,m+1} + T^+_{1,m+1} T^-_{2,m} \over T_{1,m} }  \, , & \quad &  U_{3,m} =  { T^-_{3,m-1} T_{2,m+1} + T^+_{3,m+1} T^-_{2,m} \over T_{3,m} }   \,  ,\\   V_{1,m} =  { T^+_{1,m-1} T_{2,m+1} + T^-_{1,m+1} T^+_{2,m} \over T_{3,m} } \, ,  & \quad & V_{3,m} =  { T^+_{3,m-1} T_{2,m+1} + T^-_{3,m+1} T^+_{2,m} \over T_{1,m} }  \,  . \label{eqforV}  \eea
The four contractions in (\ref{ads5-trace2}) are then defined as
\bea
W_{1,m}(\zeta) := \langle s_{-2}, s_{0} , s_{m+1}, s_{m+3}\rangle^{[-m-1]} ,  & \  &
W_{3,m}(\zeta) := \langle s_{-1}, s_{1} , s_{m+2}, s_{m+4}\rangle^{[-m-3]} , \\
W_{2,m}(\zeta) := \langle s_{-1}, s_{0} , s_{m+1}, s_{m+4}\rangle^{[-m-2]} ,  & \  &
{\overline W}_{2,m}(\zeta) := \langle s_{-2}, s_{1} , s_{m+2}, s_{m+3}\rangle^{[-m-2]} , \  \  \ \ \
\eea
which satisfy
\bea &&  W_{1,m} =  {V^-_{1,m} U^+_{3,m} - T^-_{1,m} T^+_{1,m} \over T_{2,m} }  \, , \qquad \qquad  W_{3,m} =  {V^-_{3,m} U^+_{1,m} - T^-_{3,m} T^+_{3,m} \over T_{2,m} }  \, , \\ && W_{2,m} =  {U^+_{3,m+1} U_{3,m} - T^-_{2,m} T^+_{2,m+2} \over T_{2,m+1} }  \, , \qquad  {\overline W}_{2,m} =  {V^-_{1,m+1} V_{3,m} - T^+_{2,m} T^-_{2,m+2} \over T_{2,m+1} }  \,  .  \qquad \eea
Using (\ref{eqforU})-(\ref{eqforV}), all $W$-functions can be written in
terms of $T$-functions. 

The main lesson  in this subsection is that
any small solution contraction can be expressed in terms
of $T$-functions, therefore it is enough to focus on the
$T$-functions.

\subsection{Y-system for $AdS_5$ form factors}

While it is straightforward to introduce the trace relations  to truncate the Hirota system, it is more challenging to obtain a Y-system.

We find it mostly convenient to introduce three new ${\overline Y}_a$ functions as follows:
\be \label{ads5_defofYbar} {\overline Y}_1 \ :=\  A_1\, { T_{3,n-2} \over T_{2,n-1} } \, , \qquad  {\overline Y}_3 \ := \ A_3 \, { T_{1,n-2} \over T_{2,n-1} } \, , \qquad  {\overline Y}_2 \ := \ A_2 \, {T_{2,n-2} \over T_{1,n-1} T_{3,n-1}} \, , \ee
where
\be \label{def_A} A_1 \ := \ B^{[-n]} \, , \qquad A_3 \ := \  (B^{[-n+4]})^{-1} \, , \qquad A_2  \ := \ {B^{[-n+1]} \over B^{[-n+3]} } \, . \ee
It is interesting to notice the relations
\be {A_2^+ A_2^- \over A_1 A_3} \ = \ {A_1^+ A_3^- \over A_2} \ = \ {A_1^- A_3^+ \over A_2} \ = \ 1 \, , \ee
which have appeared for the amplitudes of $n=4K$ cases for three special combinations of Y-functions (see page 27 of \cite{AMSV10}).

These ${\overline Y}$ functions satisfy the following nice equations
\bea && { {\overline Y}_1^+ {\overline Y}_3^- \over {\overline Y}_2 } \ = \  { 1+ Y_{3,n-2} \over 1+ Y_{2,n-1} } \, , \qquad\quad
 { {\overline Y}_1^- {\overline Y}_3^+ \over {\overline Y}_2 } \ = \  { 1+ Y_{1,n-2} \over 1+ Y_{2,n-1} } \, , \\
&&  { {\overline Y}_2^+ {\overline Y}_2^- \over {\overline Y}_1 {\overline Y}_3 } \ = \ { 1+ Y_{2,n-2} \over (1+ Y_{1,n-1})  (1+ Y_{3,n-1}) } \, . \eea

Notice that functions $Y_{a,n-1}$ appear on the right-hand side of equations. To have a closed system, one needs to solve them in terms of other $Y$-functions. This can be done by noticing the following relations\,%
\footnote{This also explains why we introduce $\overline Y_a$ functions in the above form. }
\be Y_{a,n-1} \ = \ (A_a^{-1} \, T_{a,n})  {\overline Y}_a \, , \qquad a =1,2,3 \, ,  \ee
while $A_a^{-1} \, T_{a,n}$  can be solved directly using the trace relations
given before. Since the trace functions of $\Omega$
are normalization independent, the trace equations (and therefore
$Y_{a,n-1}$) are guaranteed to be able to be written in terms of $Y$-functions: ${\overline Y}_a$ and $Y_{a,m}, m=1, \ldots, n-2$. This will be shown explicitly later in the three-point case.

\begin{figure}[t]
\begin{center}

\includegraphics[height=5.3cm]{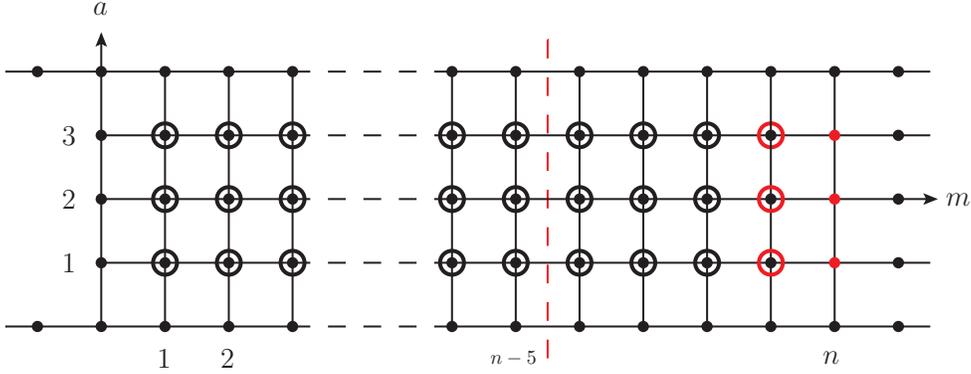}
\caption{\it Lattice picture for the T/Y-system of $AdS_5$ form
factors. Dots with black circles correspond to Y-functions $Y_{a,m}$,
while dots with red circles are for $\overline Y$. Red dots are for $T_{a, n}$ which can be solved via the trace relations.  For  amplitudes,
the Y-functions are truncated to the left hand side of the red
dashed line.} \label{ads5_lattice}

\end{center}
\end{figure}

The full Y-system for form factors in $AdS_5$ can be summarized as
\bea && {Y^-_{a,m}Y^+_{4-a,m} \over Y_{a+1,m}Y_{a-1,m}} \ = \
{(1+Y_{a,m+1})(1+Y_{4-a,m-1}) \over (1+Y_{a+1,m})(1+Y_{a-1,m})} \, , \\
&&  { {\overline Y}_a^+ {\overline Y}_{4-a}^- \over {\overline Y}_{a+1} {\overline Y}_{a-1} } \ = \ { 1+ Y_{a,n-2} \over (1+ Y_{a+1,n-1})  (1+ Y_{a-1,n-1}) } \, , \eea
where $a=1,2,3, \ \  m=1, \ldots, n-2$, and $Y_{a,n-1}$ can be expressed in terms of other $Y$-functions appearing in the equations. Therefore one obtains a closed finite system in terms of $3(n-1)$ $Y$-functions, as shown in Figure \ref{ads5_lattice}.

Comparing to amplitudes where there are $3(n-5)$ $Y$-functions, in form factors there are $12$ more (for $n>4$). This can be understood as follows. Form factor contains a $4\times 4$ unitary monodromy matrix $\Omega$ which has 15 independent components. Minus three traces, 12 degrees of freedom are left, giving 12 new $Y$-functions.
On the other hand, this does not match with the spacetime picture of a periodic Wilson line in $AdS_5$, where it only has $3n-7$ independent degrees of freedom\,%
\footnote{This can be obtained by a counting of symmetries as $3(n-5)+4+4 = 3n-7$.
$3(n-5)$ is the degrees of freedom for an $n$-cusp Wilson loop, while 
a periodic $n$-cusp Wilson line would break 4 special conformal
transformation symmetries, and there is also an off-shell momentum
$q$ which gives the other 4.}. 
 Note this is different from the $AdS_3$ case, where the number of $Y$-functions matches with the number of the degrees of freedom.  It implies  that in $AdS_5$, the monodromy matrix of  a short operator is not arbitrary but with extra four constraints. 
 In practice this is not a problem, since the WKB approximation of $Y$-functions are  determined in the same way by the $P(z)$ polynomial.  For general operators, it would require further information.\\

\noindent {\bf A proposal for the free energy}

\noindent Following the result in $AdS_3$ \cite{MZ10}, a natural proposal for the free energy is
\bea A_{\rm free} &=& \sum_s { m_s \over 2\pi}
\int d\theta \cosh\theta \log \left[(1+
Y_{1,s})(1+ Y_{3,s})(1+ Y_{2,s})^{\sqrt{2}}\right]  \nonumber\\ && + c \,  { \overline m \over 2\pi}
\int d\theta \cosh\theta \log \left[(1+
{\overline Y}_{1})(1+ {\overline Y}_{3})(1+ {\overline Y}_{2})^{\sqrt{2}}\right] , \eea
where $c$ is an integer factor which may be fixed by studying some simple limits.

\subsection{Reduction to $AdS_4$ and $AdS_3$}

We consider the reduction following the discussion for amplitudes in \cite{AMSV10}.
The reduction to the $AdS_4$ case is simply given by taking
\be T_{1,s}(\zeta) \ = \ T_{3,s}(\zeta) \, , \qquad Y_{1,s}(\zeta) \ = \ Y_{3,s}(\zeta) \,  . \ee
Therefore for form factors in $AdS_4$, there are only two trace-relations ${\rm Tr}[\Omega]$ and ${\rm Tr}^{(2)}[\Omega]$ to consider. This reduction will be used in the three-point case.

We consider further to reduce the system to $AdS_3$.  Besides the
relation $T_{1,s}=T_{3,s}$, the linear problem splits into two
decoupled problems denoted by $left$ and $right$ problems. In an
appropriate gauge
\be s_{2k} \ = \ \begin{pmatrix} s^R_{k} \\ 0 \end{pmatrix} , \qquad
s_{2k+1} \ = \ \begin{pmatrix} 0 \\ s^L_{k+1}  \end{pmatrix} \, ,\ee
where $s^L$ and $s^R$ are the small solution of the left and right
$AdS_3$ problems respectively. The left and the right problems are related
by a rotation in the spectral parameter
 \be \langle s^R_{i}, s^R_{j}
\rangle \ = \ \langle s^L_{i}, s^L_{j}\rangle^{[2]} \, .
 \ee

The small solution contraction in $AdS_5$ is reduced to
 \be \langle s_{2i}, s_{2k+1}, s_{2j}, s_{2l+1}\rangle \ = \ - \, \langle
 s^R_{i}, s^R_{j} \rangle \, \langle s^L_{k+1}, s^L_{l+1} \rangle \, .
 \ee
 One can choose a normalization $ \langle s^L_{i}, s^L_{i+1} \rangle =1 $, this corresponds to an
 unusual normalization $\langle s_i, s_{i+1},
s_{i+2}, s_{i+3}\rangle =-1$ in $AdS_5$.
Most  equations in $AdS_5$  become identically satisfied
except for  the nodes of $T_{2,2k}$. For these, they reduce to the Hirota equations in
$AdS_3$.

The monodromy matrix can be decomposed  as 
\be
\begin{pmatrix} s_1 \\ s_{-1} \\ s_0 \\ s_{-2} \end{pmatrix}(ze^{2\pi
i},\zeta) \ \sim \ \begin{pmatrix} \Omega_L(\zeta)
& 0 \\ 0 & \Omega_R(\zeta)\end{pmatrix} \begin{pmatrix} s_1 \\
s_{-1} \\ s_0 \\ s_{-2} \end{pmatrix}(z,\zeta) \, , \ee
where using the relation $s_{R,a}=s_{L,a}^{[2]}$, one can get
\be \Omega_R \ = \ \Omega_L^{[-2]} \, . \ee
One can check that the three traces-relations in  $AdS_5$ exactly
reduce to the single relation in  $AdS_3$. In deriving it
one needs to use the relation
\be \rm Tr[\Omega] \ =\ - (\rm Tr[\Omega_L] + \rm Tr[\Omega_R]) \, , \ee
while the minus sign is due to the normalization $\langle s_i, s_{i+1},
s_{i+2}, s_{i+3}\rangle =-1$.

\begin{figure}[t]
\begin{center}
\includegraphics[height=4.3cm]{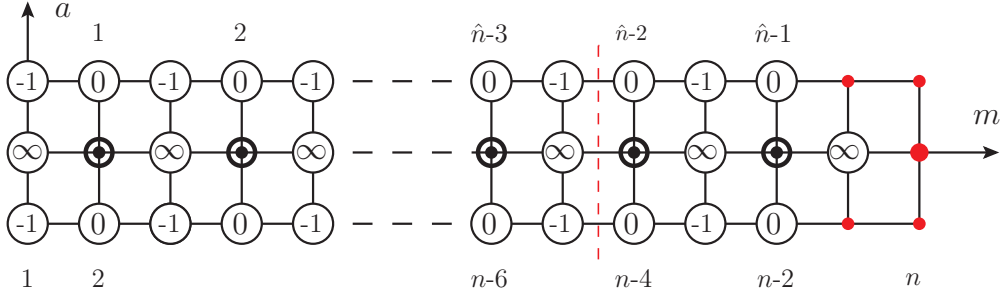}
\caption{\it The reduction of $AdS_5$ $Y$-functions to $AdS_3$. Most $Y$-functions become trivial (i.e. -1, 0 or $\infty$ as shown in the figure), while half of $Y_{2,m}$-functions are reduced to $AdS_3$ $Y$-functions. Interestingly, the three $\overline Y_a$-functions do not reduce to $\overline Y$-functions in $AdS_3$. Red dots are related only to $T_{\hat n}$-functions in $AdS_3$.} 
\label{reduction_lattice}
\end{center}
\end{figure}

The reduction of $Y$ functions can be summarized as in Figure \ref{reduction_lattice}. 



\subsection{Spacetime picture}

As discussed in the $AdS_3$ case, the spacetime monodromy can be understood as a spacetime conformal transformation. In $AdS_5$ it is convenient to consider the map for twistor variables, and in this representation the conformal group is $SU(4)$. An introduction of twistor variables is given in Appendix \ref{app-twistor}.

Spacetime monodromy can be defined by mapping four twistors $\{ \lambda_{n-2}, \lambda_{n-1} , \lambda_{n}, \lambda_{n+1} \}$ to $\{ \lambda_{-2}, \lambda_{-1}, \lambda_0, \lambda_1 \}$
\be \label{omegahatdef} \lambda_{i+n} \ \propto \ \hat\Omega \,  \lambda_i \, , \ee
where $i =  -2, -1, 0,1$.
Note that the map is only {\it projective}, and an arbitrary proportionality constant is allowed for each $i$. Since $\hat\Omega \in SU(4)$, one has $\det(\hat\Omega)=1$.
Like for small solutions, we also define
\be \bar\lambda_i^a \ := \ \epsilon^{a b c d}\, \lambda_b\, \lambda_c\, \lambda_d \, , \ee
and the corresponding monodromy $\hat {\bar \Omega}$ is defined as $\bar\lambda_{i+n} \propto \hat{\bar\Omega} \,  \bar\lambda_i$, where $i =  -2, -1, 0, 1$.

The single trace relation written in terms of spacetime variables corresponds to
\bea && {\rm Tr}[\hat\Omega] \, \langle \lambda_{-2}, \lambda_{-1}, \lambda_0, \lambda_1 \rangle \ = \ \langle \hat\Omega \lambda_{-2}, \lambda_{-1}, \lambda_0, \lambda_1 \rangle + \langle \lambda_{-2}, \hat\Omega \lambda_{-1}, \lambda_0, \lambda_1 \rangle \\ && \hskip 4.5cm + \langle \lambda_{-2}, \lambda_{-1}, \hat\Omega \lambda_0, \lambda_1 \rangle + \langle \lambda_{-2}, \lambda_{-1}, \lambda_0, \hat\Omega \lambda_1 \rangle  \, ,  \nonumber\eea
and similar from ${\rm Tr}[\hat{\bar\Omega}]$. For double trace one has 
\bea
 &&  {\rm Tr}^{(2)}[\hat\Omega] \, \langle \lambda_{-2}, \lambda_{-1}, \lambda_0, \lambda_1 \rangle \ = \  \langle \hat\Omega \lambda_{-2}, \hat\Omega \lambda_{-1}, \lambda_0, \lambda_1 \rangle + \langle \lambda_{-2}, \lambda_{-1}, \hat\Omega \lambda_0, \hat\Omega \lambda_1 \rangle \qquad \\ && \hskip 4.5cm + \langle \hat\Omega \lambda_{-2}, \lambda_{-1}, \hat\Omega \lambda_0, \lambda_1 \rangle + \langle \lambda_{-2}, \hat\Omega \lambda_{-1}, \lambda_0, \hat\Omega \lambda_1 \rangle \nonumber\\ && \hskip 4.5cm  + \langle \lambda_{-2}, \hat\Omega \lambda_{-1}, \hat\Omega \lambda_0, \lambda_1 \rangle + \langle \hat\Omega \lambda_{-2}, \lambda_{-1}, \lambda_0, \hat\Omega \lambda_1 \rangle  \, ,  \nonumber \eea
where
\bea {\rm Tr}^{(2)}[\hat\Omega] \ := \ \sum_{1\leq i<j \leq4} \hat\Omega_{ii}  \,  \hat\Omega_{jj} \ = \ {1\over2} \left( {\rm Tr}[\hat\Omega]^2 - {\rm Tr}[\hat\Omega^2] \right) . \eea

One can solve for the monodromy in terms of the Lorentz variables. We will focus on  short operators which correspond to periodic boundary conditions.
The transformation between twistor variables and Lorentz variables is reviewed in Appendix \ref{app-twistor}. The derivation of the monodromy may be given in two different ways.

Firstly one has the relations (\ref{xpmtotwistor})
\be  x^-_j \, = \, i\,{ \lambda^j_{[ 1} \, \lambda^{j+1}_{4 ]} \over \lambda^j_{[ 1} \,\lambda^{j+1}_{2 ]} } \, , \qquad  x^+_j \, = \, i\,{ \lambda^j_{[ 2} \, \lambda^{j+1}_{3 ]} \over \lambda^j_{[ 1} \, \lambda^{j+1}_{2 ]} } \, , \ee
where $x^\pm = x^1 \pm x^0$. Without loss of generality, one can choose the periodic direction to be along $x^1$-direction
\be x^{\mu}_{j+n} \ = \ x^{\mu}_{j} + q \, \delta^{1,\mu} \, , \qquad x^\pm_{j+n} \ = \ x^\pm_j + q  \, . \ee
Using (\ref{omegahatdef}), one can obtain
\bea  x_{j+n}^- &=& i\, { \hat\Omega_{1a} \, \hat\Omega_{4b} \, \lambda^j_{[ a} \, \lambda^{j+1}_{b ]} \over \hat\Omega_{1a} \, \hat\Omega_{2b} \, \lambda^j_{[ a} \, \lambda^{j+1}_{b ]} } \ = \ i\, { \lambda^j_{[ 1} \, \lambda^{j+1}_{4 ]} -i \, q \, \lambda^j_{[ 1} \, \lambda^{j+1}_{2 ]}  \over \lambda^j_{[ 1} \, \lambda^{j+1}_{2 ]} }  \ , \\  
x_{j+n}^+ & = & i\,{ \hat\Omega_{2a} \, \hat\Omega_{3b} \, \lambda^j_{[ a} \, \lambda^{j+1}_{b ]} \over \hat\Omega_{1a} \, \hat\Omega_{2b} \, \lambda^j_{[ a} \, \lambda^{j+1}_{b ]} } \ = \ i\, { \lambda^j_{[ 2} \, \lambda^{j+1}_{3 ]} - i\,q \, \lambda^j_{[ 1} \, \lambda^{j+1}_{2 ]}  \over \lambda^j_{[ 1} \, \lambda^{j+1}_{2 ]} }   \, , \eea
for all $j$.
These relations fix the monodromy uniquely as
\be \label{spacetimeOmegasolution} \hat\Omega \ = \ \begin{pmatrix} 1 & 0 & 0 & 0 \\ 0 & 1 & 0 & 0 \\ i\,q & 0 & 1 & 0 \\ 0 & -i\,q & 0 & 1  \end{pmatrix} . \ee

There is  another simpler way to find the monodromy.
From the definition of twistor variable (\ref{defmomtwistor}), one can obtain
\be \lambda_j \ \propto \ (\Lambda_j\, , \, \mu_j) \, , \qquad \lambda_{j+n} \ \propto (\Lambda_j \, , \, \mu_j - i\,  q\cdot \Lambda_j) \, ,  \ee
or equivalently,
\be \label{lambda_periodic} \lambda_{j+n} \ = \ b_j \, \hat\Omega\, \lambda_j \, , \ee
where $\hat\Omega$ is given by the same matrix as above, when $q$ is along $x^1$-direction\,%
\footnote{For general $q^\mu$, the bottom-left $2\times2$ block of $\hat\Omega$ in (\ref{spacetimeOmegasolution}) is replaced by $i \,\epsilon^{\alpha\beta}q_{\alpha\dot\alpha}$.}.
A proportionality constant $b_i$ is introduced explicitly, which plays a similar role as the $B$ factor defined for small solutions. Note the relation $\Lambda_{n+i} = b_i \, \Lambda_i$. 

One obtains that
\be \label{trOmegavalue}  {\rm Tr}[\hat\Omega] ={\rm Tr}[\hat{\bar\Omega}] = 4 \, , \qquad {\rm Tr}^{(2)}[\hat\Omega] = 6  \, . \ee
As in \cite{MZ10}, one can shown that the corresponding traces of worldsheet monodromy take the same value, which can be taken as an input of the Y-system.

Next we consider to write $\overline Y_a$ function in terms of spacetime variables. To do this, one needs first to write them in a form independent of normalization. To make the derivation simpler, one may recall the relation ${\overline Y}_a = Y_{a,n-1} / (A_a^{-1} \, T_{a,n})  $.  Since $Y_{a,n-1}$ is a normal cross ratios, one can focus on $(A_a^{-1} \, T_{a,n})$, which is similar to the $AdS_3$ case.

Consider first the $\overline Y_1$ case.
One can define $\tilde s_{-2}(z) = s_{-2}(z e^{-i2\pi}) = B^{[-2]} s_{n-2}(z)$, and therefore $\langle  \tilde s_{-2} , s_{n-1} , s_n ,  s_{n+1} \rangle = B^{[-2]} \langle  s_{n-2} , s_{n-1} , s_n ,   s_{n+1} \rangle$.
One has the correspondence
\be (A_1^{-1} T_{1,n}) (\zeta) \ \rightarrow \ { \langle
s_{-2}, s_{n-1}, s_{n}, s_{n+1} \rangle \over \langle
\tilde{s}_{-2},s_{n-1},s_n,s_{n+1} \rangle}(\zeta e^{-i\pi{n-2 \over 4}}) \, . \ee
Together with the expression of $Y_{1,n-1}$, this gives 
\be {\overline Y}_1 (\zeta) \ = \ {\langle
\tilde{s}_{-2},s_{n-1},s_n,s_{n+1} \rangle \langle
s_{-1},s_{n-2},s_{n-1},s_n \rangle \over \langle
s_{-2},s_{-1},s_{n-1},s_n \rangle\langle s_{n-2},s_{n-1},s_n,s_{n+1}
\rangle}(\zeta e^{-i\pi{n-2 \over 4}}) \, , \ee
which looks almost like a cross ratio.
It also makes it obvious that the WKB lines of $\overline Y$ form a closed contour which contains the singular point corresponding to the insertion of operator.

The normalization-independent form can be written directly in terms of spacetime coordinates as
\bea {\overline Y}_1 (\zeta = i^{{n-2 \over 2}}) &=& {\langle
\hat\Omega\lambda_{-2},\lambda_{n-1},\lambda_n,\lambda_{n+1} \rangle
\langle \lambda_{-1},\lambda_{n-2},\lambda_{n-1},\lambda_n \rangle
\over \langle \lambda_{-2},\lambda_{-1},\lambda_{n-1},\lambda_n
\rangle\langle \lambda_{n-2},\lambda_{n-1},\lambda_n,\lambda_{n+1}
\rangle} \, . \eea
One can obtain the other two $\overline Y$-functions in the same way. The normalization independent forms are
\bea
{\overline Y}_3 (\zeta ) &=& {\langle
\tilde{s}_{-3},\tilde{s}_{-2},\tilde{s}_{-1},s_n \rangle \langle
s_{-2},s_{-1},s_{0},s_{n-1} \rangle \over \langle
s_{-2},s_{-1},s_{n-1},s_n \rangle\langle s_{-3},s_{-2},s_{-1},s_{0}
\rangle}(\zeta e^{-i\pi{n-2 \over 4}}) \, , \nonumber \\
 \qquad {\overline Y}_2  (\zeta ) &=&
{\langle \tilde{s}_{-2},\tilde{s}_{-1},s_{n},s_{n+1} \rangle \langle
s_{-1},s_{0},s_{n-1},s_{n} \rangle \over \langle
s_{-2},s_{-1},s_{0},s_n \rangle\langle s_{-1},s_{n-1},s_{n},s_{n+1}
\rangle}(\zeta e^{-i\pi{n-1 \over 4}}) \, ,  \eea
and in terms of spacetime coordinates they are
\bea
{\overline Y}_3 (\zeta = e^{i\pi{n-2 \over 4}} ) &=& {\langle
\hat\Omega\lambda_{-3},\hat\Omega\lambda_{-2},\hat\Omega\lambda_{-1},\lambda_n
\rangle \langle \lambda_{-2},\lambda_{-1},\lambda_{0},\lambda_{n-1}
\rangle \over \langle
\lambda_{-2},\lambda_{-1},\lambda_{n-1},\lambda_n \rangle\langle
\lambda_{-3},\lambda_{-2},\lambda_{-1},\lambda_{0}
\rangle} \, , \nonumber \\
 \qquad {\overline Y}_2  (\zeta = e^{i\pi{n-1 \over 4}} ) &=&
{\langle
\hat\Omega\lambda_{-2},\hat\Omega\lambda_{-1},\lambda_{n},\lambda_{n+1}
\rangle \langle \lambda_{-1},\lambda_{0},\lambda_{n-1},\lambda_{n}
\rangle \over \langle
\lambda_{-2},\lambda_{-1},\lambda_{0},\lambda_n \rangle\langle
\lambda_{-1},\lambda_{n-1},\lambda_{n},\lambda_{n+1} \rangle} \, .
\eea
Using the results in Appendix \ref{app-twistor}, it is also easy to write them in terms of Lorentz variables. 

\section{Three-point form factor \label{sec-3ptFF}}

In this section we study more explicitly the three-point case. This case is interesting because of its potential connection to QCD quantities as reviewed in the introduction. It also provides an example to  show explicitly how to write the truncation relations in terms of only Y-functions.

The three trace equations for the $n=3$ case are given  as
\bea
{\rm Tr}[\Omega] &=& {\cal B}_{11}\, T^{[3]}_{1,3} + \,  {\cal B}_{33} \, T^-_{1,1} -  {\cal B}_{22} \,R_{1,1}^+  \, , \\ {\rm Tr}[\overline\Omega] &=& \overline{\cal B}_{11}\, T^{[3]}_{3,3}  + \,  \overline{\cal B}_{33} \, T^-_{3,1} -  \overline{\cal B}_{22} \, R_{3,1}^+   \, , \\   {\rm Tr}^{(2)}[\Omega] &=& {\cal B}_{11}\, {\cal B}_{22}\, T_{2,3}^{[2]} + {\cal B}_{22}\, {\cal B}_{33}\, T_{1,2}  + {\cal B}_{11}\, {\cal B}_{44}\, T_{3,2}^{[4]} - {\cal B}_{11}\, {\cal B}_{33}\, W_{1,1}^{[2]}  \, , \qquad \eea
where, using the relations in section \ref{subsec-recursion},
\bea && R_{1,1} \ = \ { T_{2,1}^- \, T_{1,2}^+ + T_{2,2} \over T_{1,1} }  \, ,  \qquad R_{3,1} \ = \ { T_{2,1}^- \, T_{3,2}^+ + T_{2,2} \over T_{3,1} }  \, , \qquad \\ && W_{1,1} \ = \  - \, { T^-_{1,1} \, T^+_{1,1} \over T_{2,1} } + { (T_{2,2}^+ + T_{2,1} T_{3,2}^{[2]}) ( T_{2,2}^- + T_{2,1} T_{1,2}^{[-2]}) \over T_{2,1} \, T_{3,1}^+ \, T_{3,1}^- } \, .   \eea
These provide the truncation for the Hirota system by expressing $T_{a,3}$ through $T_{a,1}$, $T_{a,2}$ and the traces.

To construct the Y-system, one can notice that using the definition of Y-functions,
the T-functions can be solved in terms of  $Y$-functions (with also $A$ factors defined in (\ref{def_A})):
\bea && T_{1,1} = {A_1\over Y_{2,1} {\overline Y}_1} \, , \qquad T_{3,1} = {A_3 \over Y_{2,1} {\overline Y}_3} \, , \qquad T_{2,1} = {A_2 \over Y_{1,1} Y_{3,1} {\overline Y}_2}   \, , \\  &&  T_{1,2} = {A_2 \over Y_{3,1} {\overline Y}_2} \, , \qquad T_{3,2} = {A_2 \over Y_{1,1} {\overline Y}_2} \, , \qquad T_{2,2} = {A_1 A_3 \over  Y_{2,1} {\overline Y}_1 {\overline Y}_3}    \, . \qquad \eea

One can then substitute these expressions for $T$ functions into the trace equations.
The only thing which may cause trouble are the $A$ factors. They need to be cancelled since the expressions should be gauge invariant. Indeed, after a little calculation, one can express the trace conditions explicitly in terms of {\it only} Y-functions:
\bea \label{Y12}  {\rm Tr}[\Omega] &=& {Y_{1,2}^{[3]} \over {\overline Y}_1^{[3]}} + {1\over Y_{2,1}^- {\overline Y}_1^-} - {1\over {\overline Y}_3^+} \left( 1+ {{\overline Y}_1^+ {\overline Y}_3^+ \over {\overline Y}_2 {\overline Y}_2^{[2]} } {Y_{2,1}^+ \over Y_{1,1} Y_{3,1}^{[2]}} {1\over Y_{3,1} } \right)  , \\
\label{Y32} {\rm Tr}[\overline \Omega] &=& {Y_{3,2}^{[3]} \over {\overline Y}_3^{[3]}} + {1\over Y_{2,1}^- {\overline Y}_3^-} - {1\over {\overline Y}_1^+} \left( 1+ {{\overline Y}_1^+ {\overline Y}_3^+ \over {\overline Y}_2 {\overline Y}_2^{[2]} } {Y_{2,1}^+ \over Y_{3,1} Y_{1,1}^{[2]}} {1\over Y_{1,1} } \right)  , \\
\label{Y22} {\rm Tr}^{(2)}[\Omega] &=& {Y_{2,2}^{[2]} \over {\overline Y}_2^{[2]}} + {1\over Y_{3,1} {\overline Y}_2} + {1\over Y_{1,1}^{[4]} {\overline Y}_2^{[4]} } -  {{\overline Y}_3^+ {\overline Y}_3^{[3]} \over {\overline Y}_2 {\overline Y}_2^{[2]}  {\overline Y}_2^{[4]} } {Y_{2,1}^+ \over Y_{1,1}^{[2]} Y_{3,1}} {Y_{2,1}^{[3]} \over Y_{1,1}^{[4]} Y_{3,1}^{[2]} }   \\ &&
\hskip 0.1cm  - \, {{\overline Y}_3^+  \over {\overline Y}_1^{[3]}  {\overline Y}_2 } {Y_{2,1}^+ \over Y_{3,1} } - {{\overline Y}_3^{[3]}  \over {\overline Y}_1^{+}  {\overline Y}_2^{[4]} } {Y_{2,1}^{[3]} \over Y_{1,1}^{[4]} }  - {{\overline Y}_2^{[2]}  \over {\overline Y}_1^{+}  {\overline Y}_1^{[3]} } \left(Y_{1,1}^{[2]} Y_{3,1}^{[2]} - {Y_{1,1}^{[2]} Y_{3,1}^{[2]}  \over Y_{2,1}^+ Y_{2,1}^{[3]} } \right) . \nonumber \eea
All $A$ factors are cancelled exactly. This provides a non-trivial consistency check for our construction. As claimed before, $Y_{a,2}$ can be expressed in terms of other Y-functions and the trace functions.
The final Y-system is a closed system in terms of six Y-functions: three $Y_{a,1}$ and three $\overline Y_a$.

\subsection{Reduction to $AdS_4$}

A three-cusp periodic Wilson line  can be always embedded in an $AdS_4$ subspace of $AdS_5$. Therefore, one can simplify the system further to $AdS_4$.
As mentioned before, in $AdS_4$ one has $Y_{1,m} = Y_{3,m}$. The Y-system equations are given as
\bea \label{Yeqs1_n=3} {Y^-_{1,1}Y^+_{1,1} \over Y_{2,1}} \ = \
{1+Y_{1,2} \over 1+Y_{2,1}} \, , &\quad& {Y^-_{2,1}Y^+_{2,1} \over Y_{1,1}^2} \ = \ {1+Y_{2,2} \over (1+Y_{1,1})^2} \, , \\ \label{Yeqs2_n=3}
 { {\overline Y}_1^+ {\overline Y}_1^- \over {\overline Y}_{2} } \ = \ { 1+ Y_{1,1} \over 1+ Y_{2,2} } \, , &\quad& { {\overline Y}_2^+ {\overline Y}_2^- \over {\overline Y}_1^2 } \ = \ { 1+ Y_{2,1} \over (1+ Y_{1,2})^2 } \, , \eea
where by the trace condition and also using (\ref{trOmegavalue}),
\bea \label{Y12_n=3} Y_{1,2} &=& {\overline Y}_1 \left[ 4 - {1\over Y_{2,1}^{[-4]} {\overline Y}_1^{[-4]}} + {1\over {\overline Y}_1^{[-2]}} \left( 1+ { {\overline Y}_1^2 \over {\overline Y}_2^- {\overline Y}_2^+ } {Y_{2,1} \over Y_{1,1}^- Y_{1,1}^+} {1\over Y_{1,1}^- } \right)^{[-2]} \right] , \\
\label{Y22_n=3} Y_{2,2} &=& {\overline Y}_2 \left[ 6 - {1\over Y_{1,1}^{[-2]} {\overline Y}_2^{[-2]}} - {1\over Y_{1,1}^{[2]} {\overline Y}_2^{[2]} } +  {{\overline Y}_1^- {\overline Y}_1^+ \over {\overline Y}_2 {\overline Y}_2^{[-2]}  {\overline Y}_2^{[2]} } {Y_{2,1}^- \over Y_{1,1}^{[-2]} Y_{1,1}} {Y_{2,1}^+ \over Y_{1,1} Y_{1,1}^{[2]} } \right.  \\ &&
\hskip 0.5cm \left. + {{\overline Y}_1^-  \over {\overline Y}_1^+  {\overline Y}_2^{[-2]} } {Y_{2,1}^- \over Y_{1,1}^{[-2]} } + {{\overline Y}_1^+  \over {\overline Y}_1^-  {\overline Y}_2^{[2]} } {Y_{2,1}^+ \over Y_{1,1}^{[2]} }  + {{\overline Y}_2  \over {\overline Y}_1^-  {\overline Y}_1^+ } \left( Y_{1,1} ^2 - { Y_{1,1}^2  \over Y_{2,1}^- Y_{2,1}^+ } \right) \right] . \nonumber \eea
This is the Y-system which has potential connection to strong coupling {\it leading transcendental piece}\,%
\footnote{This is in the sense of first taking a summation of the {\it perturbative} leading transcendental results which is then evaluated at the strong coupling saddle point.} of Higgs-to-3-gluons amplitudes in QCD. Note that one may also use (\ref{Yeqs1_n=3})-(\ref{Yeqs2_n=3}) to rewrite them into other forms, in particular to change the phase shift of some $Y$ functions. 

The WKB approximation of Y-functions is determined only by the $P(z)$ function\,%
\be P(z) \ = \ {a_{-1} \over z} +  {1\over z^2} \, , \ee
which will be discussed in more details in section \ref{sec-ppoly}. The degrees of freedom also match: the complex number $a_{-1}$ provides two real parameters, while the three-cusp periodic Wilson line has also two independent ratios variables.

The equations (\ref{Y12_n=3}) and (\ref{Y22_n=3}) looks a little complicated. In particular, a new feature is that some functions  have large phase-shift which is beyond the physical strip $(-\pi/4, \pi/4)$. This will make it a little more complicated to write them in the form of integral equations, as the extra pole contributions need to be carefully considered. We leave this problem to another study.

Finally, we consider to express the Y-functions in terms of  spacetime coordinates.
As in the weak coupling, it is convenient to consider following variables
\be u \ := \ {p_{12}^2 \over q^2} , \qquad v \ := \ {p_{23}^2 \over q^2 } , \qquad w \ := \ {p_{31}^2 \over q^2}  \, ,
\ee
where $p_{ij}:=p_i + p_j$. There are only two independent variables since
\be q^2 \ = \ p_{12}^2 + p_{23}^2 + p_{31}^2 \, , \qquad u+v+w \, = \, 1 \, . \ee
The $Y$ functions in terms of these variables can be obtained as (see Appendix \ref{app-twistor})
\be \overline Y_{1} (\zeta= e^{i\pi/4}) \ = \ {
\langle \lambda_{-1},\lambda_{1},\lambda_{2},\lambda_3 \rangle
\over b_{-2} \, \langle \lambda_{-2},\lambda_{-1},\lambda_{2},\lambda_3
\rangle}  =  {1 \over 1/(1-w)+1} \, , \ee
\be \overline Y_{2} (\zeta= i) \ = \ {\langle
\lambda_1, \lambda_2,\lambda_{3},\lambda_{4}
\rangle \langle \lambda_{-1},\lambda_{0},\lambda_{2},\lambda_{3}
\rangle \over b_{-2}\, b_{-1}\, \langle
\lambda_{-2},\lambda_{-1},\lambda_{0},\lambda_3 \rangle\langle
\lambda_{-1},\lambda_{2},\lambda_{3},\lambda_{4} \rangle}  =  {v \over u\, w} \, , \ee
\be Y_{1,1} (\zeta= i) \ = \ {\langle \lambda_{-2}, \lambda_{-1}, \lambda_{0}, \lambda_{3} \rangle \langle \lambda_{-1}, \lambda_{0}, \lambda_{1}, \lambda_{2} \rangle \over \langle \lambda_{-1}, \lambda_{0}, \lambda_{2}, \lambda_{3} \rangle\langle \lambda_{-2}, \lambda_{-1}, \lambda_{0}, \lambda_{1} \rangle}  = - {u\, w \over v} \, , \ee
\be Y_{2,1} (\zeta= e^{i\pi/4}) \ = \ {\langle \lambda_{-2}, \lambda_{-1}, \lambda_{2}, \lambda_{3} \rangle \langle \lambda_{-1}, \lambda_{0}, \lambda_{1}, \lambda_{2} \rangle \over \langle \lambda_{-1}, \lambda_{1}, \lambda_{2}, \lambda_{3} \rangle\langle \lambda_{-2}, \lambda_{-1}, \lambda_{0}, \lambda_{2} \rangle}  =  {w \over 1-w} \Big( {1\over 1-w}+1\Big) \, . \ee

One can compare them with the interesting set of variables necessarily appearing  at weak coupling\cite{BTY} in constructing functions via the so-called symbol technique \cite{GSVV10}:
\be
\left\{ u,v,w,1-u,1-v,1-w,1-\frac{1}{u},1-\frac{1}{v},1-\frac{1}{w}, -\frac{u v}{w}, -\frac{v w}{u},-\frac{w u}{v} \right\} .
\ee
One can see that similar combinations appear in $Y$ functions. This is like the six-gluon amplitude case, where the variables in the symbol construction \cite{GSVV10} correspond to the Y-functions at strong coupling\,%
\footnote{There is also an intriguing relation between the symbol of three-point form factor and six-gluon amplitude at two-loop at weak coupling \cite{BTY, DDH11}. It would be interesting to study this further at strong coupling, although this is not obvious by naively looking at the Y-system equations.}. 
The three-point form factor provides a further evidence that the ``correct" variables for constructing functions from symbols at weak coupling, which is  hard to know (usually only through guess work), may be read directly from Y-functions.

\section{Form factors with multi-operator insertions \label{sec-ffmulti} }

In this section we consider form factors with multi-operator insertions
\be F(q_1, \cdots, q_l; p_1, \cdots, p_n) \ = \ \prod_{k=1}^l \, \int d^4 x_l \, e^{iq_k \cdot x_k} \, \langle {\cal O}(x_1) \cdots {\cal O}(x_l) \, | \, p_1 \cdots p_n \rangle \, . \ee
We first propose a dual picture for such observables. Then we construct Y-system for the $AdS_3$ case, with arbitrary number of operator insertions.

\subsection{Evidence at weak coupling \label{multi-spacetime-picture}}

We  first recall the picture of form factors with a single operator insertion. After T-duality, the picture involves a periodic null Wilson line boundary condition. The period is defined by the momentum of the operator. A duality between form factors and periodic Wilson lines was also found at weak coupling at one-loop \cite{BSTY10}. A dual MHV rule description was proved  for tree and one-loop form factors and also proposed to higher loops in \cite{BGMTY11}.

\begin{figure}[t]
\begin{center}
\includegraphics[height=4.8cm]{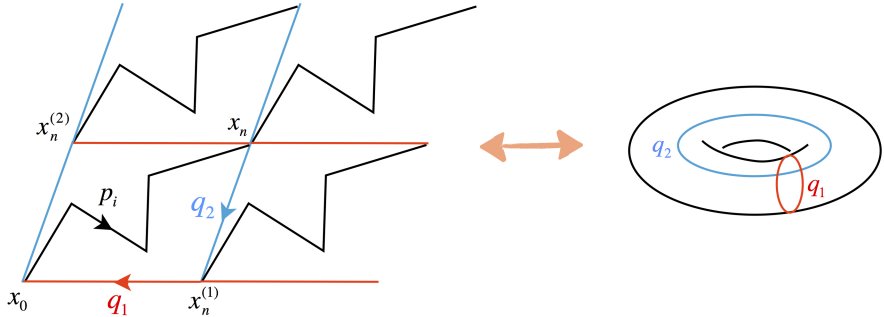}
\caption{\it The dual momentum space configuration for  form factors with two-operator inserted. There are two periodic directions which give a periodic two dimensional lattice picture. This is  equivalent to a torus.}
\label{fig-torus}
\end{center}
\end{figure}

How about a form factor with more than one operators? A natural generalization is  that in the T-dual picture, every operator will generate a periodic direction. For operator  ${\cal O}_i(q_i)$, the corresponding period is $q_i$. Such picture for form factor with two-operator inserted is shown in Figure \ref{fig-torus}.  It is given by a two dimensional periodic lattice and  is associated to a Torus topology. For general $m$-operator insertions, the corresponding topology is ${\mathds T}^m$.
The momentum space can be parametrized by introducing new coordinates $x^{(k)}_i$, which are defined as
\bea &&  x_{i+1} - x_i =  x^{(k)}_{i+1} - x^{(k)}_{i} = p_i \, , \ \qquad \ x^{(k)}_{i+n} - x_i = q_k    \, , \\ 
&& x_{i+n} - x_{i} = x^{(k)}_{i+n} - x^{(k)}_{i} = Q \, , \ \qquad \ Q : = \sum_{k=1}^l q_k \, , \eea
where $k= 1, \ldots, l$, and $l$ is the number of operators. See Figure \ref{fig-torus} for the $l=2$ case.

One particular support of this dual picture is that the dual MHV rule description \cite{BGMTY11} also applies to such generalized configuration, which provides an evidence that this dual picture may apply more generally.

In next subsection we will consider the worldsheet picture, where the introduction of multi-monodormy seems to be a natural generalization of the single insertion case. As we will discuss later in subsection \ref{muli-spacetime}, the monodromies in terms of the above spacetime coordinates can be also naturally defined.

\subsection{Small solution and multi-monodromy}

\begin{figure}[t]
\begin{center}
\subfigure{
\includegraphics[height=6.cm]{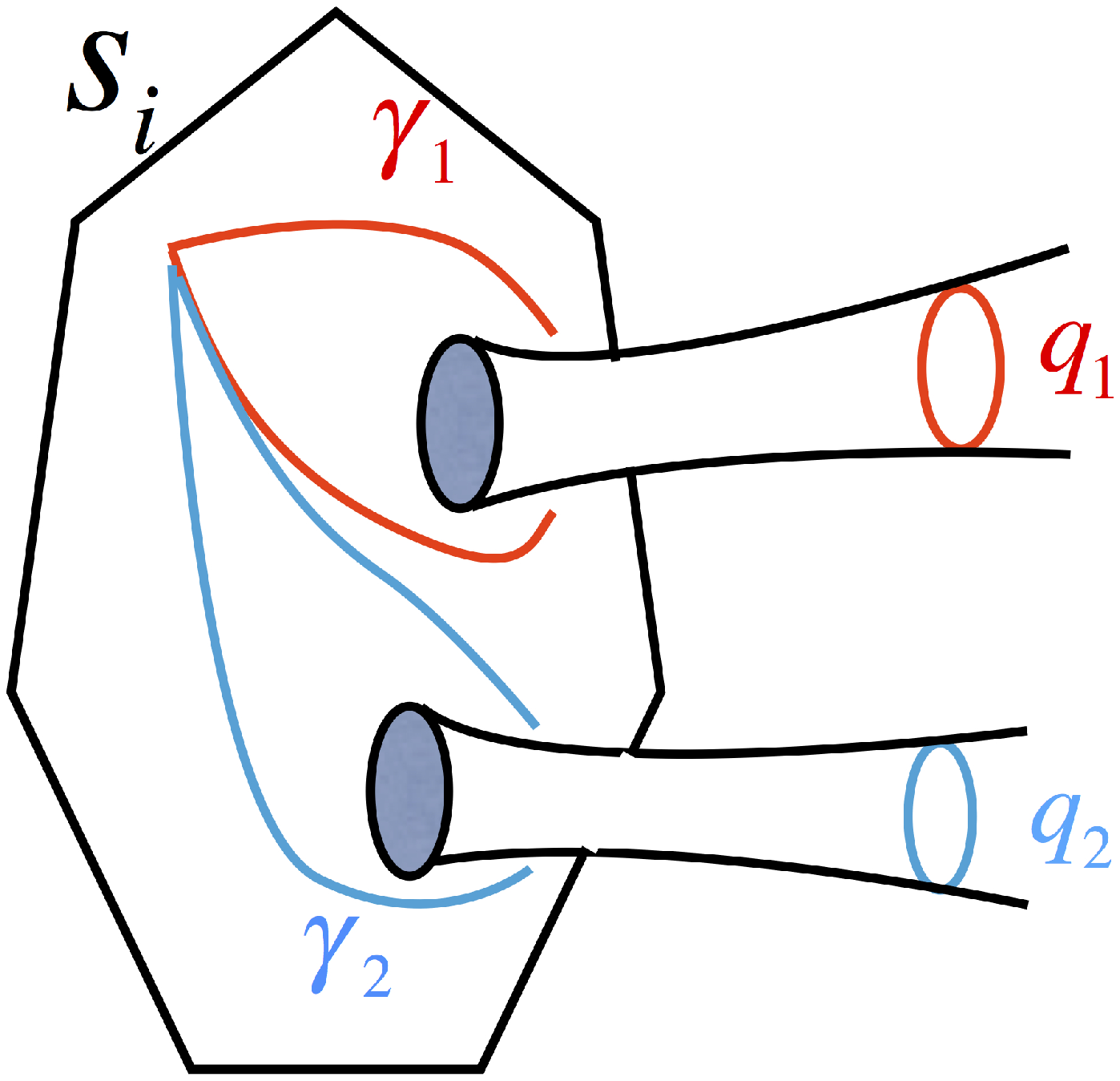}
} \hskip 1cm
\subfigure{
\includegraphics[height=5.5cm]{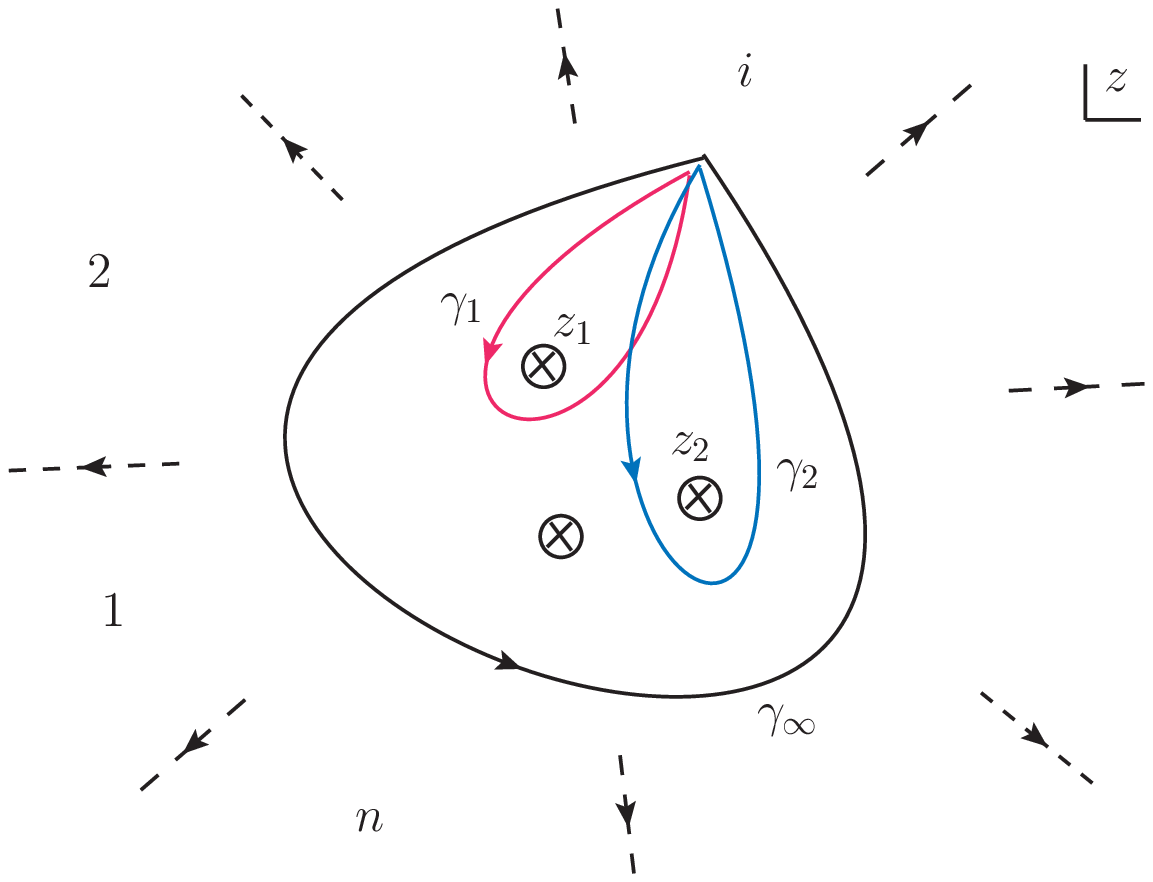}
}
\caption{\it Multi-monodromy picture on the string world-sheet.}
\label{fig-path}
\end{center}
\end{figure}

We first introduce the picture of path $\gamma_k$ on the worldsheet, as shown in Figure \ref{fig-path}. $\gamma_k$ is defined as a path that goes around the singular point $z_k$ where the operator ${\cal O}_k$ is inserted. A special path is $\gamma_\infty$ which surrounds all poles.  This special path is similar to that of the single-operator case,  $z e^{\oint\gamma_\infty} \sim z e^{i2\pi}$, in the sense that effectively one can take the combination of all operators as  one composite operator.

Next we introduce small solutions $s_i$ and $s_i^{(k)}$. This is inspired by the spacetime picture of $x_i$ and $x_i^{(k)}$ considered in last subsection, see Figure \ref{fig-torus}.

The set of small solutions $s_i$ is related to the special path $\gamma_\infty$. As mentioned above, they behave similarly as those of the form factor with a single operator inserted, as one can take all operators effectively as a single operator. Therefore one has the same relations 
\be \label{multicase-B-infinity}  s_{\hat n}(z, \zeta) \ = \ B(\zeta) \, s_0(z e^{-\oint \gamma_\infty}, \zeta) \, , \ee
\be
\begin{pmatrix} s_{1} \\ s_0
\end{pmatrix}(ze^{\oint\gamma_\infty},\zeta) \ =\ \Omega(\zeta)\begin{pmatrix} s_1 \\
s_0
\end{pmatrix}(z,\zeta)\, , \ee
as the form factor studied in section \ref{sec-ffads3}.
Using this set of small solutions, one can also define the $T$- and $Y$-functions and construct the Y-system equations in exactly the same way. The monodromy $\Omega$ should correspond to the product of the monodromies of all operators.

The small solutions $s^{(k)}_i$ is 
similar to the dual coordinate $x_i^{(k)}$ in spacetime. 
Because of the periodic structure, each set of small solutions $s^{(k)}_i$ with {\it fixed $k$} is not different from the set of small solutions $s_i$.  One has  $\langle s^{(k)}_i, s^{(k)}_j \rangle = \langle s_i, s_j \rangle$. Similarly the $Z_2$ automorphism relations also apply such that
$s^{(k)}_{i+1} (\zeta) = i \sigma_3 \, s^{(k)}_i(e^{i\pi} \zeta)$.

The information of monodromy is encoded in the relation between $s_i$ and $s^{(k)}_j$.

Recall that by definition,  $s_{\hat n+i}$ is the small solution in the same sector as $s_i$ but after going around the complex $z$-plane (more exactly the path $\gamma_\infty$ which surrounds all operators) once, therefore they should be proportional with each other as given in (\ref{multicase-B-infinity}).
Similarly, $s^{(k)}_{\hat n+i}$ is defined as a small solution in the
same sector as $s_i$ but after going around the path $\gamma_k$ once,  
as shown in Figure \ref{fig-path}. One also has the proportionality relation $s^{(k)}_{\hat n+i}(z,\zeta) \propto s_i(z e^{-\oint \gamma_k}, \zeta)$.  One can introduce proportionality constants $B^{(k)}$ as
\be s^{(k)}_{\hat n}(z,\zeta) \ = \ B^{(k)} \, s_0(z e^{-\oint \gamma_k}, \zeta) \, .  \ee
%

Because of the insertion of operators, small solutions are not single-valued. For each path $\gamma_k$, one can define a corresponding monodromy matrix as
\be
\begin{pmatrix} s_1 \\ s_0
\end{pmatrix}(ze^{\oint\gamma_k},\zeta) \ = \ \Omega^{(k)}(\zeta)\begin{pmatrix} s_1 \\
s_0
\end{pmatrix}(z,\zeta)  \, , \ee
which is similar to the single insertion case.

At the same worldsheet point, one has
\be \begin{pmatrix} s^{(k)}_{\hat n+1} \\
s^{(k)}_{\hat n} \end{pmatrix} (z, \zeta) \ = \ {\cal B}^{(k)}(\zeta) \cdot (\Omega^{(k)})^{-1}(\zeta) \begin{pmatrix} s_1 \\
s_0 \end{pmatrix} (z, \zeta) \, , \ee
where
\be {\cal B}^{(k)}(\zeta) \ = \ \begin{pmatrix} (B^{(k)})^{[2]}(\zeta) & 0 \\ 0 & B^{(k)}(\zeta) \end{pmatrix} , \qquad \det[ {\cal B}^{(k)}(\zeta) ] = 1 \, , \ee
and similar relation between the $s_{i+\hat n}$ and $s_{i}$ as (\ref{ads3-basic}).
Since $e^{\oint \gamma_\infty} = \prod_{k=1}^l e^{\oint \gamma_k}$, one has
\be   \Omega \ = \ \prod_{k=1}^l  \, \Omega^{(k)} \, . \ee
Note that the order of operators should be not important, which implies that $\Omega^{(k)}$ should commute with each other. This requirement imposes further constraints on the monodromy matrices as will be discussed later\,%
\footnote{The commutativity of monodromy matrices is also implied by the spacetime monodromy explicitly derived in section \ref{muli-spacetime} for short operators. It makes the counting of the degrees of freedom look also more consistent, as discussed at the end of section \ref{sec-ysystem-mult}. For general operators this constraint  to monodromy matrices may not apply. It is also important to understand further the relation of this constraint to the fundamental group of Riemann surfaces which is in general not Abelian. We would like to thank Till Bargheer for discussion on this point.}.

\subsection{New T and Y functions}

Now we consider the definition of $T$ and $Y$ functions and their relations.
As we mentioned before, if one just focuses on the set of small
solution $s_i$, one obtains a Y-system  exactly
the same as the single operator case, with the total monodromy $\Omega$.
The same Y-system can be constructed  for the set of small solutions
$s^{(k)}_i$ with {\it fixed} $k$, since $\langle s_i, s_j \rangle =
\langle s^{(k)}_i, s^{(k)}_j \rangle$.
The new degrees of freedom due to multi-operator insertions are contained in the interplay between small solutions $s_i$ and $s^{(k)}_i$, which would necessarily involve the monodromy $\Omega^{(k)}$. 

We define new $T$-functions as
\be  \begin{array}{lll} T^{(k)}_{1,2m+1} \ := \ \langle s_{-m-1}, s^{(k)}_{m+1}\rangle, & \quad\qquad &
T^{(k)}_{1,2m} \ := \ \langle s_{-m-1}, s^{(k)}_{m}\rangle^+,
 \\
T^{(k)}_{0,2m} \ := \ \langle s_{-m-1}, s^{(k)}_{-m}\rangle,  \begin{array}{ll} ~ \\ ~ \end{array} & \quad\qquad &
T^{(k)}_{0,2m+1} \ := \ \langle s_{-m-2}, s^{(k)}_{-m-1}\rangle^+,
 \\
T^{(k)}_{2,2m} \ := \ \langle s_{m}, s^{(k)}_{m+1}\rangle, & \quad\qquad &
T^{(k)}_{2,2m+1} \ := \ \langle s_{m}, s^{(k)}_{m+1}\rangle^+ \, ,
 \end{array} \ee
and $T_{a,m}^{(k)}=0$ if $a\neq 0,1,2$.
Note that $T^{(k)}_{0,m}, T^{(k)}_{2,m}, T^{(k)}_{1,0}$  are not normalized to be 1, and  $T^{(k)}_{1,-1} = \langle s_0, s^{(k)}_0 \rangle \neq 0$. Using $Z_2$ automorphism, one gets the shifting relation $\langle s_{i+1}, s_{j+1}^{(k)} \rangle = \langle s_i, s_j^{(k)} \rangle^{[2]}$.

Despite the difference,  one still has the
Hirota equations by using  Schouten identity (see appendix \ref{app-funeqs})
\be T^{(k)+}_{a,m} \, T^{(k)-}_{a,m} \ = \ T^{(k)}_{a,m-1} \, T^{(k)}_{a,m+1} + T_{a-1,m} \, T_{a+1,m} \, , \ee
where $a=1,2,3$. Similarly, $Y$-functions can be defined as
\be Y^{(k)}_{a,m} \ = \ {T^{(k)}_{a,m-1}T^{(k)}_{a,m+1} \over T_{a-1,m}T_{a+1,m}} \, . \ee
The Hirota equations give the equations for $Y$-functions as
\be {Y^{(k)+}_{a,m}Y^{(k)-}_{a,m} \over
Y_{a-1,m}Y_{a+1,m}} \ = \ {(1+Y^{(k)}_{a,m-1})(1+Y^{(k)}_{a,m+1}) \over
(1+Y_{a-1,m})(1+Y_{a+1,m})} \, , \ee
where $a=1$. In the normalization
$\langle s_i, s_{i+1}\rangle \ = \ \langle s_i^{(k)}, s_{i+1}^{(k)}\rangle =1$, the above equations are simplified as
\be Y^{(k)+}_m Y^{(k)-}_m \ = \ (1+Y^{(k)}_{m-1})(1+Y^{(k)}_{m+1}) \, , \ee
where $Y^{(k)}_m := Y^{(k)}_{1,m}$.

\subsection{Truncations and Y-system \label{sec-ysystem-mult}}

The main challenge is to construct a finite integrable system.
We will firstly consider the truncation of Hirota equations, and then show how to write them into a gauge invariant Y-system.

Similar to the single-insertion case (\ref{TrOmega-ads3}), one can introduce the trace relation
\be {\rm Tr}[\Omega^{(k)}(\tilde\zeta)] \ = \ B^{(k)}(\tilde\zeta) \,T^{(k)}_{\hat n}(\zeta) - (B^{(k)})^{-1}(\tilde\zeta)\, T^{(k)}_{\hat n -2}(\zeta)  \, , \ee
where $\tilde\zeta = e^{-i ({\hat n}+1) \pi/2}$.
$T^{(k)}_{\hat n}$ can be expressed in terms of $T^{(k)}_{\hat n-2}$ and ${\rm Tr}[\Omega^{(k)}]$, which provides a truncation for the chain of Hirota equations from the right-hand side. 

One can then  define
\be {\overline Y}^{(k)}(\zeta) \ := \ (B^{(k)})^{-1}(\tilde\zeta)\, T^{(k)}_{\hat n -2}(\zeta)  \, , \ee
and obtain
\bea T^{(k)}_{\hat n}(\zeta) & = & (B^{(k)})^{-1}(\tilde\zeta) \left[ {\rm Tr}[\tilde\Omega^{(k)}](\zeta) + \overline Y^{(k)}(\zeta) \right] \, , \\ Y_{\hat n-1}^{(k)} & = & {\rm
Tr}[\tilde\Omega^{(k)}] \overline{Y}^{(k)}+[\overline{Y}^{(k)}]^2 \, , \eea
where $\tilde\Omega^{(k)}(\zeta) := \Omega^{(k)}(\tilde\zeta)$.
From this one has the equations
\bea  Y^{(k)+}_{{\hat n}-2} Y^{(k)-}_{{\hat n}-2} &=& (1+ Y^{(k)}_{{\hat n}-3}) \big[ 1 + {\rm Tr}(\tilde\Omega^{(k)}) \, \overline Y^{(k)} +  (\overline Y^{(k)})^2 \big] \, , \\
\overline Y^{(k)+} \overline Y^{(k)-} &=& 1 + Y^{(k)}_{{\hat n}-2}  \, . \eea
Naively, one may introduce $Y^{(k)}_{{\hat n}-2}$ and $\overline Y^{(k)}$ for each new insertion of operator. However, the equation for $Y^{(k)}_{\hat n-2}$ contains $Y^{(k)}_{{\hat n}-3}$, whose equation would then involve $Y^{(k)}_{{\hat n}-4}$ and so on. This means that we also need to ``truncate" the equations from the left-hand side, so that to formulate the equations into a finite system.

We find it convenient to apply the relation
\be \langle s_1, s^{(k)}_{m -1}\rangle \, \langle s^{(k)}_m, s^{(k)}_{m+1}\rangle \ = \ \langle s_1, s^{(k)}_m \rangle \, \langle s^{(k)}_{m -1}, s^{(k)}_{m+1}\rangle - \langle s_1, s^{(k)}_{m +1}\rangle \, \langle s^{(k)}_{m-1}, s^{(k)}_m\rangle \, ,  \ee
which written in terms of T-functions is (up to a phase shift)
\be T^{(k)}_{m} \ = \ T_{m+1}^{(k)+} \ T_1^{[m+3]} -T_{m+2}^{(k)[2]} \, .  \ee
This provides a recursion relation for $T^{(k)}_{m}$, and one can
express all $T^{(k)}$ in terms of only two $T^{(k)}$ functions and
$T_1$. For our purpose it is enough to consider\,%
\footnote{One may also use this relation to solve for $T_{\hat n}^{(k)}$ in terms of $T_{\hat n-2}^{(k)}$ and $T_{\hat n-1}^{(k)}$, and then substitute it into $Y_{\hat n-1}^{(k)} = T_{\hat n-2} T_{\hat n}$. This will give the same equation as (\ref{introduce_Z}).}
\be T^{(k)}_{\hat n -3} \ = \ T_{\hat n-2}^{(k)+}\ T_1^{[\hat n]} -T_{\hat n-1}^{(k)[2]} \ .  \ee
Together with the trace relation involving $T^{(k)}_{\hat n}$, one can truncate the chain of Hirota equations which involve  only $T_{\hat n-2}^{(k)}$ and $T_{\hat n-1}^{(k)}$.

We need  to further write the truncated Hirota system into  a gauge invariant Y-system.
To do this, one can first write $Y^{(k)}_{\hat n-2}$ as
\bea Y_{\hat n-2}^{(k)} &=&
T_{\hat n-3}^{(k)}T_{\hat n-1}^{(k)}=(T_{\hat n-2}^{(k)+}T_1^{[\hat n]}-T_{\hat n-1}^{(k)[2]})T_{\hat n-1}^{(k)}
\nonumber \\
 &=&Z^{(k)} -1- Y_{\hat n-1}^{(k)+}  \ , \label{introduce_Z} \eea
where a new function $Z^{(k)}$ is introduced as
\be Z^{(k)} \ := \ T_{\hat n-2}^{(k)+} \, T_{\hat n-1}^{(k)} \, T_1^{[\hat n]} \, . \ee
The advantage of introducing $Z^{(k)}$ is that it is straightforward to obtain the equation
\be Z^{(k)+}\, Z^{(k)-}  \ = \
\left( Z^{(k)} - Y_{\hat n-1}^{(k)+} \right)^+ \left( Y_{\hat n-1}^{(k)}+1 \right) \left( Y_{1}^{[\hat n]}+1 \right) \, .
 \ee
In particular, $Y^{(k)}_{{\hat n}-3}$ no longer appears, and one gets a closed set of equations.
Therefore, rather than use $Y_{\hat n-2}^{(k)}$, we will use function $Z^{(k)}$.

To summarize, one obtains a closed  $Y$-system with the functions $Y_m$, $\overline Y$,  $Z^{(k)}$, and $\overline{Y}^{(k)}$:
\bea  Y_m^+Y_m^- &=& (1+Y_{m+1})(1+Y_{m-1})\, , \qquad m=1,..., \hat n-3 \, , \\  Y_{{\hat n}-2}^+ Y_{{\hat
n}-2}^- &=& (1+ Y_{{\hat n}-3}) ( 1 + {\rm Tr}[\tilde\Omega] \overline Y + \overline Y^2 )\, , \\
\overline Y^+ \overline Y^- &=& 1 + Y_{{\hat n}-2} \, , \eea
%
%
\bea
\label{eqforZ} Z^{(k)+}\, Z^{(k)-}  &=&
\left( Z^{(k)+}-Y_{\hat n-1}^{(k)[2]} \right) \left( Y_{\hat n-1}^{(k)}+1 \right) \left( Y_{1}^{[\hat n]}+1 \right)  \, , \\
\overline{Y}^{(k)+} \overline {Y}^{(k)-} &=& Z^{(k)}  - Y_{\hat n-1}^{(k)+}\, ,  \eea
where
\be Y_{\hat n-1}^{(k)} \ = \ {\rm
Tr}[\tilde\Omega_k] \overline{Y}^{(k)}+[\overline{Y}^{(k)}]^2 \, , \ee
and $k=1, \ldots , l-1$.
One can see the function  $Y_{\hat n-1}^{(k)}$ in (\ref{eqforZ}) has phase shift which is beyond the physical strip $(-\pi/2, \pi/2)$. To write it into an integral equation, the extra pole contribution should be considered.

We  comment on the degrees of freedom. For each operator ${\cal O}_k$, two new functions are introduced.  One may understand this using the same argument of the single insertion case: the new $2\times 2$ unitary matrix $\Omega^{(k)}$ subtracting the trace ${\rm Tr}[\Omega^{(k)}]$ leaves two independent components.  However, there are extra constraints that the monodromy matrices commute with each other. This implies the matrices should be in general in the form of 
\be \begin{pmatrix} a_k & 0 \\ b_k & 1/a_k \end{pmatrix} \, , \ee
and only one new degree of freedom is introduced for each new operator (for example, given $b_k$ then $a_k$ is fixed). This implies that  $\overline Y^{(k)}$ and $Z^{(k)}$ are not independent.

This matches with the degrees of freedom from spacetime boundary configuration that we considered in subsection \ref{multi-spacetime-picture}, where each new operator introduces a new periodic direction characterized by $q$. Note that one $Y$-function in $AdS_3$ gives two real degrees of freedom, due to the left and right hand decomposition. Since the boundary information enters into the Y-system via the WKB approximation, this is also related to the structure of $P(z)$ function which will be discussed in section \ref{sec-ppoly}.

\subsection{Spacetime picture \label{muli-spacetime}}

Now we consider monodromy in terms of spacetime variables. We first recall the dual momenta space configuration
\bea &&  x_{i+1} - x_i \ = \  x^{(k)}_{i+1} - x^{(k)}_{i} = p_i \, , \qquad x^{(k)}_{i+n} - x_i = q_k    \, , \\ 
&& x_{i+n} - x_{i} \ = \ x^{(k)}_{i+n} - x^{(k)}_{i} = Q  = \sum_{i=1}^l q_i \, . \eea
Each monodromy corresponds to a conformal transformation which maps $\{ X^{(k)}_{n}, X^{(k)}_{n+1}, X^{(k)}_{n+2} \}$ to
$\{ X_0, X_1, X_2 \}$ and can be  defined in terms of left-hand spinors as
\be \lambda^{(k),L}_{i+n} \ \propto \ \hat\Omega^{(k)} \,  \lambda^L_i \ = \ \begin{pmatrix} \hat\Omega^{(k)}_{11} \lambda^L_{i,1} + \hat\Omega^{(k)}_{12} \lambda^L_{i,2} \\ \hat\Omega^{(k)}_{21} \lambda^L_{i,1} + \hat\Omega^{(k)}_{22} \lambda^L_{i,2} \end{pmatrix}  ,  \ee
where $i = 0, 1, 2$.
One has
\be \hat\Omega^{(k)} \ = \  \begin{pmatrix} 1 & 0 \\ q_k & 1 \end{pmatrix} \, , \qquad  \hat\Omega \ = \  \begin{pmatrix} 1 & 0 \\ Q & 1 \end{pmatrix}  \, , \ee
which  are indeed commuted with each other and satisfy $\hat\Omega = \prod_{l=1}^k \hat\Omega^{(k)}$.

One can express $\overline Y^{(k)}$ and $Z^{(k)}$ in terms of spacetime variables which specify the shape of dual Wilson line configuration.  For  $\overline Y^{(k)}$ functions, one has similar to ${\overline Y}$
\be {\overline Y}^{(k)}(\zeta= i^{\hat n+1}) \ = \  -\, {\langle \lambda^{L}_1, \lambda^{L,(k)}_{\hat n} \rangle \over \langle \hat\Omega^{(k)} \lambda^{L}_1 , \lambda^{L,(k)}_{\hat n} \rangle}  \ = \ - { x_{1}^{+} - x_{\hat n}^{(k)+} \over x_{\hat n+1}^{(k)+} -x_{\hat n}^{(k)+}} \ . \ \ee
For function $Z^{(k)}$, one can first write it in a gauge
invariant form
\be Z^{(k)}(\zeta) \ = \ {T_{1,\hat n-2}^{(k)+} \, T_{1,\hat
n-1}^{(k)} \, T_{1,1}^{[\hat n]} \over T_{0,\hat n-2} \, T_{2,\hat
n-2} \, T_{2,0}^{[\hat n]}}
\ = \ {\langle s_0, s^{(k)}_{\hat n }\rangle \, \langle s_1, s^{(k)}_{\hat n }\rangle \,  \langle s^{(k)}_{\hat n-1}, s^{(k)}_{\hat n+1}\rangle \over \langle s_0, s_{1}\rangle \, \langle s^{(k)}_{\hat n-1}, s^{(k)}_{\hat n}\rangle \, \langle s^{(k)}_{\hat n}, s^{(k)}_{\hat n+1}\rangle }(\zeta e^{-i\pi{\hat n \over 2}})  \, . \ee
Then the spacetime expression can be given as
\be Z^{(k)} (\zeta= i^{\hat n}) \ = \ {( x_0- x^{(k)}_{\hat n }) \, (x_1- x^{(k)}_{\hat n }) \,  ( x^{(k)}_{\hat n-1} - x^{(k)}_{\hat n+1}) \over ( x_0 - x_{1}) \, ( x^{(k)}_{\hat n-1} -x^{(k)}_{\hat n}) \, ( x^{(k)}_{\hat n} - x^{(k)}_{\hat n+1}) } \, , \ee
which is manifestly conformally invariant.

\section{Function $P(z)$ and WKB approximation \label{sec-ppoly} }

As reviewed in section \ref{sec-review}, the boundary conditions are related to the holomorphic function $P(z)$ which is also related to the WKB approximation. In this section, we study this in more details. We will focus on the cases with short operators, which are dual to periodic Wilson line configurations. The general structure of $P(z)$ will be proposed. We will also discuss the general pattern of corresponding WKB lines.

\subsection{$P(z)$ for general form factors}

For amplitudes or null Wilson loops, $P(z)$ is a pure polynomial. For form factors, due to the insertion of operators,  pole terms are involved. This can be understood by studying the behavior of the solution near the horizon. Our discussion is for $AdS_5$ cases,  following $AdS_3$ in \cite{MZ10}.  

The main trick is that by doing a worldsheet conformal transformation, one can bring the horizon at infinity to the origin in the new coordinate. Firstly we recall the picture of the dual surface in Figure \ref{Tdual}.
Without loss of generality, one can take the periodic direction to be along $x_1$. We parametrize the worldsheet by the coordinate $\tilde z = r + i x_1$.
Near the horizon where $r\rightarrow \infty$ or $\tilde z \rightarrow \infty$, the surface is asymptotically  a straight strip. We can set $x_2 = x_3 =t = 0$ up to a translation. The induced Poincare metric takes the form
\be d s_{\rm ind}^2 \ = \ { d \tilde z d \bar {\tilde z}  \over (\tilde z + \bar {\tilde z})^2 } \ . \ee
For this simple solution, the corresponding polynomial is simply $\tilde P(\tilde z) = 0$.

One can now apply a standard coordinate transformation to map the strip to the unit disc with new coordinate $z$
\be z \ = \ e^{- \tilde z} \ . \ee
The infinity of $\tilde z$ becomes the origin of $z$. It is in this new coordinate $z$ that we discuss the  picture of the worldsheet monodromy in previous sections\,%
\footnote{For small solutions the boundary is at $|z|\rightarrow\infty$. It seems there would be a problem as $|z| \rightarrow \infty$ implies $|\tilde z| = r \rightarrow- \infty$. As our focus here is on the behavior near the horizon, one should think that above transformation is only  for the region near the horizon.}. In the new coordinate, the above induced metric takes the form
\be d s_{\rm ind}^2 \ = \ { d  z d \bar { z}  \over z \bar z \log^2(z \bar z) } \ = \ e^{-2 \alpha} d z d \bar z \ , \qquad  \alpha \ = \ -\, {1\over 2 } \log( z \bar z \log^2(z \bar z)) \ . \ee
This provides the boundary condition for the solution when $z\rightarrow 0$. In particular, one can see that unlike amplitudes, $\alpha$ is not regular near the origin, which means the worldsheet is no longer smooth. Near the cusps when $z\rightarrow \infty$, one still has $\hat \alpha\rightarrow0$.

In $AdS_3$,  the function  $p(z) \sim N\cdot \partial^2 X$ \cite{AM09} and have the transformation property
\be \tilde p(\tilde z) \ = \ \left( {\partial \tilde z \over \partial z } \right)^{-2} p(z) \ = \ z^2 \, p(z) \, . \ee
As discussed above when $\tilde z \rightarrow \infty$ (or $z\rightarrow 0$), $\tilde p (\tilde z) \rightarrow 0$, this implies that
\be p(z) \ = \ {1\over z} + {\cal O}(z^0) \, . \ee
For the $AdS_5$ case, $P(z)=\partial^2 X \cdot \partial^2 X$ which gives
\be \tilde P(\tilde z) \ = \ \left( {\partial \tilde z \over \partial z } \right)^{-4} P(z) \ = \ z^4 \, P(z) \, . \ee
The condition $\tilde P(\tilde z)=0$ when $\tilde z \rightarrow \infty$ requires that
\be P(z)  \ = \ {c_1\over z} + {c_2\over z^2} + {c_3\over z^3} + {\cal O}(z^0) \, . \ee
However,  $1/z^3$ term is not allowed. This can be understood as that near the horizon, the $AdS_5$ solution can be embedded into $AdS_3$\,%
\footnote{This is true for short operators dual to periodic Wilson lines, but not necessary for more general operators, where the pole structure could be more complicated.}, which must then satisfy $P(z) \propto p(z)^2$.

One concludes that for form factors in $AdS_5$, $P(z)$ has the
following general structure
\be P(z) \ = \ a_{n-4} z^{n-4}  + \cdots  + a_1 z + a_0 + {a_{-1} \over z} +  {1\over z^2} \ . \ee
For the $n=2$ case (which can be always embedded in $AdS_3$), one has
\be P(z) \ = \ p(z)^2 \ = \ {1\over z^2} \, . \ee
For the three-point case it is given as
\be P(z) \ = \  {a_{-1} \over z} +  {1\over z^2} \, . \ee

For a general $n$-point form factor, the number of parameters  from the coefficients $a_i$ is
\be 2(n-3) + 2 \ . \ee
For the $AdS_4$ case this matches exactly with the degrees of freedom from a counting of the symmetries of a periodic null Wilson line configuration. For $AdS_5$, there should be further $(n-3)$ parameters from gauge connection, as in the case of scattering amplitudes \cite{AMSV10}. In total the number of parameters is $3n - 7$, which is also consistent with the counting of symmetries.

For multi-operator insertions, a natural proposal is that each operator introduces one new pole term. For example, for the case with two cusps and two operators we would have
\be \label{pforF22} {1\over z} + {a^{(2)} \over z - z_0^{(2)}} \, , \ee
where we have used  scaling and translational symmetries of the worldsheet theory to set $a^{(1)}=1$ and $z_0^{(1)}=0$. In the limit that $z_0^{(2)}\rightarrow0$, it reduces to the single operator form, which is consistent with the picture in section \ref{sec-ffmulti}.   

In this proposal, the degree of the polynomial is related to the number of cusps, and the number of poles corresponds to the number of operators inserted. Each $P(z)$ function defines an algebraic curve, or a Riemann surface. The numbers of genera and singularities are related to the numbers of cusps and operators.  It also produces a consistent WKB line picture as shown in an example in the next subsection.

\begin{table}[t]
\begin{center}
\begin{tabular}{ | l | c | }
\hline
$\begin{matrix} \, \\ \,   \end{matrix}$ &   $AdS_3$ \quad (${\hat n} = n/2$)   \cr \hline
$\begin{matrix} \, \\ \,   \end{matrix}$ Amplitudes   & $p(z) = z^{{\hat n}-2} + a_{{\hat n}-4} \, z^{{\hat n}-4} + \cdots + a_1\, z + a_0$    \cr \hline
$\begin{matrix} \, \\ \,   \end{matrix}$   One-operator  &  $p(z) = z^{{\hat n}-2} + a_{{\hat n}-4} \, z^{{\hat n}-4} + \cdots + a_0 +  {a_{-1} \over z - z_0}$  \cr \hline
$\begin{matrix} \, \\ \,   \end{matrix}$ Multi-operator   &  $p(z) = z^{{\hat n}-2} + a_{{\hat n}-4} \, z^{{\hat n}-4} + \cdots + a_0 + \sum_{i=1}^m {a^{(i)}_{-1} \over z - z^{(i)}_0}$   \cr \hline
$\begin{matrix} \, \\ \,  \end{matrix}$ & $AdS_5$ and $AdS_4$   \cr \hline
$\begin{matrix} \, \\ \,   \end{matrix}$  Amplitudes  & $P(z) = z^{n-4} + a_{n-6} \, z^{n-6} + \cdots + a_1 \, z + a_0$    \cr \hline
$\begin{matrix} \, \\ \,   \end{matrix}$  One-operator   &  $P(z) = z^{n-4} + a_{n-6} \, z^{n-6} + \cdots + a_0 +  {a_{-1}  \over z - z_0 } + { a_{-2} \over (z - z_0 )^2} $  \cr \hline
$\begin{matrix} \, \\ \,   \end{matrix}$  Multi-operator   &  $P(z) = z^{n-4} + a_{n-6} \, z^{n-6} + \cdots + a_0 + \sum_{i=1}^m  {a^{(i)}_{-1} \,  z + a^{(i)}_{-2} \over (z - z^{(i)}_0 )^2} $  \cr \hline
\end{tabular}
\caption{\it $P(z)$ function for amplitudes and form factors in $AdS_3$ and $AdS_5$. $n$ is the number of cusps, $m$ is the number of operators inserted. They apply to short operators which are dual to periodic configurations.
\label{tab-polyP}
}
\end{center}
\end{table}

However, this doesn't seem to be the full story. The problem is that the remaining two complex parameters in (\ref{pforF22}) give four degrees of freedom, which do not match with the T-dual spacetime picture which has only 2 degrees of freedom. This implies that one may need to impose extra constraints on the coefficients $a_{-1}^{(k)}$ and $z_0^{(k)}$ for each insertion.  This seems to require a better knowledge of the T-dual picture of the minimal surface from which one may do a similar study as for the single insertion case. 

There is also another possibility. Although the function $P(z)$ contains more parameters, the final area may be independent of these extra degrees of freedom. In other words, some of the parameters in $P(z)$ may be taken as ``gauge-like" degrees of freedom, and one may change them without changing the physical area. This picture seems more natural but needs to be checked through a detailed study of the area. 

In either case, we believe that the general structures of  $P(z)$ functions are correct.
We summarize them in Table \ref{tab-polyP}.

\subsection{WKB approximation}

The asymptotic behavior of $Y$-functions can be determined by $P(z)$ through WKB approximation. The WKB lines are defined by the parametric line $z(t)$ as:
\be {\rm Im}\Big( { x \over \zeta} \, {d z(t) \over dt} \Big) \ = \ 0  \, ,\ee
where 
\be AdS_3: \quad p(z) = x^2  \, , \qquad AdS_5: \quad P(z) = x^4 \, . \ee
The $AdS_3$ case corresponds to the $SL(2)$ Hitchin system which has been
studied in details in \cite{GMN09}. As $\theta$ changes ($\zeta = e^{i\theta}$), the WKB lines
will change correspondingly which is related to the
wall-crossing phenomenon in ${\cal N}=2$ theory. Although the physical context
looks quite different here, the mathematics is basically the same.
Below we  summarize the general patterns of WKB lines for both $AdS_3$
and $AdS_5$ cases.

\begin{figure}[t]
\begin{center}
\subfigure{
\includegraphics[height=4.2cm]{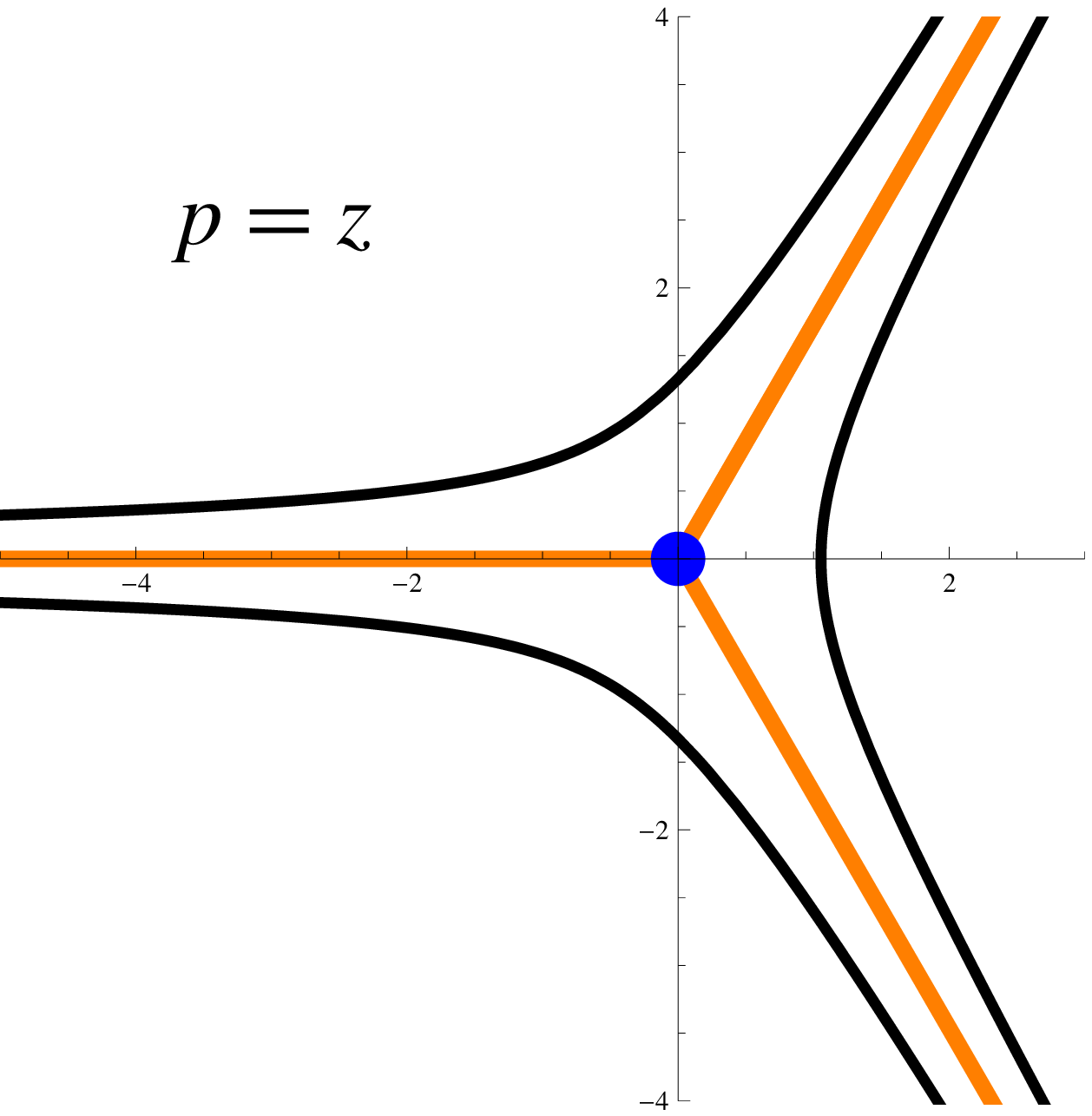}
} \hskip 4.cm
\subfigure{
\includegraphics[height=4.2cm]{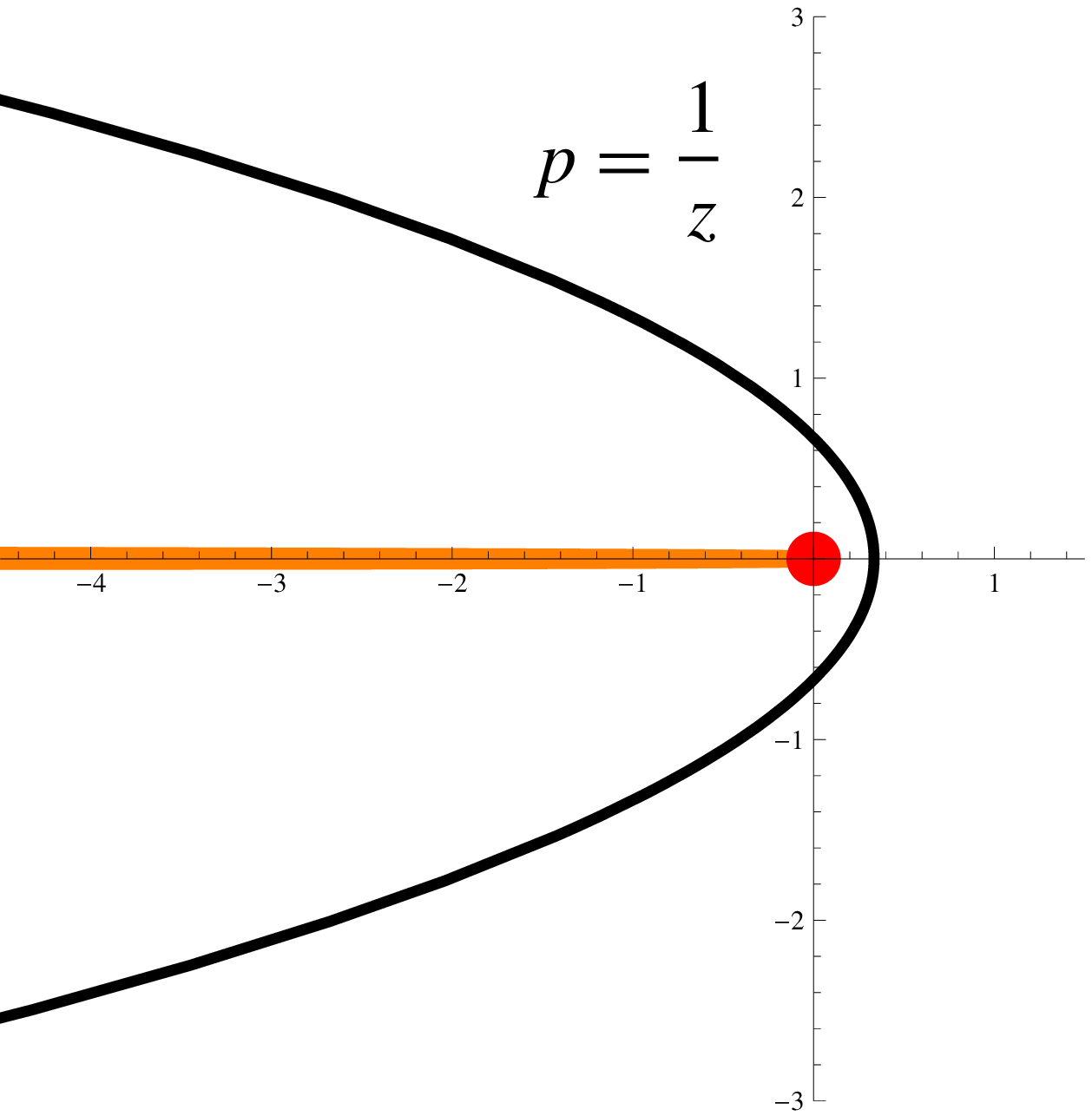}
}
\caption{\it Patterns of WKB lines in $AdS_3$, where we choose $\theta=\pi/2$. The two figures illustrate the behavior of WKB lines near zero and simple pole respectively.   WKB lines which end on zeros or poles are shown in orange color. Zeros are shown as blue points, while poles are denoted as red points.}
\label{fig-WKB-ads3}
\end{center}
\end{figure}

\begin{figure}[t]
\begin{center}
\subfigure{
\includegraphics[height=3.6cm]{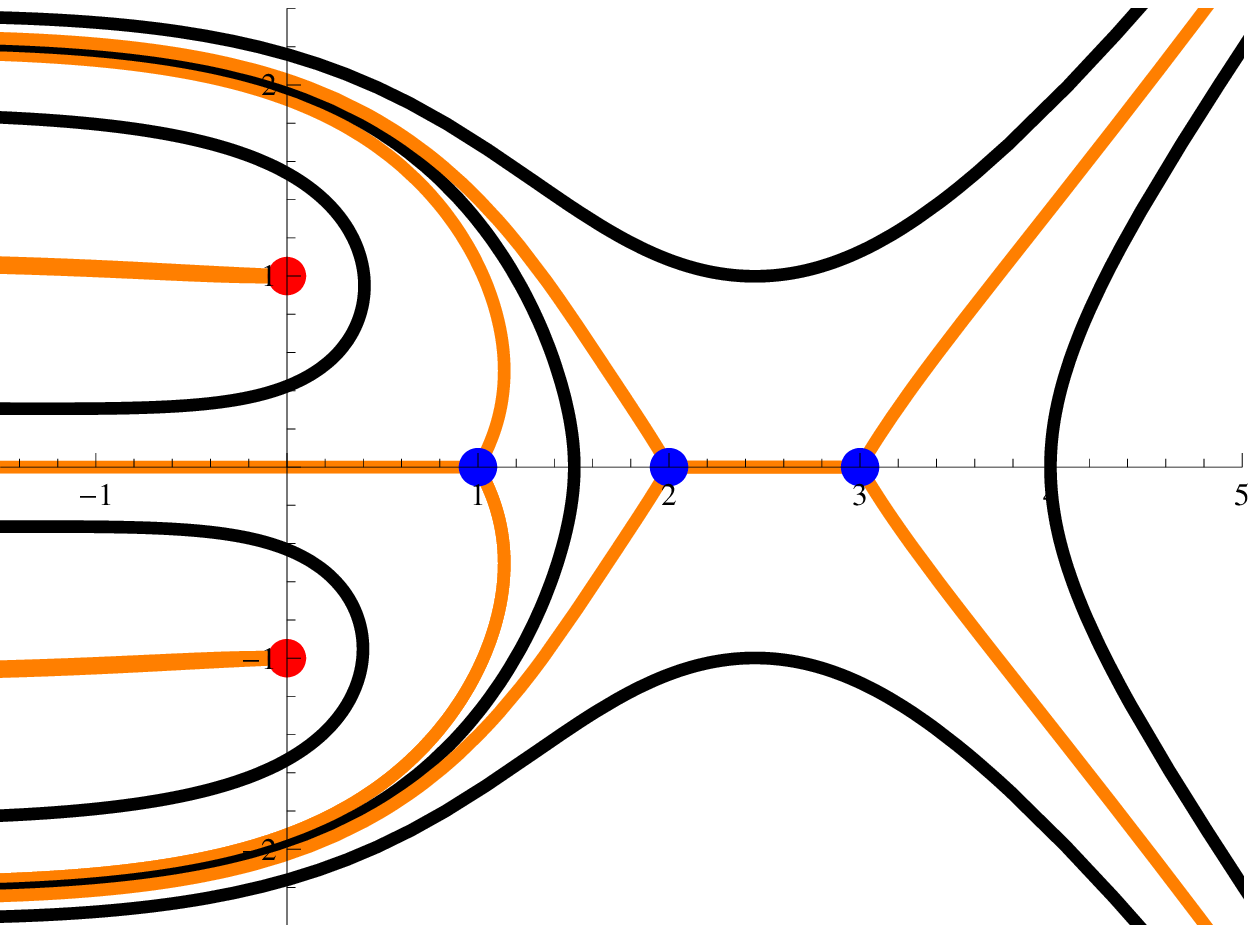}
} \hskip 2.5cm
\subfigure{
\includegraphics[height=3.6cm]{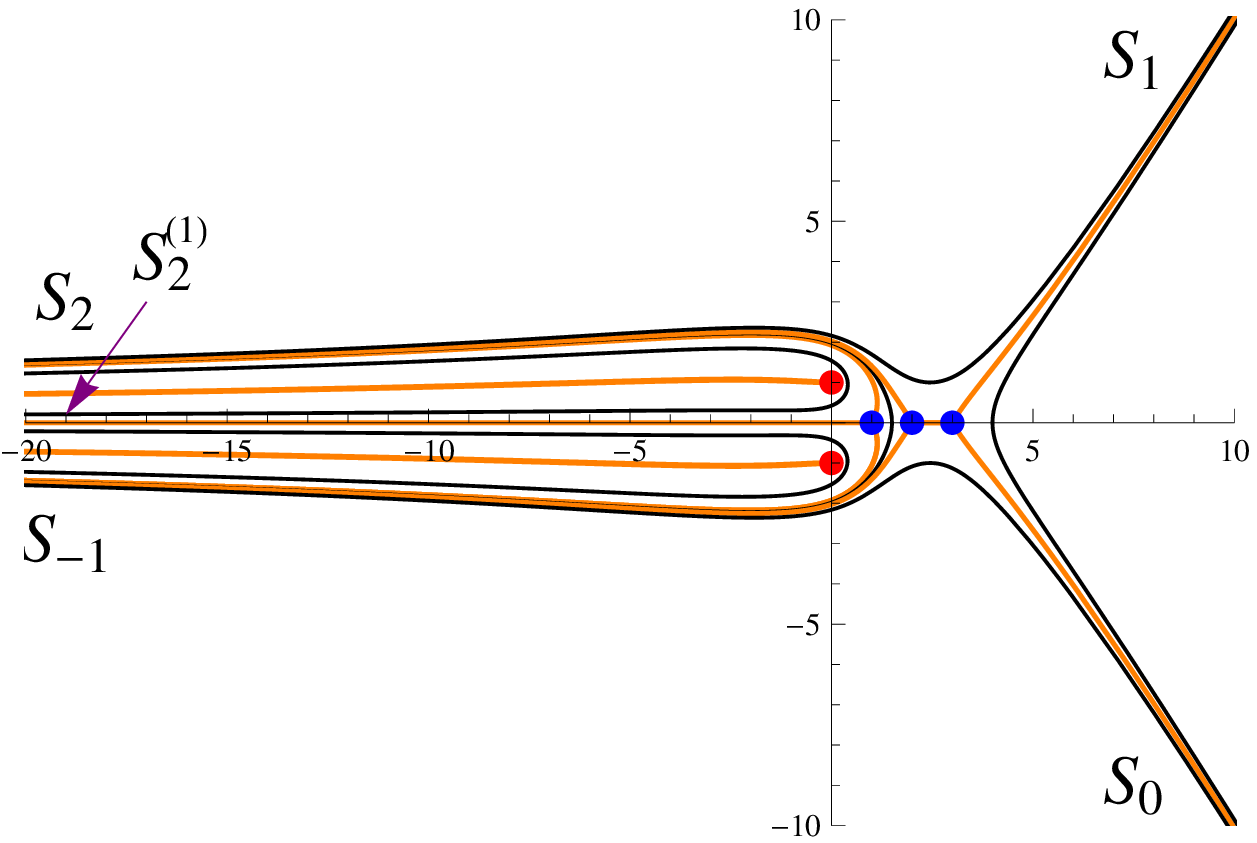}
}
\caption{\it WKB lines in $AdS_3$ for the polynomial $p(z)$=${(z-1)(z-2)(z-3) \over (z+i)(z-i)}$,  $\theta=\pi/2$. It corresponds to a six-point (${\hat n}$=3) form factor with two operators inserted. The second figure plots the WKB lines in a much larger range, which makes it obvious that there are three cusps at infinity.}
\label{fig-WKB-ads3-full}
\end{center}
\end{figure}

\begin{figure}[t]
\begin{center}
\subfigure{
\includegraphics[height=3.9cm]{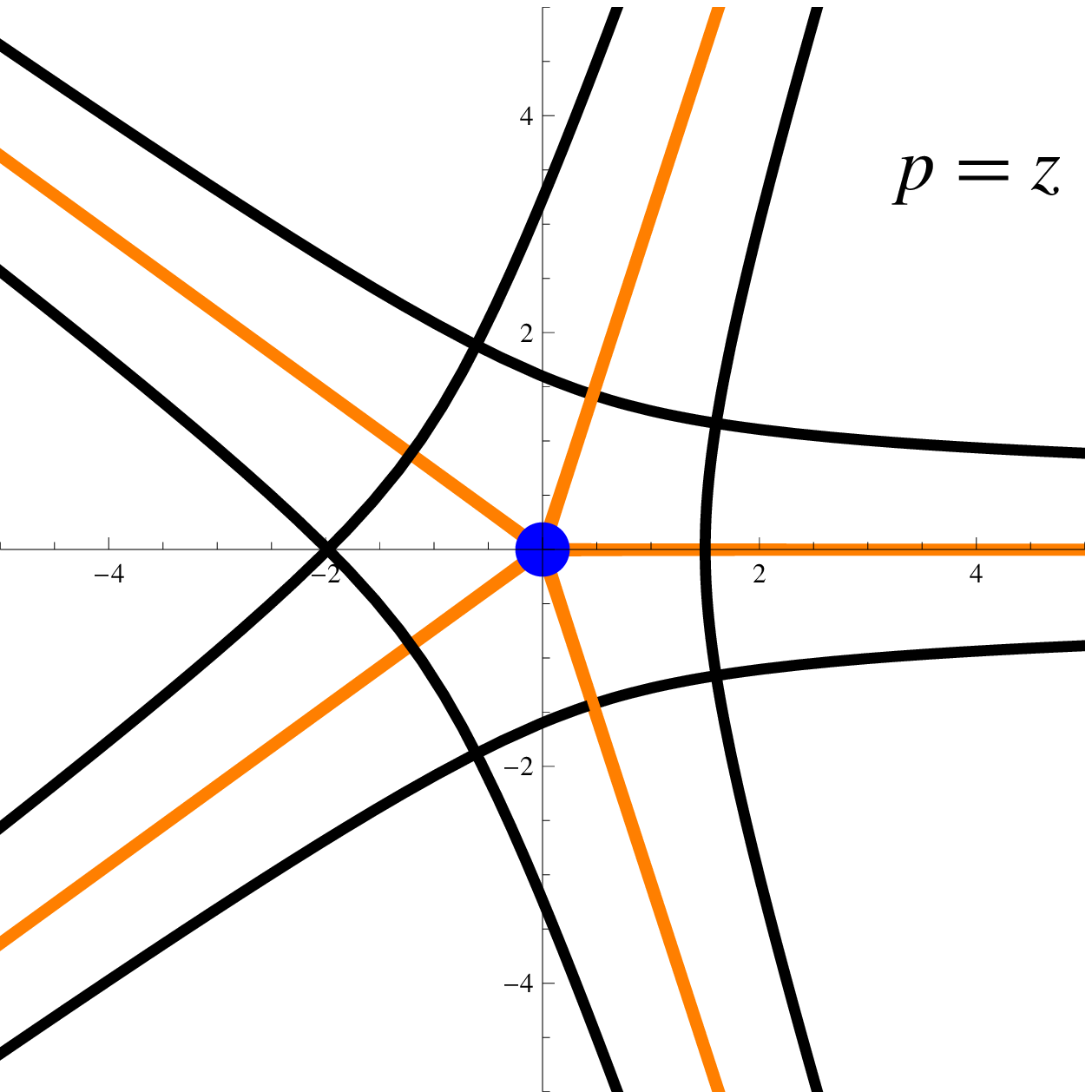}
} \hskip 1.2cm
\subfigure{
\includegraphics[height=3.9cm]{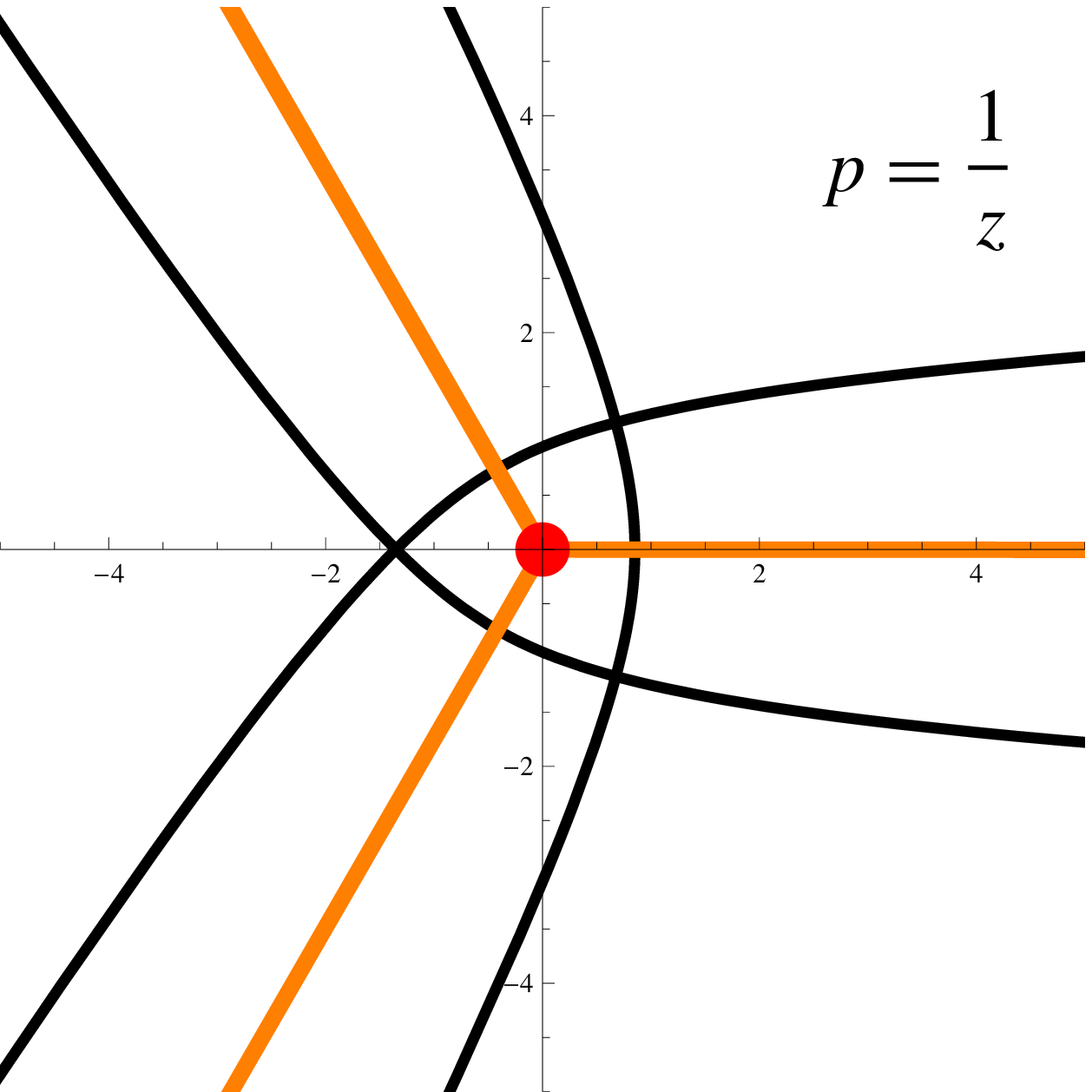}
}  \hskip 1.2cm
\subfigure{
\includegraphics[height=3.9cm]{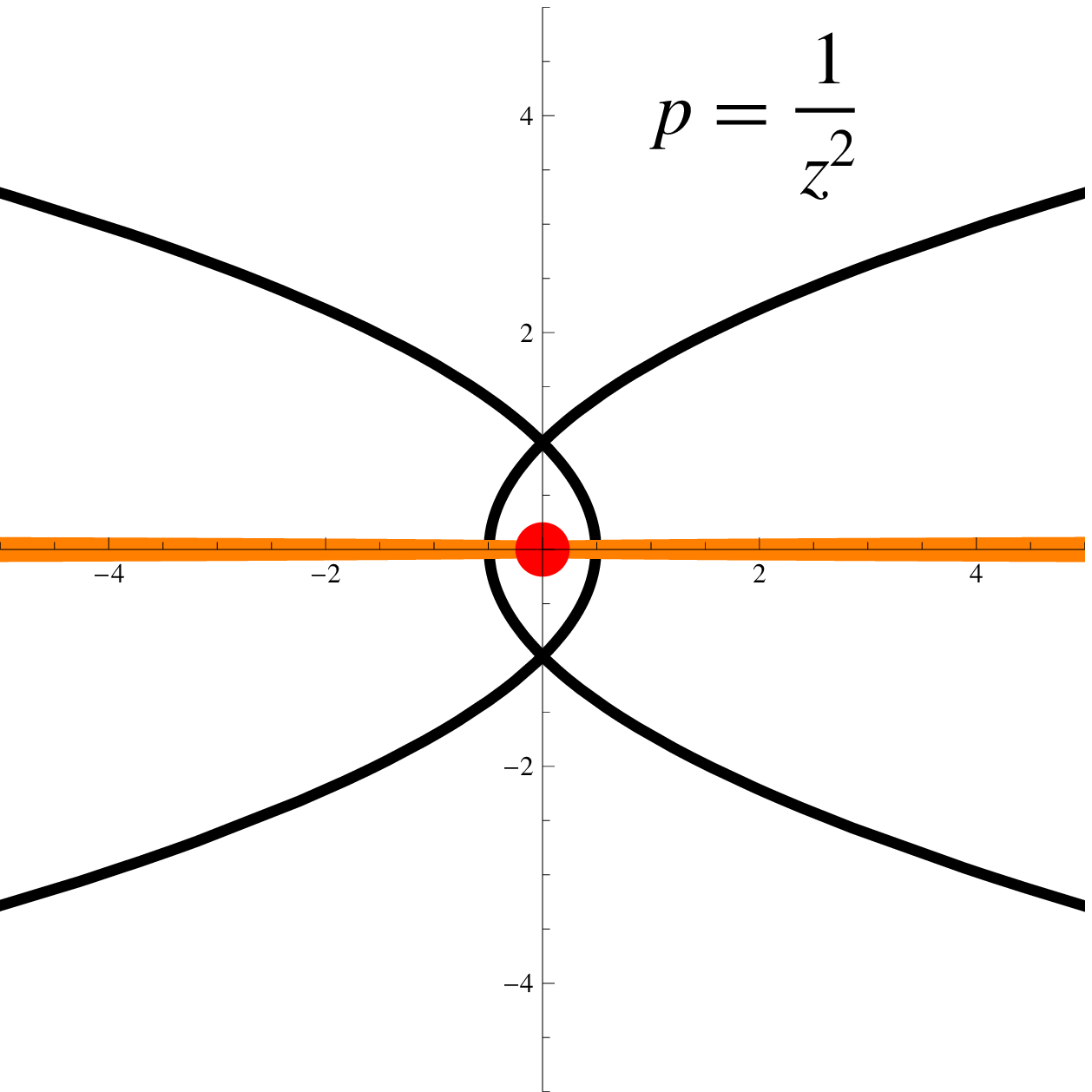}
}
\caption{\it Patterns of WKB lines in $AdS_5$, where we choose $\theta=0$. The three figures illustrate the behavior of WKB lines near zero, simple pole and double pole respectively. Zeros are shown as blue points, while poles are denoted as red points. Orange lines are special WKB  lines which end on zeros or poles.}
\label{fig-WKB-ads5}
\end{center}
\end{figure}

\begin{figure}[t]
\begin{center}
\subfigure{
\includegraphics[height=4.2cm]{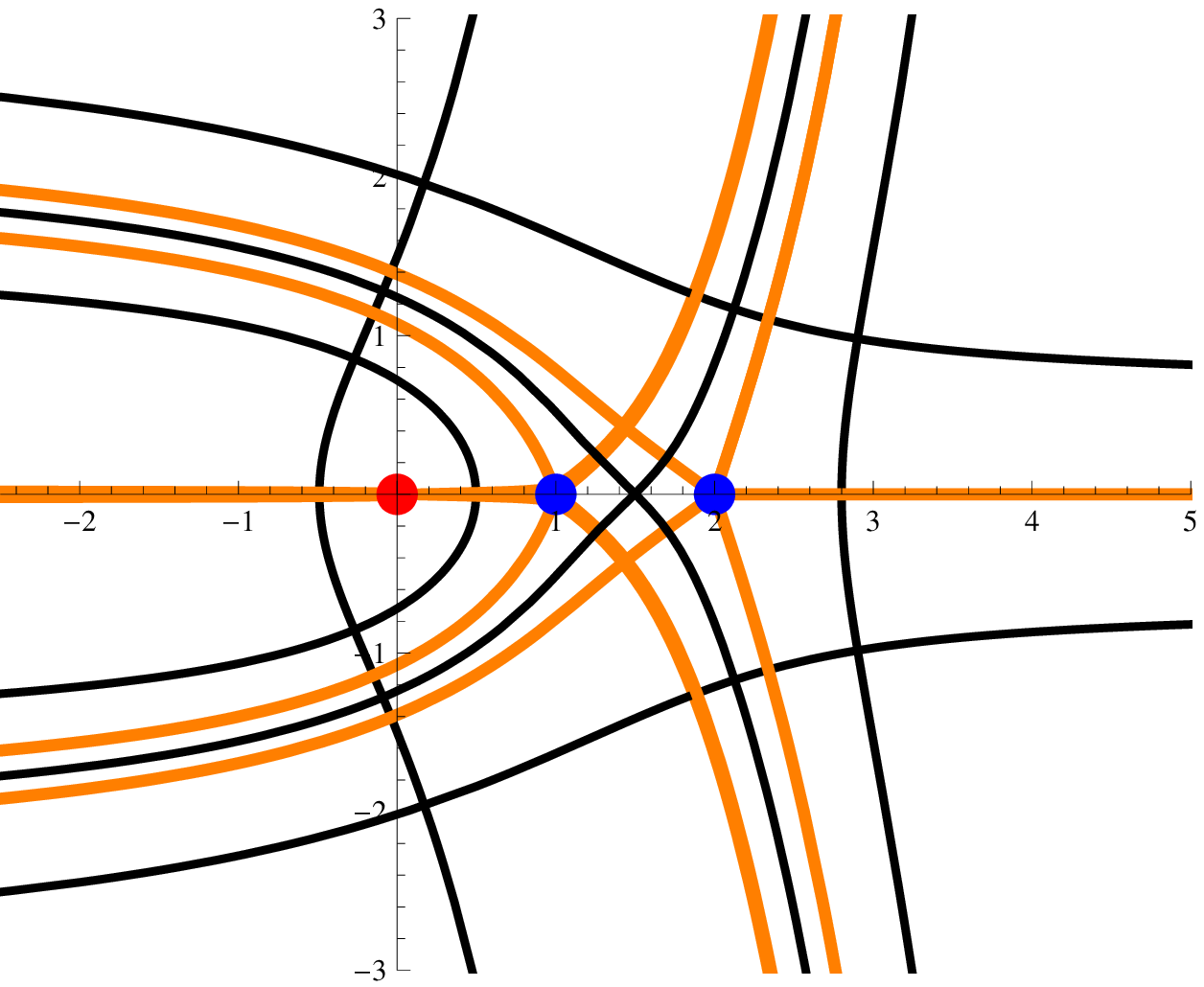}
} \hskip 2.5cm
\subfigure{
\includegraphics[height=4.2cm]{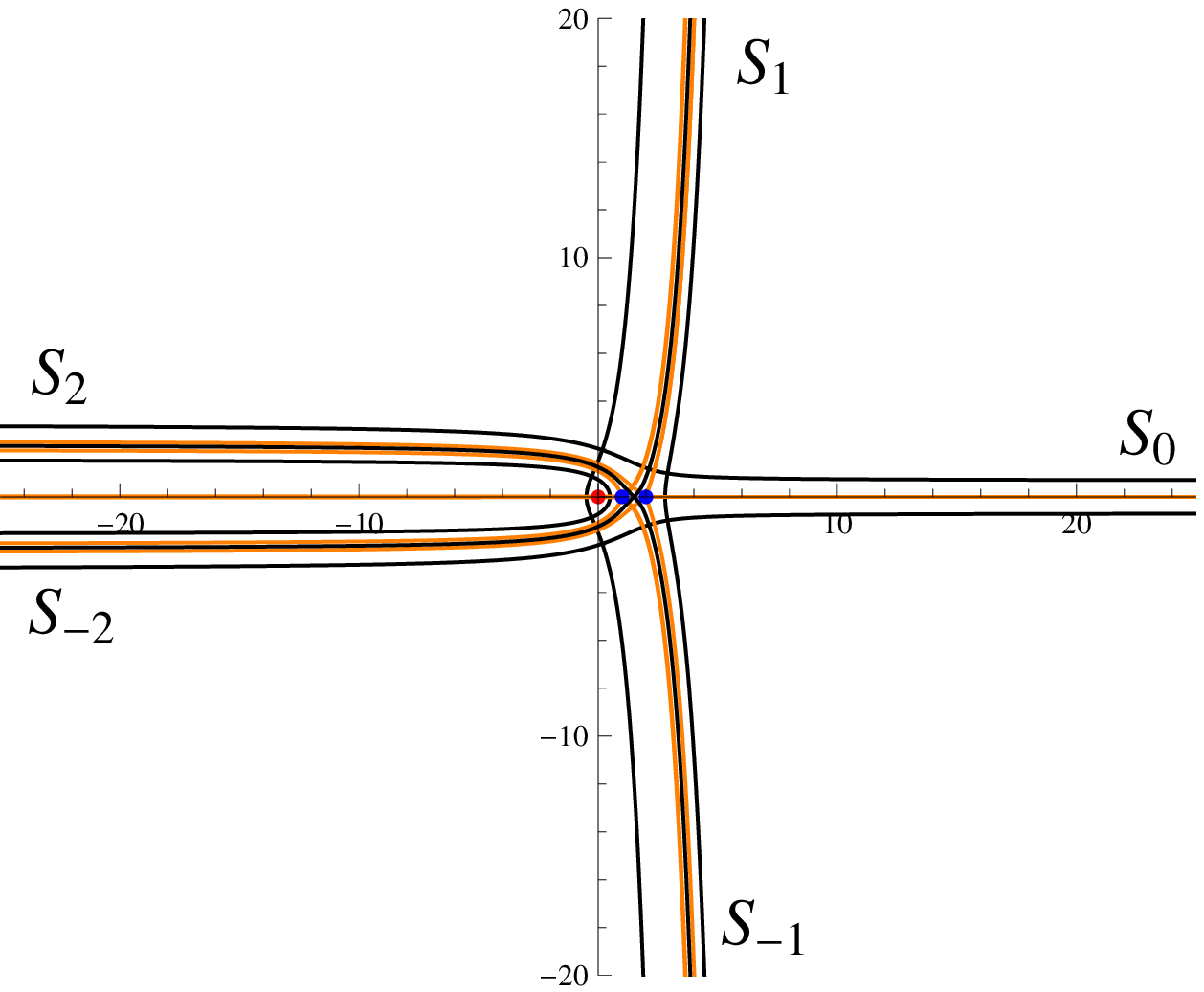}
}
\caption{\it WKB lines in $AdS_5$ for the polynomial $p(z)$=${(z-1)(z-2) \over z^2}$,  $\theta=0$. It corresponds to a four-point form factor. The second figure plots the WKB lines in a much larger range, which makes it obvious that there are four cusps at infinity.}
\label{fig-WKB-ads5-full}
\end{center}
\end{figure}

The WKB lines in $AdS_3$ have the following structure. For a general
point, there is only one WKB line going through. The special points
are the zeros and poles. There are three lines ending on each zero,
and one line ending on each (simple) pole. These are shown in Figure
\ref{fig-WKB-ads3}. The full WKB lines for the six-point form factor with
two-operator inserted are shown in Figure \ref{fig-WKB-ads3-full}.

The $AdS_5$ case corresponds to a $SU(4)$ Hitchin system. The WKB lines have more
complicated structures. For a general point, there are two WKB lines going
through\,%
\footnote{To be more precise, this is the picture projected on a single $z$-plane. $P(z)= x^4$ defines a Riemann surface with four branch covers. On each sheet there is only one WKB line going through each point. Four sheets give actually four lines. Two of them overlap with the other two (but with different orientations) \cite{AMSV10}. Projectively one gets the figures shown here. Similar picture applies for the $AdS_3$ case.}. There are five lines ending on each zero, three lines
ending on each simple pole and two lines ending on each double pole.
The WKB patterns are show in Figure \ref{fig-WKB-ads5}. The WKB lines for a four-point
form factor are given  in Figure
\ref{fig-WKB-ads5-full}.

One can associate small solutions to the asymptotic WKB lines, as labeled in Figure \ref{fig-WKB-ads3-full} and  \ref{fig-WKB-ads5-full}. Integrals over WKB lines (show as black lines in the figures) that connect different small solutions will provide the WKB approximation for the contraction of  small solutions.  For Y-functions, the corresponding WKB lines always form a closed contour. Therefore, the WKB approximation of Y functions in the limit of $\zeta\rightarrow0, \infty$ is given by cycle integrals, which are related to the mass parameters \cite{AMSV10}. They are related to the coefficients $a_i$ appearing in $P(z)$, and also implicitly related to the shape and periods of Wilson lines.

\section{Summary and discussions \label{sec-summary}}

In this paper, we study form factors in ${\cal N}=4$ SYM at strong coupling in $AdS_5$ and with multi-operator insertions. These are two non-trivial generalizations of the $AdS_3$ form factors studied in \cite{MZ10}.

The generalization to $AdS_5$  involves  new technical problems comparing to the $AdS_3$ case. The main challenge is how to introduce the truncation conditions with a more complicated $4\times4$ monodromy matrix, and how to write the system in a gauge invariant form, i.e. in terms of Y-functions. We clarify and solve these problems. The Y-system of  three-point case is constructed  explicitly, which potentially would have a connection to the strong coupling Higgs-to-3-gluon amplitudes in QCD.

The second generalization to the multi-operator insertion cases
provides a more interesting picture and we would like to make a few further
comments. The main hope is to provide a new technique to study
correlation functions. In doing so, we take an unconventional point
of view at strong coupling: to apply the on-shell techniques to compute
off-shell observables. Although similar ideas have
been studied in the weak coupling side, this point of view seems
have not been taken seriously at strong coupling.

According to the picture we proposed, adding one operator
corresponds to introducing a new monodromy matrix, which is 
taken as a condition imposed on small solutions related to null
cusps. The techniques developed for amplitudes or null Wilson loops
can be applied to the computation of such more general class of observables.
We construct the Y-system explicitly for the $AdS_3$ case with arbitrary
number of operator-insertions. The construction should in principle
be generalizable to $AdS_5$ based on the prescription of
the single-operator result developed in this paper.

The derivation of the Y-system is expected to be applicable for general
operators. Different operators would correspond to different kinds
of monodromies. The simplest case is for short operators which are dual
to light string states as being studied. For them, the monodromy depends only on the
momenta of the operators. 
This is actually very interesting, considering that normally it is  hard to study correlation functions with purely light operators at strong coupling besides the perturbative Witten diagram techniques, partially due to the complexity of string vertex operators in $AdS_5$ (see some ideas in \cite{Minahan12}). In our setup, the vertex operator information is in some sense encoded in the geometry via T-duality, and the problem becomes totally classical.
It would also be interesting to construct the monodromy for more general
operators, in particular classical solutions such as the GKP string
\cite{GKP02}.

Although our construction relies on the on-shell structure
of the observables, the multi-operator structure should in principle
contain all kinds of information in correlation functions.  In particular
from the general structure of OPE
\be {\cal O}_1(x) \, {\cal O}_2(y) \ = \ \sum_\alpha \, c_{12 \alpha} \, f(x-y, \partial_y) \, {\cal O}_\alpha(y) \, , \ee
an immediate step would be extracting the OPE coefficients by using form factors with two operator insertions and comparing it with form factor with the single operator ${\cal O}_\alpha$ in the OPE expansion.

Although our construction of $Y$ system does not rely on an exact knowledge of the string solutions, we do not have an explicit string T-dual picture of a form factor with multi-operator insertions. As  discussed in section \ref{sec-ppoly}, this should provide a better knowledge for the function $P(z)$. The T-dual picture of short string states was studied recently in \cite{KT12}, where a similar Wilson line picture was obtained. It would be interesting to understand the picture of interacting multi-closed-string states.

While there are lots of studies on correlation functions, we would like to point out particularly \cite{JW11, KK11, KK12, CT12}, where quite similar integrability techniques have been used. However, the detailed physical pictures and the building blocks are quite different. The method in those papers is limited to classical heavy operators (where the $S^5$ geometry plays also an important role), while our prescription is focused on short operators (but in principle could be for more general operators). It would be interesting to study the possible connection between the two pictures.

There are interesting algebraic curves appearing in the construction as discussed in Section \ref{sec-ppoly}.  Similar algebraic curves (the Seiberg-Witten curves) also appear in gauge theories \cite{GMN09}. It would be interesting to study their possible connections. There are also similar spectral curves for classical string solutions such as those studied in \cite{JL12}. There is an important difference though: while the spectral curve is a curve defined on the spectral parameter $\zeta$-plane,  the algebraic curve here is on the worldsheet $z$-plane.
On the other hand, in our picture there are close interplays between the two planes. It would be  interesting to understand the connection more explicitly.

We would also like to make a few comments on the symmetries.  Unlike amplitudes, form factors do not have dual conformal symmetry\,%
\footnote{This may be understood most easily in the T-dual picture, where a periodic Wilson line does not preserve special conformal symmetry.}. However, as we have seen that there is no problem to use integrability to compute strong coupling form factors.
Technically this may be understood that through changing of variables, one can bring spacetime quantities to a picture on the worldsheet, where the symmetries are in some sense enhanced and integrability techniques can be applied. It would be very interesting to study its possible correspondence at weak coupling, for example to have a realization of small solution picture at weak coupling, see an interesting proposal along this direction in \cite{SVW12}\,%
\footnote{See also \cite{FLMPS12} for an interesting idea of introducing spectral parameters for amplitudes at weak coupling.}.  In our opinion, it would be the symmetry of the theory rather than the symmetry of observables that plays the most important role.

Finally, let us mention that there are a few technical
problems to clarify. While we have obtained the Y-system and considered the WKB approximation, we have not discussed how to find the explicit solutions. The explicit integral form of the Y-system
equations are not given. The main complexity is due to the phases of some
functions appearing in the equation are outside the physical strip,
and extra pole contributions need to be considered. It is
also necessary to show how to derive the expression for the area.
For the $AdS_5$ form factors,  a natural expression for the non-trivial free energy part is
proposed, while for the multi-operator case a further study is necessary. There are also some issue about the $P(z)$ function of the multi-operator insertion case. These problems are under investigation and we hope to report them in the near future.


\section*{Acknowledgements}

We would like to thank Till Bargheer, Rutger Boels,  Andreas Brandhuber, Davide Fioravanti, Yasuyuki Hatsuda, Gregory Korchemsky,  Andrew Neitzke, Volker Schomerus, J\"{o}rg Teschner,  Gabriele Travaglini, Arkady Tseytlin, Congkao Wen and Chuanjie Zhu for useful discussions. This work was supported by the German Science Foundation (DFG) within the Collaborative Research Center 676 ``Particles, Strings and the Early Universe". GY would also like to thank the Gauge Theory as an Integrable System (GATIS) network for the support of traveling.

\appendix
\section{Hirota and Y-system equations \label{app-funeqs}}

We start with {\it CramerÕs rule}:
\be s_{i_1} \langle s_{i_2}, s_{i_3}, \cdots, s_{i_{k+1}} \rangle - s_{i_2} \langle s_{i_1}, s_{i_3}, \cdots, s_{i_{k+1}} \rangle + \cdots + (-1)^{k}  s_{i_{k+1}} \langle s_{i_1}, \cdots, s_{i_{k}} \rangle = 0 \, , \ee
where $s_i$ is a $k$ dimensional vector and the contraction is defined as
\be \langle s_1, s_2, \ldots, s_k \rangle \ := \ \epsilon^{\alpha_1 \ldots \alpha_k} s_{1,\alpha_1} \ldots s_{k,\alpha_k} \, . \ee
{\it Pl\"{u}cker relations} can be obtained by contracting the small solutions with another set of small solutions $s_{j_1}, \cdots, s_{j_{k-1}}$
\be \langle s_{j_1}, \cdots, s_{j_{k-1}}, s_{i_1} \rangle \langle s_{i_2}, s_{i_3}, \cdots, s_{i_{k+1}} \rangle + \cdots + (-1)^k \langle s_{j_1}, \cdots, s_{j_{k-1}}, s_{i_{k+1}} \rangle \langle s_{i_1}, \cdots, s_{i_{k}} \rangle = 0 \, . \ee
When $k=2$, one gets the Schouten identity
\be \label{schouten} \langle s_i, s_j \rangle \langle s_k, s_l \rangle + \langle s_i, s_k \rangle \langle s_l, s_j \rangle +  \langle s_i, s_l \rangle \langle s_j, s_k \rangle = 0 \, . \ee
When $k=4$, one obtains useful relations for the $AdS_5$ case,  such as the Wronskian relation
\be \label{schouten-ads5} \langle s_i ,s_j,  s_a, s_b \rangle \langle s_k, s_l,  s_a, s_b \rangle + \langle s_i ,s_k , s_a, s_b \rangle \langle s_l, s_j,  s_a ,s_b \rangle +  \langle s_i, s_l,  s_a, s_b \rangle \langle s_j ,s_k,  s_a, s_b \rangle = 0 \, . \ee

Below we  give the definition of T- and Y-functions, then we apply the above relations to obtain corresponding equations.


\subsection{The $AdS_3$ case}

We use the convention (note it is different from the $AdS_5$ case):
\be f^\pm \ := \ f(e^{\pm i{\pi\over2} }\zeta) , \qquad f^{[k]} \ := \ f(e^{i{k\pi\over2}} \zeta) \ . \ee
In this notation one has from  (\ref{Z2-relation})
\be
\langle s_{i+1}, s_{j+1} \rangle \ = \ \langle s_i, s_j \rangle^{[2]} \, .  \ee
The T-functions are defined as
\be  \begin{array}{lll} T_{1,2m+1} \ := \ \langle s_{-m-1}, s_{m+1}\rangle, & \quad\qquad &
T_{1,2m} \ := \ \langle s_{-m-1}, s_{m}\rangle^+,
 \\
T_{0,2m} \ := \ \langle s_{-m-1}, s_{-m}\rangle, & \quad\qquad &
T_{0,2m+1} \ := \ \langle s_{-m-2}, s_{-m-1}\rangle^+,
 \\
 T_{2,2m} \ := \ \langle s_{m}, s_{m+1}\rangle, & \quad\qquad &
T_{2,2m+1} \ := \ \langle s_{m}, s_{m+1}\rangle^+.
 \end{array} \ee
$T_{1,m}$ is non-zero for $m=1,...,n-1$, and the normalization
$\langle s_0, s_1\rangle =1$ corresponds to a gauge choice
$T_{0,m}=T_{2,m}=1$.

Using Schouten identity (\ref{schouten}), one can obtain the so-called Hirota
equations
\be T^+_{a,m}T^-_{a,m} \ = \ T_{a,m-1}T_{a,m+1}+T_{a-1,m}T_{a+1,m} \, , \ee
where the indices $a,m$ take integer values.

Hirota equations contain huge gauge redundancies
\be T_{a,m}(\zeta) \ \rightarrow \ \prod_{\alpha,\beta=\pm}
g_{\alpha\beta}(e^{{i\pi \over 2}(\alpha a + \beta m)}\zeta)
T_{a,m}(\zeta) \, , \ee
where $g_{\alpha\beta}(\zeta)$ are four arbitrary functions. Like defining field strength in gauge theory, one can introduce gauge invariant functions, so-called Y-functions
\be Y_{a,m} \ : = \ {T_{a,m-1}T_{a,m+1} \over T_{a-1,m}T_{a+1,m}} \, . \ee
The Hirota equations become the equations of Y-functions
\be {Y^+_{a,m}Y^-_{a,m} \over
Y_{a-1,m}Y_{a+1,m}} \ = \ {(1+Y_{a,m-1})(1+Y_{a,m+1}) \over
(1+Y_{a-1,m})(1+Y_{a+1,m})} \, . \ee
In the normalization
$\langle s_i, s_{i+1}\rangle =1$ the equations of Y-functions simplify as
\be Y^+_m Y^-_m \ = \ (1+Y_{m-1})(1+Y_{m+1}) \, , \ee
where $Y_m := Y_{1,m}$\,.

\subsection{The $AdS_5$ case}

We use the convention:
\be f^\pm \ := \ f(e^{\pm i{\pi\over4} }\zeta) \, , \qquad f^{[k]} \ := \ f(e^{i{k\pi\over4}} \zeta) \, . \ee
Useful relations due to the $Z_4$ automorphism are
\bea && \langle s_{j-1} , s_j , s_{k-1}, s_k \rangle^{[2]} \ = \ \langle s_j ,
s_{j+1}, s_k ,s_{k+1} \rangle \, , \\ &&  \langle s_k, s_{j-2} ,
s_{j-1}, s_j \rangle^{[2]} \ = \ \langle s_k ,s_{k+1} ,s_{k+2} , s_j
\rangle \,
. \eea
Define the T-functions as
\bea 
& & T_{0,m}(\zeta) \ := \ \langle s_{-2}, s_{-1} , s_{0}, s_{1}\rangle^{[-m-1]} \, , \nonumber\\
& & T_{1,m}(\zeta) \ := \ \langle s_{-2}, s_{-1} , s_{0}, s_{m+1}\rangle^{[-m]} \nonumber , \\
& & T_{2,m}(\zeta) \ := \ \langle s_{-1}, s_{0} , s_{m+1}, s_{m+2}\rangle^{[-m-1]} \ ,
 \\
& & T_{3,m}(\zeta) \ := \ \langle s_{-1}, s_{m} , s_{m+1}, s_{m+2}\rangle^{[-m]} \ , \nonumber\\
& & T_{4,m}(\zeta) \ := \ \langle s_m, s_{m+1} , s_{m+2}, s_{m+3}\rangle^{[-m-1]} \, .
\nonumber
\eea
Using the Wronskian relation, one obtains the Hirota equations
\be T^+_{a,m}T^-_{4-a,m} \ = \ T_{4-a,m+1}T_{a,m-1}+T_{a+1,m}T_{a-1,m} ,
\ \ \ a=1,2,3\, .  \ee
Gauge invariant Y-functions can be defined similarly as
\be Y_{a,m} \ := \ {T_{a,m+1}T_{4-a,m-1} \over T_{a+1,m}T_{a-1,m}} \, .
\ee
The Hirota equations become the Y-system equations:
\be {Y^-_{a,m}Y^+_{4-a,m} \over Y_{a+1,m}Y_{a-1,m}} \ = \
{(1+Y_{a,m+1})(1+Y_{4-a,m-1}) \over (1+Y_{a+1,m})(1+Y_{a-1,m})} , \
\ \ a=1,2,3 \ . \ee
%

\section{Twistor variables \label{app-twistor} }

In this appendix we give a brief review on (momentum) twistor variables, see for example \cite{Hodges, MS09}. 
One technical point we would like to clarify is  how to transform twistor variables into Lorentz variables, and vice verse.

We first recall the relation between embedding and Poincar\'{e} coordinates
\be \label{embed_poincare} X^\mu \ = \ {x^\mu \over r} , \qquad   X^+ \ = \ {1 \over r}  , \qquad
X^- \ = \ {r^2 + x^\mu x_\mu \over r}\, , \ee
where ($\eta_{\mu\nu}=(-1,1,1,1)$)
\be - 1 \ = \  -X^+ X^- + X^\mu X_\mu \, ,  \qquad X^\pm \ := \ X^{-1} \pm X^4 \, .\ee

Twistor variables can be understood as in the spinor representation of the embedding $SO(2,4)$ space, which is a $SU(4)$ fundamental representation, denoted by $\lambda$,
\be X_{ab} \ = \ \Gamma^A_{ab} \cdot X_A = \lambda_{[a} \, \lambda_{b]} \, ,  \qquad \lambda_{[a} \, \lambda_{b]} \ := \ \lambda_a \lambda_b - \lambda_b \lambda_a \, .  \ee
With one explicit choice of gamma matrices, one has
\be \label{X_lambda} X_{ab} \ = \ {1\over \sqrt{2}} \, \begin{pmatrix} 0 & i\, X^+ &  X^2 + i X^3 & X^1 - X^0 \\
 * & 0 & X^1 + X^0 & -X^2 + i X^3 \\
 * & * & 0 & i\,X^-
 \\  * & * & * & 0
\end{pmatrix} \ = \ \lambda_{[a} \, \lambda_{b]} \, . \ee
Note that $\det(X)=(X \cdot X)^2/4$. Due to the freedom of choosing normalization, twistor variables $\lambda_i$ are projective coordinates in ${\rm CP}^3$.

Using (\ref{embed_poincare}) and (\ref{X_lambda}), one can obtain the relations between $\lambda_a$ and $x^\mu$.  For example, define
\be x^\pm \ = \ x^1 \pm x^0 \, , \ee
one has the  relations
\be \label{xpmtotwistor}  x^-_j \ = \  i\, { X_{j,14} \over X_{j,12}} = i \, { \lambda^j_{[ 1} \, \lambda^{j+1}_{4 ]} \over \lambda^j_{[ 1} \,\lambda^{j+1}_{2 ]} } \ , \qquad  x^+_j \ = \ i\, { X_{j,23} \over X_{j,12}} = i\, { \lambda^j_{[ 2} \, \lambda^{j+1}_{3 ]} \over \lambda^j_{[ 1} \, \lambda^{j+1}_{2 ]} } \ . \ee

Another description of so-called momentum twistors,  which was first introduced  at weak coupling \cite{Hodges}, is practically more useful\,%
\footnote{It is called momentum twistor just because it is defined in the momentum space, mathematically it is not different from usual twistor.}. Consider a null Wilson line configuration  defined in the momentum space of amplitudes or form factors
\be x_{i+1} - x_{i} \ = \ p_i \, , \qquad p_i^2 \ = \ 0 \, , \qquad p_{i,\alpha\dot\alpha} \ = \ \Lambda_{i,\alpha} \, \tilde\Lambda_{i,\dot\alpha} \, , \ee
where the left and right-hand $SU(2)$ Weyl spinors are denoted by $\Lambda$ and $\tilde\Lambda$.
We define $x_{ij} := x_i - x_j$. The Weyl spinor contractions are defined as
\bea && \langle i, j \rangle \ := \ \epsilon^{\alpha\beta} \Lambda_{i,\alpha} \Lambda_{j,\beta} \, , \quad \quad \ \ \, \ \ \qquad  [ i, j ] \ := \ \epsilon^{\dot\alpha\dot\beta} \tilde\Lambda_{i,\dot\alpha} \tilde\Lambda_{j,\dot\beta} \, , \\ && {\langle i | \, x\, y \, | j \rangle } \ := \ \Lambda_{i}^{\alpha}\, x_{\alpha\dot\alpha} \,y^{\dot\alpha\beta} \,\Lambda_{j,\beta} \, , \qquad {\langle i | \, x\, | j ] } \ := \ \Lambda_{i}^{\alpha}\, x_{\alpha\dot\beta}  \,\tilde\Lambda_{j}^{\dot\beta} \, . \eea

The momentum twistors can be explicitly defined as follows
\be \label{defmomtwistor}   \lambda_{j} \ = \ \big( \Lambda_{j,\alpha}, \,  \mu_{j,\dot\alpha} \big) , \qquad \mu_{j,\dot\alpha} \ := \ - i\,(x_j \cdot \Lambda_j)_{\dot\alpha} \ = \ -i\, \epsilon^{\alpha\beta} x_{j,\alpha\dot\alpha} \Lambda_{j,\beta} \, . \ee
Note that also $\mu_{j,\dot\alpha} = -i\, (x_{j+1}\cdot\Lambda_j)_{\dot\alpha}$\,.
The contraction of twistors is defined as
\be \langle \lambda_i, \lambda_j, \lambda_k, \lambda_l \rangle \ := \ \epsilon^{abcd} \lambda_{i,a} \lambda_{j,b} \lambda_{k,c} \lambda_{l,d} \, , \qquad a = (\alpha,\dot\alpha) \ . \ee

The geometric picture of twistor space is that, each spacetime point corresponds to a line in twistor space determined by two twistor variable,  $x_i \sim X_i \sim \lambda_{i-1} \wedge \lambda_{i}$. If two spacetime points are null separated, the corresponding two lines in twistor space  intersect with each other.
This is obvious in the above definition since null-separated $x_i$ and $x_{i+1}$ both contain $\lambda_{i}$.

To write the contractions of twistor variables in terms of Lorentz coordinates, one practically very useful formula is \cite{MS09} 
\be \label{master} {\langle i | \, x_{i, j}\, x_{j,k} \, | k \rangle } \ = \ {\langle
\lambda_i,\lambda_{j-1},\lambda_j,\lambda_k\rangle \over \langle j-1,j\rangle } \, . \ee
For example, using (\ref{master}) it is easy to obtain the relation
\be x_{i,j}^2 \ = \ {\langle \lambda_{i-1},\lambda_{i},\lambda_{j-1},\lambda_{j} \rangle \over \langle i-1,i\rangle \langle j-1, j \rangle} \, .\ee
Furthermore, any normalization independent expression of twistor contractions can be written in terms of Lorentz variables. For example, for the  ratio variables appearing in form factors, one has
\bea && {\langle \lambda_1,\lambda_2,\lambda_3, \hat\Omega \lambda_{4} \rangle \over   \langle \lambda_1,\lambda_2,\lambda_3, \lambda_4\rangle} \ = \ {\langle \lambda_1,\lambda_2,\lambda_3,\lambda_{4+n} \rangle \over b_4 \, \langle \lambda_1,\lambda_2,\lambda_3, \lambda_4\rangle} \ = \ {\langle 1 | p_2 \,(q+p_3) | 4\rangle \over \langle 1| p_2 \, p_3 | 4 \rangle} \, , \\
 && {\langle \hat\Omega\, \lambda_{1}, \lambda_2,\lambda_3,\lambda_4 \rangle \over  \langle \lambda_1,\lambda_2,\lambda_3, \lambda_4\rangle} \ = \ {\langle \lambda_{n+1},\lambda_2,\lambda_3,\lambda_4 \rangle \over b_1 \, \langle \lambda_1,\lambda_2,\lambda_3, \lambda_4\rangle} \ = \ {\langle 1 | (-q+p_{12})\, p_3 | 4\rangle \over \langle 1| p_{12} \, p_3 | 4 \rangle} \, ,
\eea
where we also use the relation (\ref{lambda_periodic}) $\lambda_{i+n} = b_i \, \hat\Omega\, \lambda_i$.

\section{Monodromy with a different basis \label{app-omegabar}}

In this section we briefly explain the monodromy defined in a different basis of small
solutions, in particular how $\overline\Omega$ is related to $\Omega$.

First recall the definition of the monodromy
\be
\begin{pmatrix} s_1 \\
s_0 \\ s_{-1} \\ s_{-2}
\end{pmatrix} (z e^{2\pi i} , \zeta) = \Omega^{-1}(\zeta) \begin{pmatrix} s_1 \\
s_0 \\ s_{-1} \\ s_{-2}
\end{pmatrix} (z , \zeta) \, , \ \
\begin{pmatrix} s_2 \\ s_1 \\
s_0 \\ s_{-1}
\end{pmatrix} (z e^{2\pi i} , \zeta)  = (\overline\Omega^{[2]})^{-1}(\zeta)  \begin{pmatrix} s_2 \\ s_1 \\
s_0 \\ s_{-1}
\end{pmatrix} (z , \zeta)   \, ,
\ee
and
\bea  \begin{pmatrix} s_{n+1} , s_n , s_{n-1} , s_{n-2}
\end{pmatrix}^T (z, \zeta) &=& {\cal B}^{-1}(\zeta)  \begin{pmatrix} s_1
,s_0 , s_{-1} , s_{-2} \end{pmatrix}^T (z e^{-2\pi i}, \zeta) \, ,
\nonumber \\
   \begin{pmatrix} s_{n+2} , s_{n+1} , s_n , s_{n-1}
\end{pmatrix}^T (z, \zeta) &=& ({\overline{\cal B}}^{[2]})^{-1}(\zeta)   \begin{pmatrix} s_2, s_1
,s_0 , s_{-1}  \end{pmatrix}^T (z e^{-2\pi i}, \zeta) \, , \eea
where using the relations of (\ref{Zautomorphism}), the proportionality
constants are given by a single $B$ function
\bea {\cal B}^{-1} &=& {\rm diag} \big\{ B , (B^{[2]})^{-1},
B^{[-4]}, (B^{[-2]})^{-1} \big\} \, , \nonumber \\
{\overline{\cal B}}^{-1} &=& {\rm diag} \big\{ (B^{[4]})^{-1} , B^{[-2]},
B^{-1}, B^{[2]} \big\} \, . \eea
One can expand $s_{-2}$ in terms of $\{s_{-1}, s_0, s_1, s_2\}$, then
one gets
\be s_{-2}\ = \ T_{1,1}^{[-1]} \,  s_{-1} - T_{2,1} \, s_0 + T_{1,1}^{[1]} \, s_1 -
s_2 \, . \ee
Similarly for $s_{n-2}$ one has the expansion
\be s_{n-2} \, = \, \langle s_{n-2}, s_n, s_{n+1}, s_{n+2}\rangle \, s_{n-1} -
T_{2,1}^{[2n-4]}  \, s_{n} +  \langle s_{n-2}, s_{n-1}, s_n, s_{n+2}\rangle \,
s_{n+1} - s_{n+2} \, , \ee
where the other two contractions are $T_{1,1}$ or $T_{3,1}$
depending on whether $n$ is even or odd.

By introducing
\bea M &=& \begin{pmatrix}
 0 & 1 & 0 & 0 \\ 0 & 0 & 1 & 0 \\  0 & 0 & 0 & 1  \\ -1 & T_{1,1}^{[1]} & -T_{2,1} & T_{1,1}^{[-1]} \end{pmatrix} , \
\\ M' &=& \begin{pmatrix}
 0 & 1 & 0 & 0 \\ 0 & 0 & 1 & 0 \\  0 & 0 & 0 & 1  \\ -1 &  \langle s_{n-2}s_{n-1}s_ns_{n+2}\rangle & -T_{2,1}^{[2n-4]} & \langle s_{n-2}s_ns_{n+1}s_{n+2}\rangle  \end{pmatrix} ,
\eea
one can obtain
\be (\overline\Omega^{[2]}) \ = \ \overline{B}^{[2]} \, M'^{-1} \, B^{-1} \, \Omega \, M \, . \ee
%


\end{document}